# PASER: A Physics-Inspired Theory for Stimulated Growth and Real-Time Optimization in On-Demand Platforms

Ioannis Dritsas

*Oxford Brookes Business School*

email: idrits@gmail.com

January 24, 2025

## Abstract

This paper introduces an innovative framework for understanding on-demand platforms by quantifying positive network effects, trust, revenue dynamics, and the influence of demand on platform operations at per-minute or even per-second granularity. Drawing inspiration from physics, the framework provides both a theoretical and pragmatic perspective, offering a pictorial and quantitative representation of how on-demand platforms create value. It seeks to demystify their nuanced operations by providing practical, tangible, and highly applicable metrics, platform design templates, and real-time optimization tools for strategic what-if scenario planning. Its model demonstrates strong predictive power and is deeply rooted in raw data. The framework offers a deterministic insight into the workings of diverse platforms like Uber, Airbnb, and food delivery services. Furthermore, it generalizes to model all on-demand service platforms with cyclical operations. It works synergistically with machine learning, game theory, and agent-based models by providing a solid quantitative core rooted in raw data, based on physical truths, and is capable of delivering tangible predictions for real-time operational adjustments. The framework's mathematical model was rigorously validated using highly detailed historical data retrieved with near 100% certainty.

Applying data-driven induction, distinct qualities were identified in big data sets via an iterative process. Through analogical thinking, a clear and highly intuitive mapping between the elements, operational principles, and dynamic behaviors of a well-known physical system was established to create a physics-inspired lens for Uber. This approach shed new light not only on its macroscopic picture but importantly on its inner workings and micro-dynamics. Subsequently, using a reductionist approach, an elegant, simple, and highly accurate mathematical model was formulated. This novel quantitative framework was named PASER (Profit Amplification by Stimulated Emission of Revenue), drawing an analogy to its physical counterpart, the LASER (Light Amplification by Stimulated Emission of Radiation).

PASER provides a highly appealing theoretical foundation not only for the understanding of but also for the accurate quantification of dynamic network effects, dynamic resource allocation, trust dynamics across the platform, demand-side induced influences, revenue generation, and profit extraction. Previously, these aspects of successful platforms like Uber and Airbnb have been studied mostly in aggregate. PASER offers a quantitative understanding that can facilitate platform engineering. As a result of its quantitative approach, the framework also provides a practical tool for platform operations optimization (with real-time capabilities) and profit maximization. It offers an in-depth understanding of on-demand digital platform aspects such as network effects, growth thresholds, consistency in profit generation, operations efficiency, real-time surge pricing, and scalability. Overall, it provides a tangible understanding of why and how digital platforms transition from traditional "lamp-like" revenue generation (very lossy mechanism) to "laser-like" revenue emission (highly coherent profit generation with minimal losses).

## Table of Contents

# 1. INTRODUCTION

This section outlines the key issue and overall aim of the paper. It introduces the fundamental motivation for using a physics-inspired quantitative framework to study positive network effects, trust, and revenue generation in on-demand B2C platforms. A discussion is provided on why a deeper, more transparent modeling approach is valuable for both current and emerging digital ecosystems.

## 1.1 KEY ISSUE AND OVERALL AIM

### 1.1.1 Key issue

This paper addresses inefficiencies in real-time, highly predictive quantitative modeling at per-second granularity for network effects, trust, and revenue on digital on-demand platforms. The study also addresses the lack of practical real-time quantitative optimization tools and insights into machine learning models. These models are perceived as black boxes, offering no tangible understanding of the platform's essence and operations. Furthermore, existing models lack holistic quantitative capability to approach platforms from a mathematically elegant and systemic angle. Viewing them through a physics-based quantitative lens would synergistically enable platform engineering and qualitative orchestration.

But why physics-based? What does this offer? A physics-based approach brings a well-established framework for modeling complex systems with interacting components, feedback loops, and emergent phenomena (exactly the kinds of features observed in on-demand digital platforms). Physics offers mathematically elegant, conservation-based principles (e.g., conservation of mass/energy) and proven formalisms for capturing thresholds, critical masses, and nonlinear interactions. By borrowing these concepts, platform operations can be analyzed more holistically, moving beyond black-box algorithms to a transparent model of how network effects, trust, and revenue generation interrelate in real time. Viewing the key issue from this perspective not only promotes platform understanding but also enables real-time predictive modeling and optimization. In doing so, it provides a rigorous theoretical foundation for engineering platform-level solutions, particularly relevant as autonomous technologies and decentralized ownership models (e.g., robotaxis) reshape future on-demand services.

The rise of autonomous technologies (e.g., robotaxis, humanoid robots) offers new socioeconomic opportunities for on-demand digital platforms. Independent ride-hailing drivers could be empowered to own autonomous vehicle fleets, addressing anticipated job displacement from automation.

### 1.1.2 Overall Aim

The overall aim is to explore the intricacies of how massively on-demand platforms, like the Uber ride-hailing platform, stimulate and sustain positive networks to monetize them. To dive into the complexities of quantifying their mechanism of revenue amplification and surface with a deterministic model that demystifies their intricate operations. This paper also aims to ground its findings in raw data and inductively emerge with a theory that interprets and predicts on-demand platform operations dynamically.

Rather than competing with machine learning, agent modeling, and game theory, this paper aims to complement and contribute to their grounding in physical truths. Meaning that it also helps filter out ML hallucinations and serves as a synthetic data generator for ML model training.

This overall aim involves proposing a physics-inspired digital platform theory that provides analytical tools to promote platform understanding and optimization. A generalizable framework is to be formulated as a general theory of on-demand platforms, enabling platform engineering and optimization, as well as real-time predictive modeling for surge pricing and platform stability. The proposed framework also aspires to predict the dynamic operations of futuristic platforms like autonomous ride hailing and domestic robot hailing.

### 1.2 BACKGROUND CONTEXT

Digital platforms have been widely characterized as digital ecosystems that enable interactions among user groups, often serving as intermediaries that harness network effects (Constantinides et al., 2018; Hagiu and Wright, 2015). They customarily employ layered, modular architectures and governance models to facilitate value creation across industries (Chen et al., 2021; Constantinides et al., 2018). It has been demonstrated by platforms such as Uber and Airbnb that conventional business models can be disrupted through positive network effects across demand and supply sides, though their rapid expansion has introduced barriers to the development of decentralized ecosystems intended to broaden efficiency and inclusion (Muzellec et al., 2015; Rietveld and Schilling, 2021).

Recent advances in robotics and autonomous vehicles are reshaping the socioeconomic landscape. They are creating new possibilities for platform operators, automation providers, and technology manufacturers (Chen et al., 2021). Operators, including Uber drivers, are enabled by these technological advances to become managing owners of autonomous fleets, thereby moderating the job displacement (Ao et al., 2024). This paper demonstrates that platforms can evolve into multimodal and cooperative ecosystems, aided by governance structures that are neither wholly centralized nor entirely decentralized, but instead strike a balance for optimal performance (Chen et al., 2021). It also develops tools for designing, optimizing, and strategically testing such multimodal platform setups, underscoring a win-win vision for the digital platform economy by quantifying platform fundamentals.

Multidisciplinary foundations are incorporated by linking a physics-inspired theoretical framework with the socioeconomic dimensions of B2C digital platforms such as Uber and Airbnb, and B2B2C arrangements like Uber Eats. This novel context encourages new environments for platform design, optimization, and strategic scenario analysis (Gawer, 2022; Constantinides et al., 2018). The "so what?" of this paper is found in its capacity to demystify digital platforms by scientifically quantifying their elements and dynamic behaviors, drawing on the explanatory power of physics reductionism (Cozzolino et al., 2021). By validating theoretical predictions with raw data retrieved at the highest certainty, a rigorous scientific lens is offered, guided by fundamental truths. An enhanced understanding and control of decisive platform observables (ranging from cross-platform network effects and governance structures to real-time revenue generation) are facilitated, thereby establishing the scientific foundation vital for digital platform engineering.

## 1.3 RESEARCH AIM/PURPOSE; RESEARCH QUESTIONS; AND OBJECTIVES
### 1.3.1 Research aim/purpose

The aim of this paper is to propose and rigorously validate a unifying, data-driven framework that quantifies network effects, trust, revenue generation, and the dynamics of supply (in the form of supplier transitions and mobilization) in on-demand platforms (Hagiu and Wright, 2015; Constantinides et al., 2018). This framework is grounded in a mathematical model inspired by theoretical physics and validated with high-certainty raw data, allowing it to emerge iteratively from empirical observations. The model adopts a pragmatic and practical orientation, enabling a powerfully predictive lens that supports real-time platform optimization and an enhanced understanding of transient behaviors in both demand and supply (Cozzolino et al., 2021; Chen et al., 2021).

The paper aims to deliver valuable tools for present-day platform managers and strategists, alongside future platform designers and scenario planners (Gawer, 2022). It aspires to achieve this with the necessary granularity and precision in its predictions. These insights can guide policymakers in responding to socio-economic challenges, including the displacement of workers by autonomous technologies such as autonomous vehicles. It is intended that this quantitative platform engineering approach will help shape decentralized, scalable, and efficient digital platform ecosystems, thereby advancing a new generation of digitally driven collaboration and value creation (Rietveld and Schilling, 2021).

### 1.3.2 Research questions

How do B2C digital platforms, such as Uber, monetize the dynamic supply-demand network patterns that drive exponential growth?

This overarching question recognizes that positive network effects, trust, revenue generation, and supply-side transitions are intricately connected. It leads to further sub-questions:

1. **Critical Parameters**

   What are the critical parameters that determine operational coherence, efficiency of revenue generation, growth thresholds, stimulated network effects, and the strength of trust within a two-sided B2C platform (Hagiu and Wright, 2015; Constantinides et al., 2018)?

2. **Model Validation**

   What model validation methods can robustly confirm the accuracy and resilience of the proposed platform theory, especially when tested against shock events (Chen et al., 2021; Ao et al., 2024)?

3. **Predictive Power**

   How accurately can the mathematical model project future states and behaviors, and what are the practical or theoretical limitations that might constrain its applicability (Muzellec et al., 2015; Rietveld and Schilling, 2021)?

4. **Generalizability**

   To what extent can this model be applied beyond ride-hailing, for instance, to lodging, food delivery, goods delivery, and emerging on-demand services such as robotaxis or domestic robots (Gawer, 2022)?

These questions underscore how a deeper understanding of platform growth dynamics can help entrepreneurs and stakeholders ensure timely and controllable exponential growth in varied market scenarios (Cozzolino et al., 2021).

### 1.3.3 Objectives

To effectively answer the aforementioned research questions, the research objectives are listed below.

**O1: Review the LIterature**

Critically review published research efforts that contextualize this paper and identify how it aligns with and enriches the field. O1 is most prominently met in Section 2, where existing scholarship is examined to contextualize the paper.

**O2: Quantify Key Catalysts**

Identify and quantitatively characterize the catalysts that trigger exponential growth in B2C digital platforms, including how they operate transiently over time. This involves determining both the operational and socio-technical parameters that prompt sustained network effects (Constantinides et al., 2018; Chen et al., 2021). O2 is addressed primarily in sections 4.1 and 4.2, which empirically identify and measure factors driving exponential growth (e.g., surges, trust, coherence).

**O3: Construct and Validate Theory**

Develop and rigorously validate a physics-inspired, data-driven theory. This includes applying real-world shock tests and micro-level validations to prove the model's accuracy in capturing transient behaviors (Ao et al., 2024). O3 is satisfied in Section 5 sections 5.3 and 5.4, which present the PASER model and its validation under normal and shock conditions.

**O4: Predictive and Policy-Relevant Modeling**

Demonstrate the model's predictive power and derive insights that inform platform design choices (surge pricing, driver mobilization, trust-building mechanisms) and policymaking. Attention is devoted to mitigating emerging challenges, such as worker displacement by automation (Gawer, 2022). O4 is emphasized in discussions around surge pricing, optimization, and scenario testing (Sections 5.6, 5.7, 5.9).

**O5: Enable Generalization**

Explore the capabilities of the validated framework to extend to other types of on-demand services beyond ride-hailing (lodging, food delivery, goods delivery, and futuristic applications such as autonomous fleets and domestic robots) thereby laying the foundation for scalable, decentralized, and efficient digital ecosystems (Rietveld and Schilling, 2021; Cozzolino et al., 2021). O5 is addressed mainly at the end of Section 5, where the model's adaptability to other on-demand services is discussed.

## 1.4 SUMMARY OF THE APPROACH TAKEN FOR THE RESEARCH

An interdisciplinary approach is employed to develop a general theoretical framework for modeling platform economies and optimizing their operations. Data-driven Induction practices are followed, allowing iterative calibration of simulation parameters. High-certainty raw big data retrieval is utilized so as to completely avoid inference-based errors, especially in data-driven theory validation. The New York City Taxi and Limousine Commission High-Volume For-Hire Vehicle (HVFHV) Trip Records database (NYC TLC, 2025) is leveraged to identify emergent patterns in drivers' and vehicles' states, trip durations, and revenue dynamics with near 100% certainty at per-second granularity. These emergent trends in Uber and Lyft form a solid foundation for a mathematically rigorous, broadly applicable platform theory.

Numerous key dynamics are modeled via analogizing from laser physics principles: Cross-platform network effects, supplier agent population inversion, trust levels, growth thresholds, stimulated revenue emission, supply agent mobilization, profitable trip pattern "modal" capture, and spatial incoherences that reduce revenue-generation efficiency. Platform structural quality parameters are quantified by drawing on the analogy of supply and demand applications acting as mirrors, reflecting and amplifying network effects to optimize revenue and build trust.

Simplicity, adaptability, mathematical elegance, and adherence to physics conservation principles are offered by the model. Rigorous validation is conducted against near 100% certainty data from millions of trips spanning various dates, including the COVID-19 pandemic period, when both Uber and Lyft experienced severe shortages and complete nighttime shutdowns before morning restarts. These shock periods enabled testing of transient network-effects thresholds and critical mass build-ups at per-second granularity.

Great care was taken to ensure that, although the framework focuses on temporal dynamics for overall NYC operations, localized simulations can be adaptively run for small areas where spatial influx and outflux of drivers and vehicles require modeling. Micro simulations are inherently supported, and spatial awareness is embedded through macroscopic parameters that reflect spatial incoherence in completed trips. Calculations down to per-second or longer intervals can be handled. The framework generalizes readily to other platforms, such as Airbnb, and to natural evolutions of Uber (like food and product deliveries) by drawing on multimodal insights from laser physics. The theory powerfully illustrates the transition from lamp-like (heavy-asset, revenue-lossy) businesses to laser-like (light-asset, revenue-efficient) digital platforms.

**1.5 STRUCTURE OF THE PAPER**

The structure of this paper is as follows: Section 2 reviews relevant literature, Section 3 outlines the research methodology, Section 4 presents the findings, Section 5 discusses the implications of these findings, and Section 6 concludes with key insights and recommendations.

**1.6 SECTION SUMMARY**

This section introduced the motivation behind the direction taken in this paper for examining digital platforms through a physics-based lens. The key issue addressed involves deterministically achieving high predictive power of platform real-time operations at per-minute granularity through a scientific lens. Grounded in raw data with minimal inference, such an approach seeks to complement existing machine learning models while offering tangible insights into network effects, real-time optimization, and trust dynamics. The next section will review the relevant literature to contextualize and support this research direction.

**2.     LITERATURE REVIEW**

The evident significance of two-sided digital platforms has generated a lot of interest among researchers who sought to understand the mechanisms behind stimulated positive network effects as the foundation of multisided platform ecosystems. In this section, a physics-inspired real-time approach is taken in pursuing the understanding of how asset light platforms manage to orchestrate network effects and monetize them efficiently. It is shown that a physics-inspired framework can fill interpretive gaps in  the literature (this addresses O1).

Through network theory, Zeng, Khan and De Silva (2019) underscores how trust formation requires continuous, context-specific network building, a process not easily captured by existing real-time optimization frameworks. Wang and Yang (2019) review ridesourcing systems, emphasizing spatiotemporal demand forecasting and network equilibrium. Their approach lacks a unifying model. Non-equilibrium dynamic pricing is proposed by Nourinejad and Ramezani (2020) revealing how transient imbalances reduce waiting times. They do not provide a deeper view into the resilience of ride-hailing systems when they experience shocks in their operations.

Advanced data-driven techniques have enhanced short-term demand prediction. Chen, Thakuriah and Ampountolas (2021) achieve higher accuracy via deep neural networks. Their effort does not provide a macro-level framework that would link the micro-level predictions to the performance stability. The driver repositioning has been studied with deep reinforcement learning (Jiao et al., 2021) and game-based models of mixed fleets (Ao, Lai and Li, 2024) but without offering a broader lens that for the critical mass not the system-wide coherence. The importance of real-time operational strategies is highlighted by these works but the coupled interplay of trust, supplier mobilization are not cohesively modeled.

The platform viability is centered around trust and consumer behaviour. The trust mechanisms of urgent rescue, escrow, provider certification influence significantly the ridesharing adoption according to Alamoudi et al. (2023). Mao et al. (2020) highlight that platform-level signals shape a mix of cognitive, and calculation-based processes from which trust formation emerges in Airbnb. This signifies that intangible factors like trust drive network effects in a critical way. The work from (McIntyre et al., 2021; Cohen and Zhang, 2022) reinforce the notion that the platform participant behaviour is orchestrated at platform system level. This points towards the direction that there might be parallels to physical systems to explore. The system's coherence appears to be shaped by complex feedback loops. The paper from Kim and Yoo (2019) define platform stages of entry, growth, expansion, and maturity but do not mathematically specify threshold conditions under which network effects become explosive. Wang and Yang (2019) point to the importance of supply-demand balance. (Zhan, Szeto and Chen, 2022) attempt quantitative optimization. They show near-optimal solutions for ride-hailing operations. At the same time the underlying mechanisms that sustain self-reinforcing or stimulated dynamics remain elusive.

Research on multi-sided competition (Cozzolino, Corbo and Aversa, 2021) and data-driven models (Poniatowski et al., 2022) explores how incumbents either collaborate or compete to capture value. Gawer (2022) identifies challenges in governance and scaling. These contributions though do not integrate the reported real-time spatio-temporal modeling with intangible elements like trust. Agent-based or dynamic programming approaches (Al-Kanj, Nascimento and Powell, 2020; Gao et al., 2024) successfully solve localized repositioning and surge pricing. But do not unify these with macroscopic systemic phenomena such as abrupt collapses or intangible cost factors (Bi et al., 2020).

The unification of local and global insights has been the focus of paperers examining shock scenarios. Albert, Ruch and Frazzoli (2019) use stochastic modeling of mobility-on-demand to incorporate feedback control for spatial imbalance. Qian et al. (2020) analyze real-time matching and surge-chasing behaviors from New York City Uber data. Fluid-inspired or queueing-based perspectives (Xu et al., 2021) capture spatiotemporal heterogeneity but are generally based on aggregate assumptions. COVID-19 caused dramatic disruptions for Airbnb. This reflects the kind of critical mass losses that ride-hailing platforms also faced according to Kourtit et al. (2022). It is made evident that trust reconstitution and threshold effects operate at city-wide scales.

In summary, major methodological advances have been made: Deep learning (Chen, Thakuriah and Ampountolas, 2021), dynamic programming (Al-Kanj, Nascimento and Powell, 2020), evolutionary game theory (Wang et al., 2020), and multi-objective optimization (Zhan, Szeto and Chen, 2022). Yet a comprehensive theory unifying intangible trust, real-time stimulation of network effects, resources allocation, dynamic revenue generation, platform design quality, remains elusive. Existing studies recognize that cross-platform network effects, dynamic dispatch, strategic pricing, and socio-behavioral signals collectively underpin platform growth. But they do not fully explain transitions between subcritical and supercritical states or how local disruptions can extinguish broader system-wide progress. A deeper theoretical foundation has been widely called for.

This paper proposes a physics-inspired perspective on platform growth. It draws from laser systems analogies that incorporate threshold conditions, feedback loops, and coherent or incoherent modes in statistical distributions. While queueing-based models (Xu et al., 2021) and deep learning methods (Wen, Li and Lau, 2024) address real-time system dynamics, they do not provide a physics-like lens for critical mass, coherence, and stimulated network effects that that would offer valuable insights to the system's operation. Trust formation models (Mao et al., 2020; Alamoudi et al., 2023) demonstrate the platform's role in shaping loyalty but do not integrate dynamic stimulation of network effects. Multi-objective optimization (Zhan, Szeto and Chen, 2022; Ao, Lai and Li, 2024) provide local solutions without clarifying how entire systems remain stable or collapse abruptly.

The proposed PASER framework addresses the aforementioned gaps by allowing a laser-inspired system of coupled dynamic equations to be naturally deduced from high-volume raw data sets. The framework unifies a multitude of platform properties while seamlessly scales and adapts to model diverse on-demand platforms.This integrated approach aims to explain how critical mass can trigger explosive network effects or sudden collapses, as well as to offer real-time optimization tools for surge pricing, resource allocation, and scenario testing. By unifying socio-behavioral insights, dynamic resource allocation, and rigorous data analysis, PASER responds to repeated calls in the literature for a more robust theoretical foundation that accommodates intangible elements and operational dynamics alike.

This review demonstrated how empirical, theoretical, and methodological advances have paved the way for an integrated lens on digital platforms. It also reveals where this paper can contribute: by delivering a physics-inspired, data-validated model that captures local-to-global feedback, formalizes intangible trust within real-time mobilization decisions, and embeds short-term predictions in a unifying view of network effects. The next section outlines the proposed methodology, which harnesses near 100% certainty raw data through interdisciplinary reasoning, thereby operationalizing phenomena that remain only partially understood in the scholarly discourse on platform ecosystems.

## 3.    RESEARCH METHODOLOGY

This paper focuses on on-demand two-sided and three-sided digital platforms (primarily B2B and B2B2C). Uber and Lyft serve as the principal case studies. The ride-hailing aspect of Uber is viewed as a B2C sociotechnical system linking drivers and vehicles (supply or business side) with riders (demand or customer side), leveraging technology for real-time matching and resource allocation.

Uber's operations are highly dynamic. In May 2024 alone, nearly 20.7 million non-shared trips were logged in NYC. Figure 1 plots a single day's ride durations contributing an average revenue of approximately USD 20 per ride. That day's 559,224 rides amounted to almost USD 10 million generated on May 2, 2024, as shown in Figure 4. Figures 2 to 5 illustrate the hourly trip volumes, total revenue, and per minute average revenue, trip costs, revealing transient, high-frequency behaviors. Figures 6 to 10 provide parallel metrics for Lyft (181,797 rides), demonstrating similar patterns.

Platform leaders such as Uber's CEO emphasize technology, economics, and social interactions to stimulate cross-platform network effects, build trust, and shape urban mobility and labor markets. Despite generating revenue in asset-light ways, such platforms remain only partly understood, even by their own analysts who rely on black-box machine learning algorithms. A deeper, transparent grasp of these complex systems is thus pursued here.

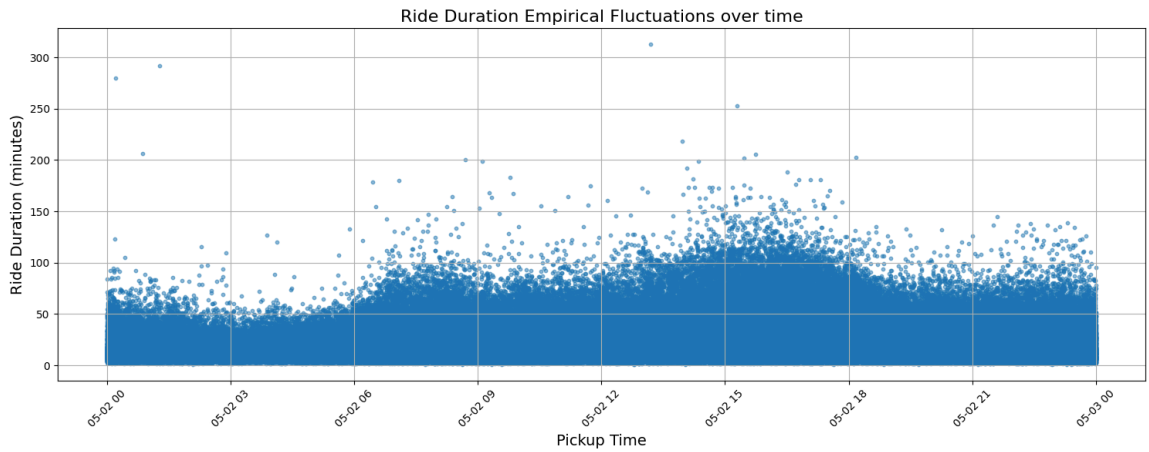

Figure 1. Uber - 559224 non-shared Uber rides with positive ride durations on May 2, 2024

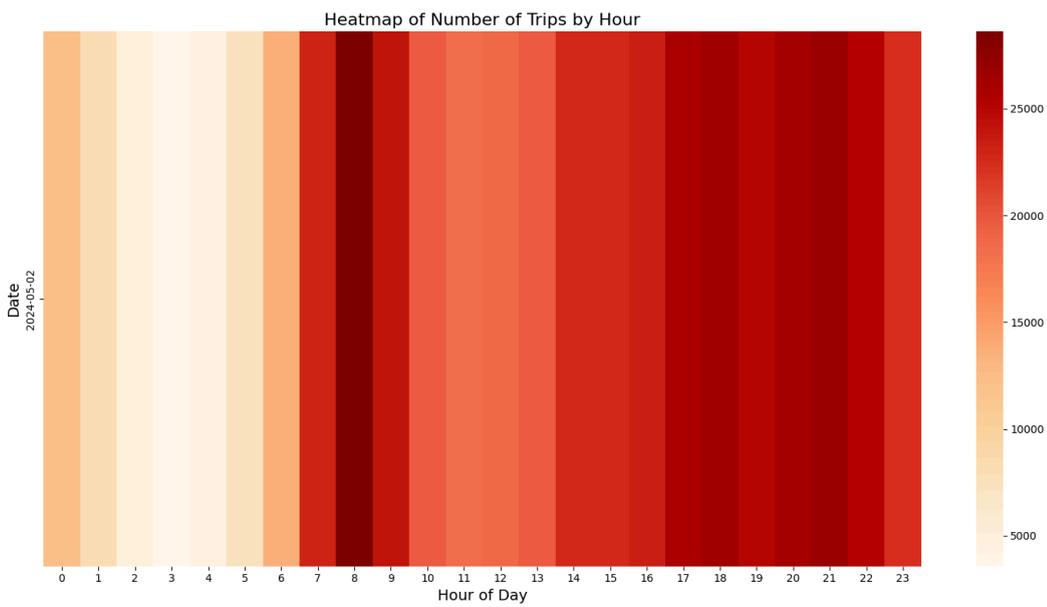

Figure 2 - Uber hourly trip volumes on May 2, 2024

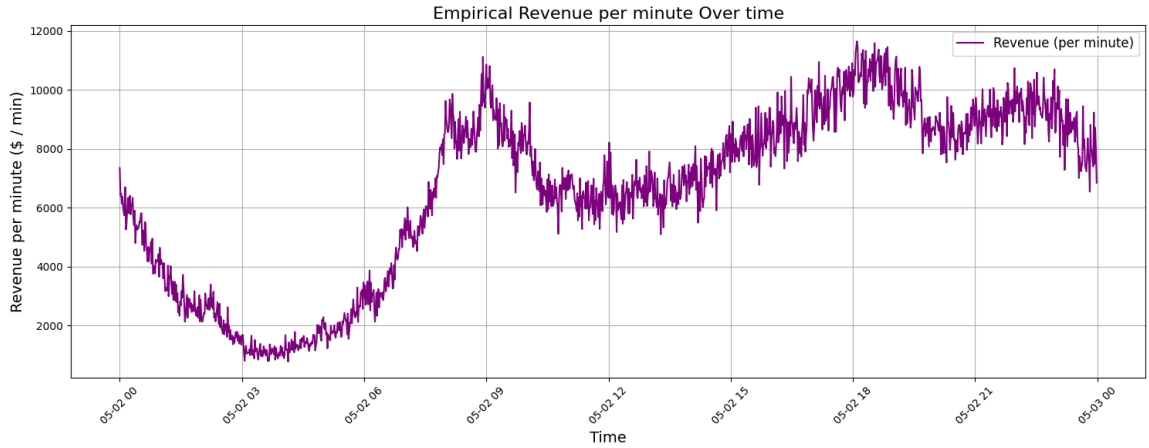

Figure 3 - Uber minute-level average revenue on May 2, 2024

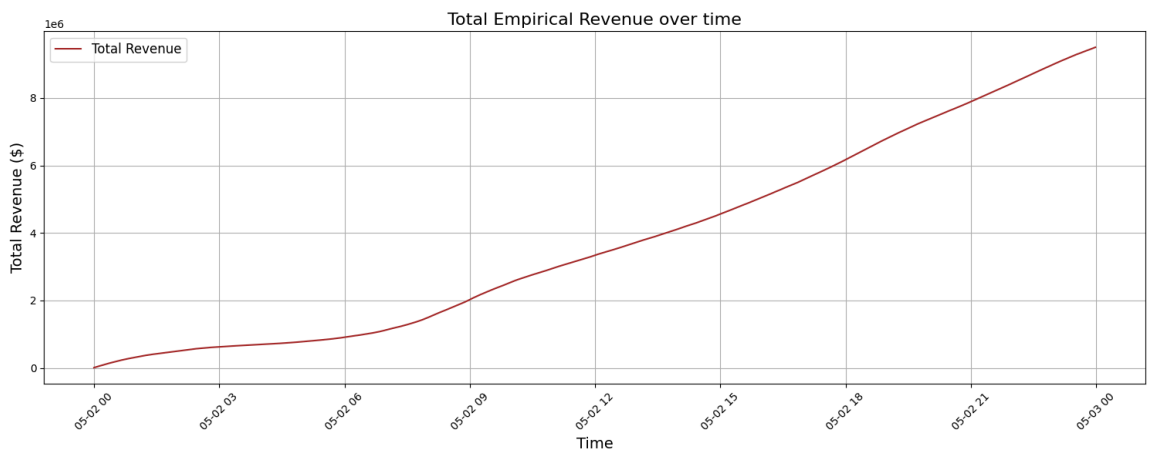

Figure 4 - Uber total revenue on May 2, 2024

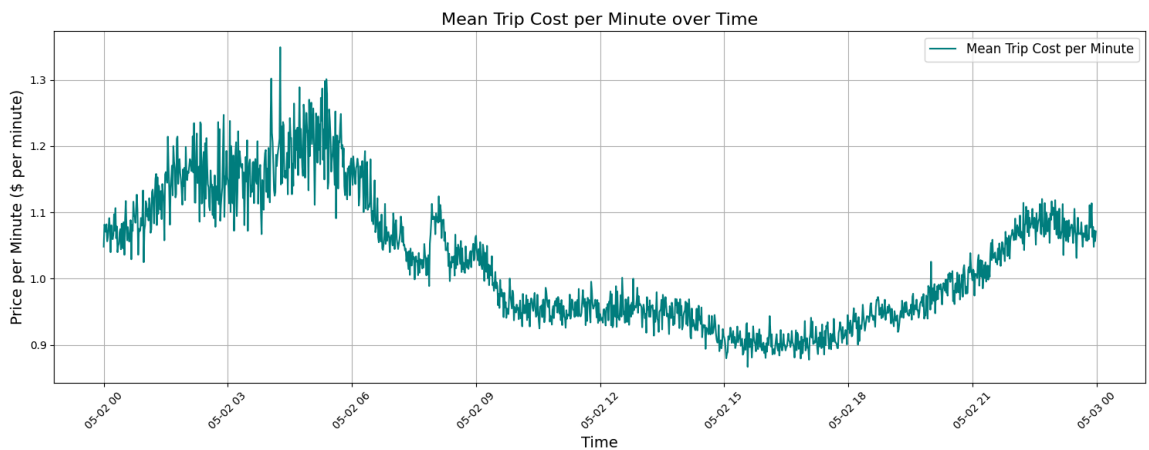

Figure 5 - Uber average per-minute trip cost on May 2, 2024

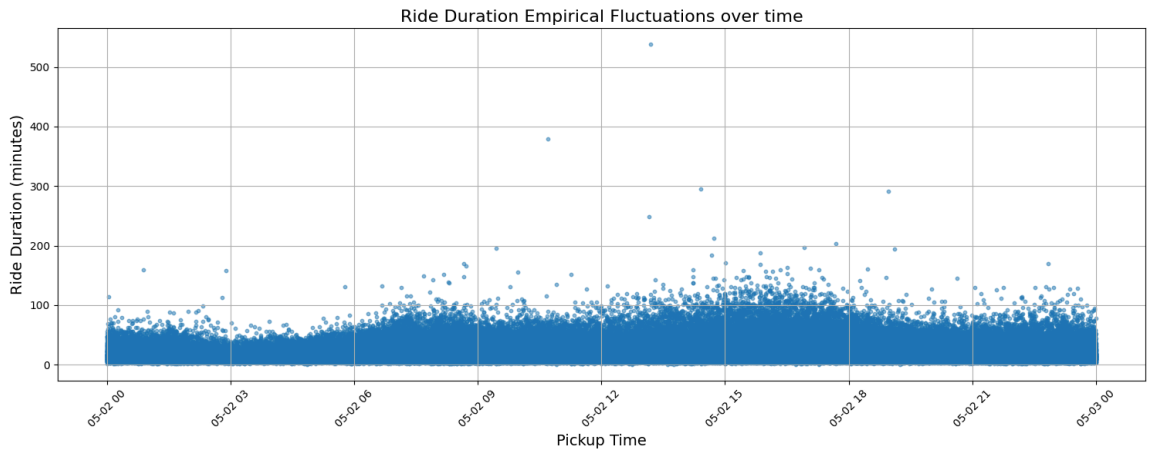

Figure 6 - Non-shared Lyft rides with positive ride durations on May 2, 2024

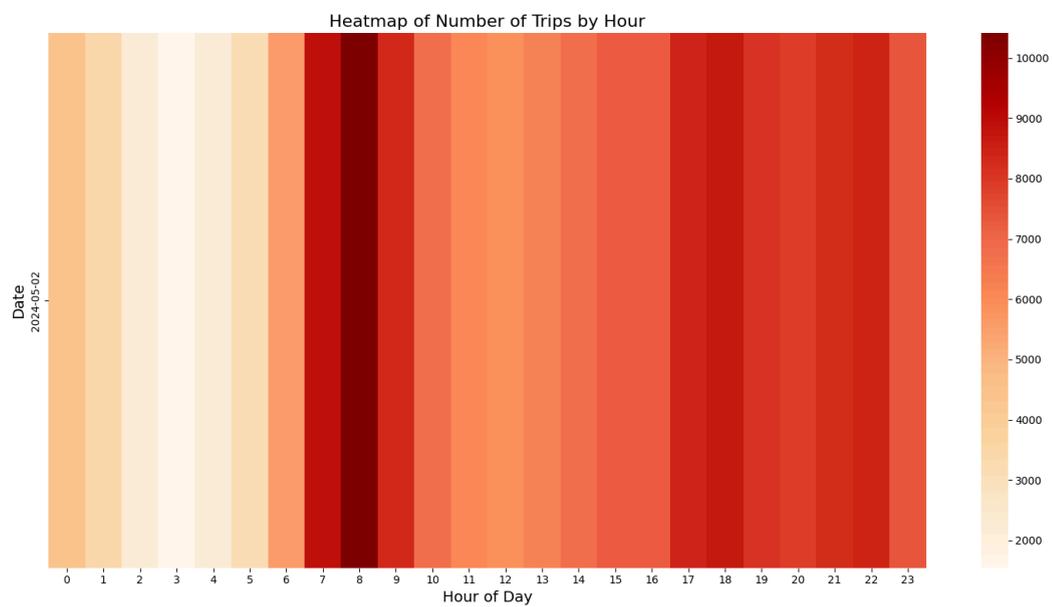

Figure 7 - Lyft hourly trip volumes on May 2, 2024

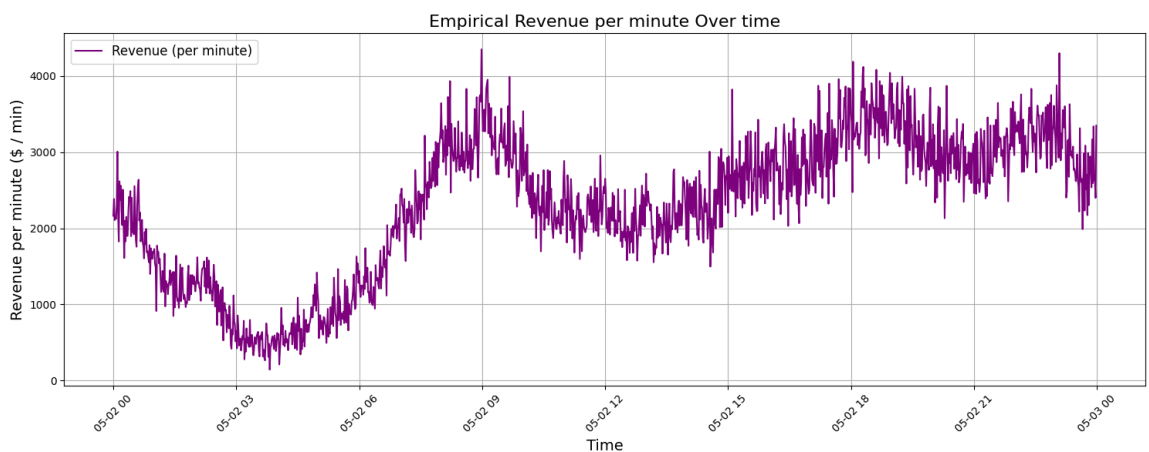

Figure 8 - Lyft minute-level average revenue on May 2, 2024

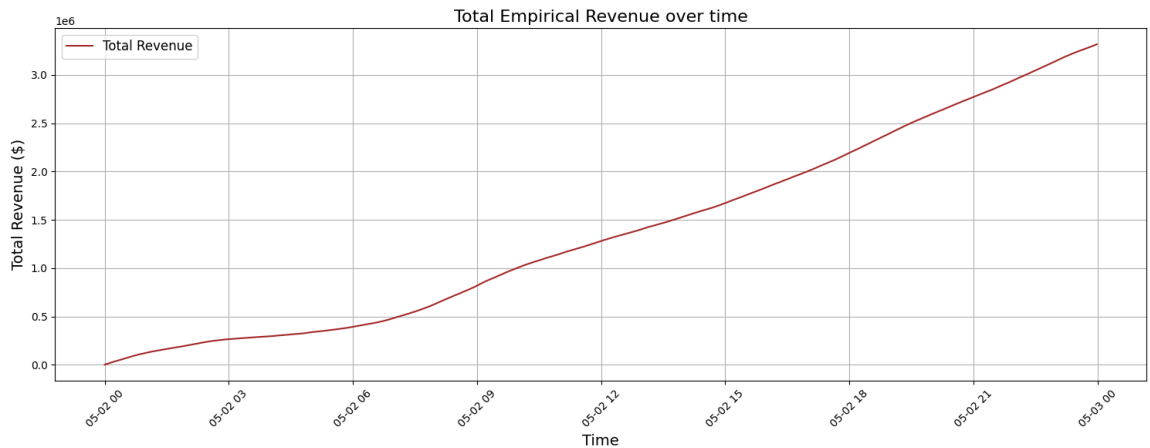

Figure 9 - Lyft total revenue on May 2, 2024

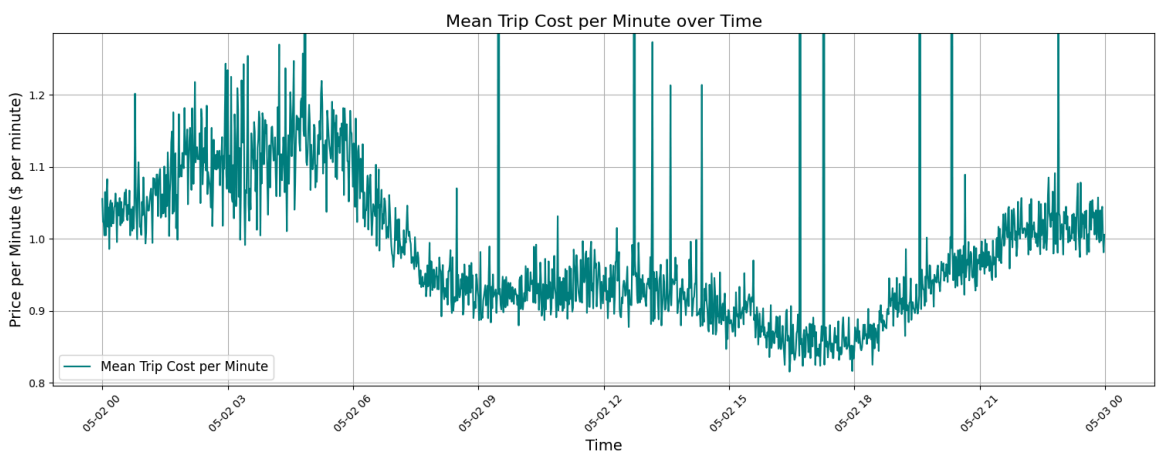

Figure 10 - Lyft average per-minute trip cost on May 2, 2024

## 3.1 METHODOLOGY: FROM EXPLORATION TO GENERALIZATION

In alignment with Seuring et al.'s (2021) adaptation of the Saunders et al. (2019), "Research Onion," this study follows an integrative approach comprising six concentric layers: Techniques and Procedures → Time Horizon → Strategy → Methodological Choice → Approach to Theory Development → Philosophy. Each layer is described below in the order employed in this paper.

**Techniques and Procedures: Big Data Mining**

The first layer involves big data mining from the high-volume New York City Taxi and Limousine Commission datasets (NYC TLC, 2025), which contain per-second time-stamped logs spanning five years when it comes to high volume (quite detailed) datasets. The raw data collection and cleaning procedures are tailored to retrieving the key platform dynamics (e.g., enroute, on-scene, on-trip states) with near 100% certainty. Exploratory Data Analysis (EDA) then uncovers emergent patterns and provides empirical grounding.

The raw data collection and cleaning procedures are tailored to retrieving the key platform dynamics (e.g., enroute, on-scene, on-trip states) with near 100% certainty.

**Time Horizon: Longitudinal**

A longitudinal approach tracks the platform network effects dynamics and transient behaviour of the revenue over extended periods. Observations are made in multiple windows (hourly, daily, weekly, monthly) to capture persistent or emerging trends within the five-year data collection period. This long-range view supports a deeper understanding of how platform behaviors evolve under varying conditions (e.g., weekend vs. weekday, normal vs. shock events like COVID-19 restrictions).

A longitudinal approach tracks the platform network effects dynamics and transient behaviour of the revenue over extended periods.

**Strategy: Experiment**

From a strategic perspective, this research is classified as an experiment. It numerically tests hypotheses about platform operations by applying derived mathematical models driven by real data. Real life historical scenarios are replicated via simulation (e.g., curfew shut-down shock events) verify how well the model predicts observed behaviors.

**Methodological Choice: Multi-Method Quantitative**

The study researches diverse facets of platform dynamics at high temporal resolution. It achieves this by combining descriptive statistics, time-series analyses, and reductionist modeling in a cohesive research design. Large-scale data retrieval is guided by statistical modeling while the retrieved data drives the simulations in replicating the real world scenarios under which the data were logged.

**Approach to Theory Development: Induction**

The approach to theory development is predominantly inductive. Emergent empirical observations provide hints as to what might be the deeper system properties like the distinct transitions of the drivers between different states of waiting for a match event, driving to the customer's location, picking up, and dropping off. These empirical data patterns then inform an analogical thinking approach that discovers strongly coupled analogies with physical systems behaviors. The analogies discovered act then as surrogate observations that lead to the theory development inductively.

**Philosophy: Critical Realism**

The philosophical approach to research is clearly that of critical realism. It acknowledges both the objective mechanisms behind platform operations and the socially constructed nature influencing the interpretation of those mechanisms. Critical realism supports the view that empirical data can uncover true to life underlying structures (e.g., cross platform network effects, driver mobilization as a causality effect driven by rider requests) while recognizing that user or analyst perspectives shape how these structures are understood and modeled.

By following these six layers (big data mining, longitudinal, experiment, multi-method quantitative, induction, and critical realism) the methodology systematically combines empirical rigor with theoretical innovation. Grounded in the "Research Onion" approach as interpreted by Seuring et al. (2021), the study progresses from data-driven exploration to theory development and validation, culminating in a predictive framework that illuminates the essential mechanics of on-demand platform economies.

### 3.1.1 Raw Data Mining (Data Collection and Preparation)

TLC data spanning five years, including monthly sets of millions of trip records time-stamped per second, form the foundational dataset. Near 100% certainty was prioritized (figure 11) by excluding invalid trip entries. Python libraries (Pandas, NumPy, Matplotlib, Seaborn) handled large-scale data manipulation and visualization at per-second to per-minute resolution.

Time indices were created to mitigate edge effects. Driver/vehicle states (enroute, on-scene, on-trip) were identified via timestamps, allowing real-time population tracking with near-total certainty. Missing or incomplete fields (e.g., on-scene timestamps for Lyft) were imputed or merged. Lack of unique vehicle IDs required approximate assumptions that were excluded from model validation when uncertainty could not be mitigated (N_available and N_waiting in figure 17).

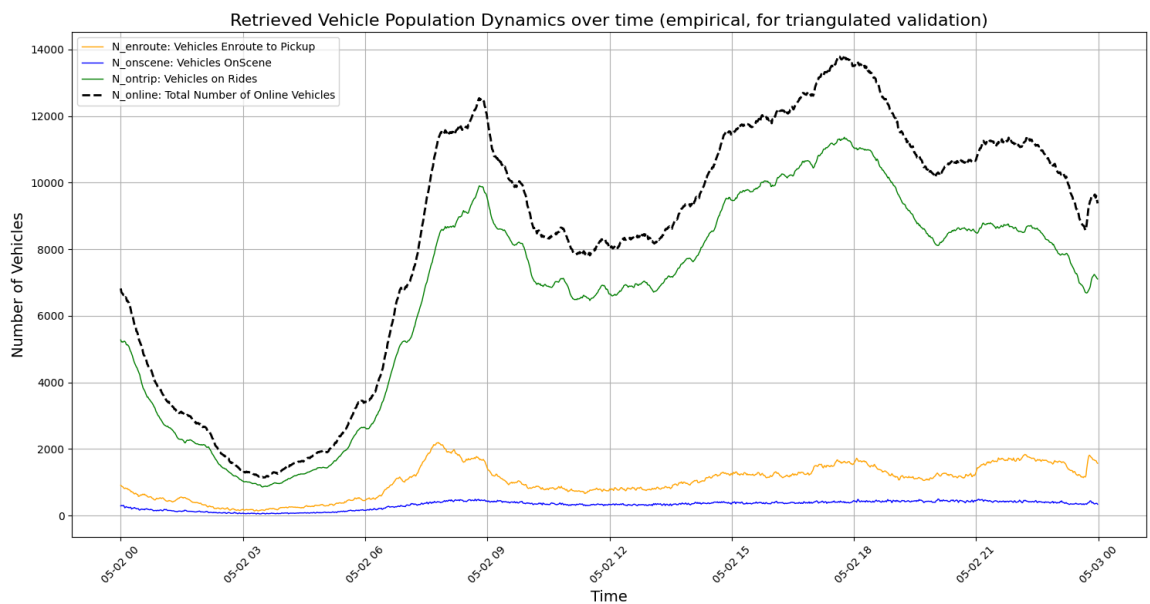

Figure 11. Retrieved driver/vehicle average per minute populations with 100% certainty

### 3.1.2 Exploratory Data Analysis (EDA): Understanding the data's characteristics

High-certainty retrieval of raw ride requests, driver states, and revenue facilitated robust validation. Visualization techniques (scatter, box, violin, heatmap) revealed emergent patterns and statistical distributions, assisting in recognizing cross-platform network effects. These insights laid the groundwork for a physics-informed theoretical framework. Figures 12 to 16 provide hints as to what qualities and systemic behavior may emerge from the data.

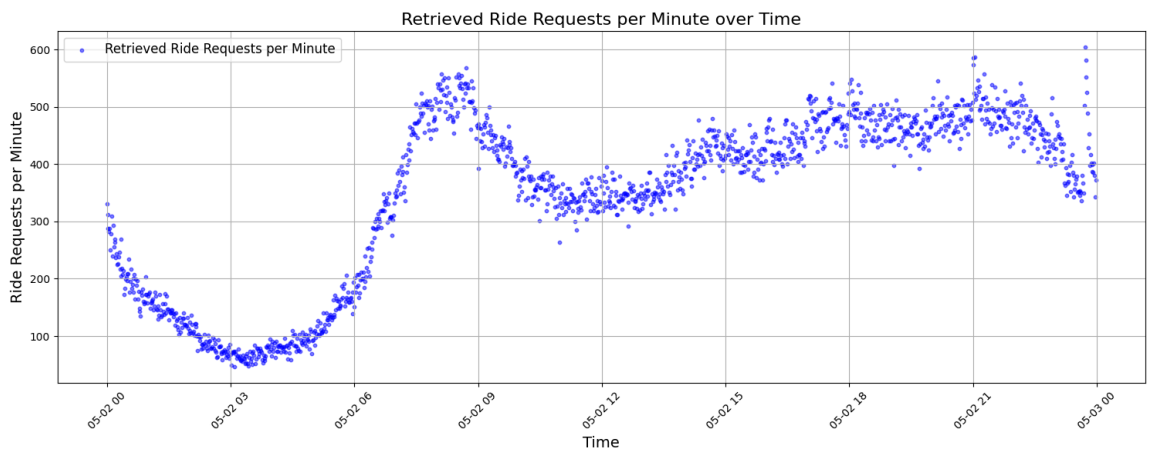

Figure 12. The ride requests that mobilized the drivers and shaped their transient behavior, depict dynamic network effet in conjunction with Figure 11 (on May 2, 2024).

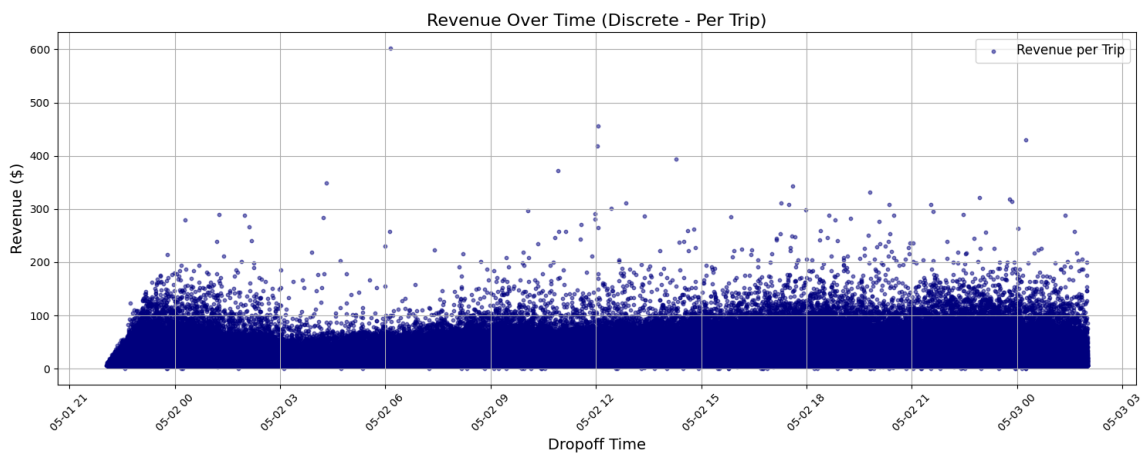

Figure 13. The revenue per trip fluctuations corresponding to figure 1.

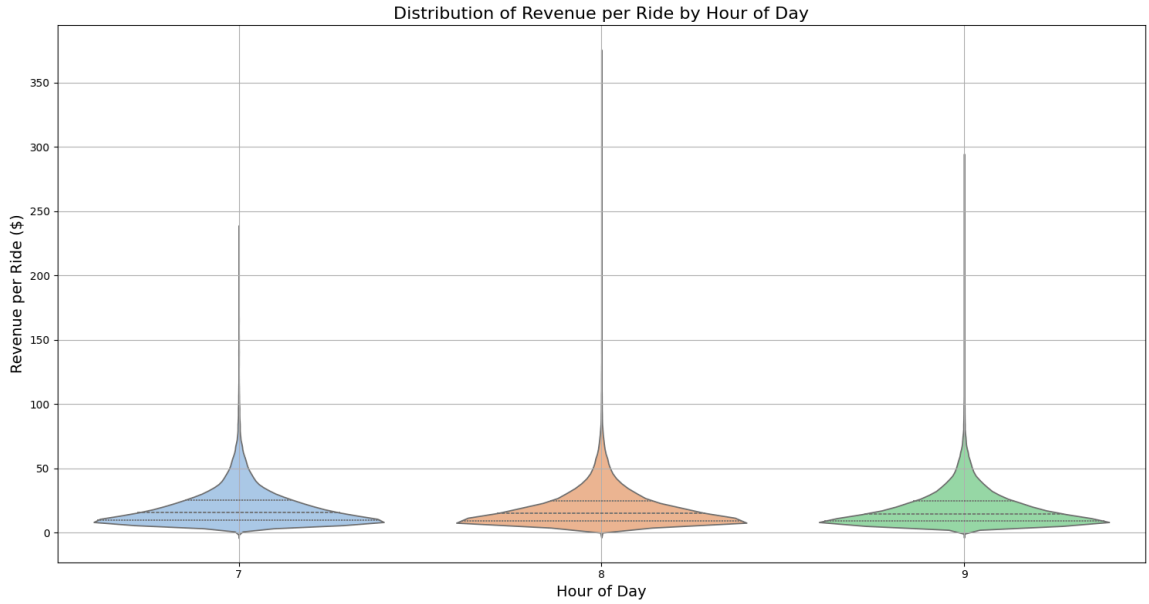

Figure 14. Statistical distribution denoting consistency in ride cost

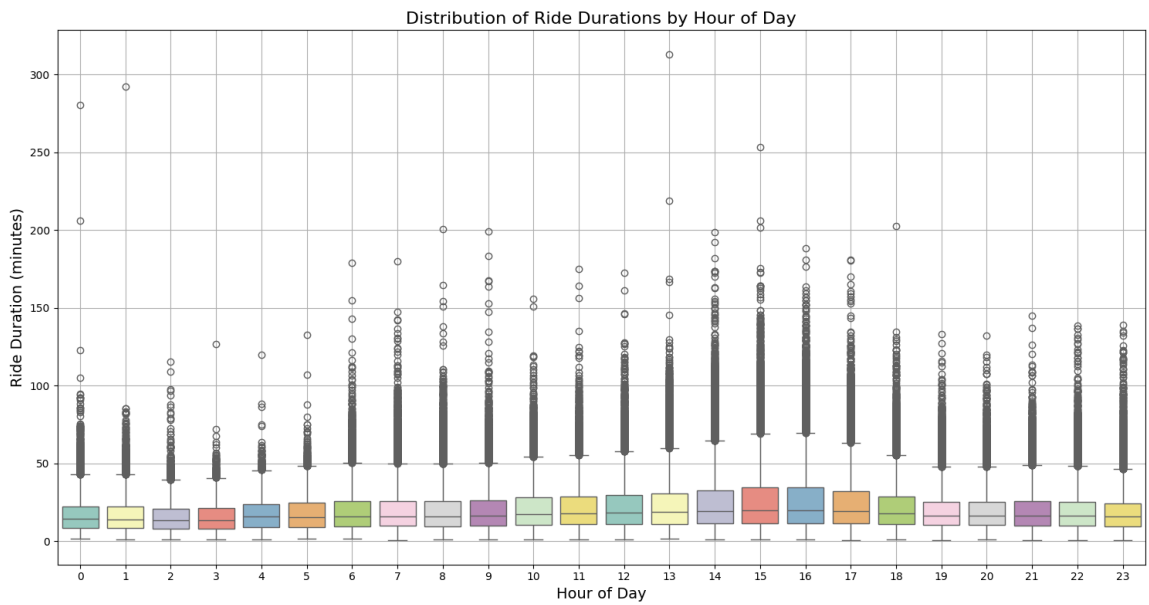

Figure 15. Statistical distribution denoting consistency in ride duration

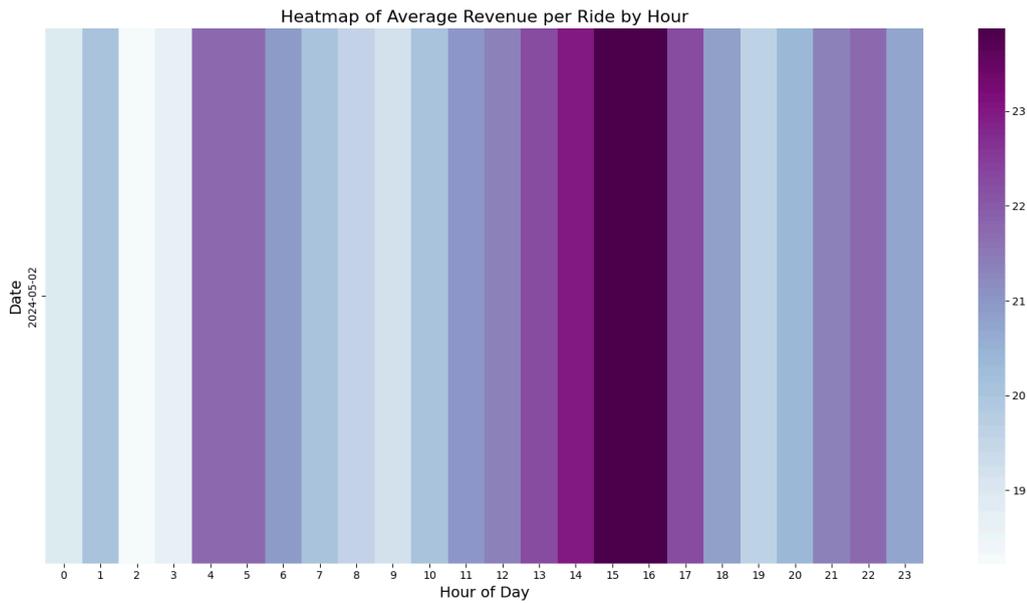

Figure 16. The revenue per ride by hour peaks before the late afternoon rush hour

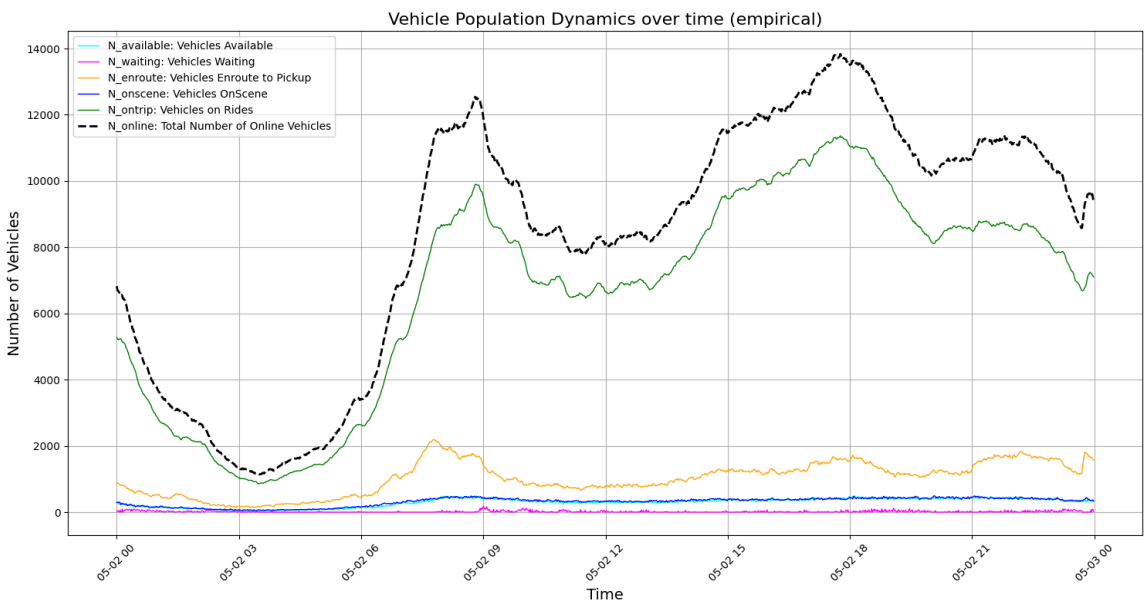

Figure 17. Retrieved driver/vehicle average per minute populations at different states (on May 2, 2024). Include both 100% certainty populations and inferred populations ('waiting' and 'available' drivers)

### 3.1.3 Inductive Analysis: Deriving Insights from Quantitative Data

Surges and other fluctuations in driver states were uncovered in patterns when exploring the temporal behaviour of drivers in high volumes of trips. These inductive findings confirmed that the time-series data retrieved contain cyclical phenomena. The induced insights pointed to the need for a unifying model that could potentially couple those together in a harmonious way.

### 3.1.4 Analogical Thinking: Drawing Parallels with Physics

The observed dynamic patterns resembled the transient behavior of known physical systems. Drivers flowing among states seemed to follow conservation principles (like conserving their overall population from one transition to the next). The ride requests though were causing surges in the total population of the drivers. Extensive iterations between different time windows confirmed consistent analogies, showing promise for elegant, coupled mathematical descriptions.

### 3.1.5 Reductionist Modeling: Quantifying Platform Dynamics

The analogical thinking stage acted as a bridge that led to the discovery of fundamental building blocks which, if systemically coupled together, were promising to synthesize a complex system greater than the sum of its constituent parts. The process involved identifying the fundamental components of the system, the states in which they were found, the rates of change in those states, and then described those with symbolic maths holistically. This would provide a unifying language to scientifically describe the complex system and provide a deep understanding of it at the same time. A crucial step was the coupling of the revenue dynamics in the symbolic expressions that fused the network effect dynamics thus elegantly describing how they are monetised by the complex system of the platform as a whole.

### 3.1.6 Validation: Testing the Model Against Real-World Data

It was of highest importance to validate the model by driving it with near 100% certainty data. Equally important was to sensitivity test it as well as subject it to true to life shock like the OVID-19 curfew shock-periods. These periods embedded in the raw data offered valuable transient stability testing. Populations of enroute, on-scene, and on-trip drivers were predicted alongside revenue and surge behavior, demonstrating alignment between the model and observed data. Figures 18-21 offer more clues about Uber's response to decline and extreme shock periods in preparation for the findings section.

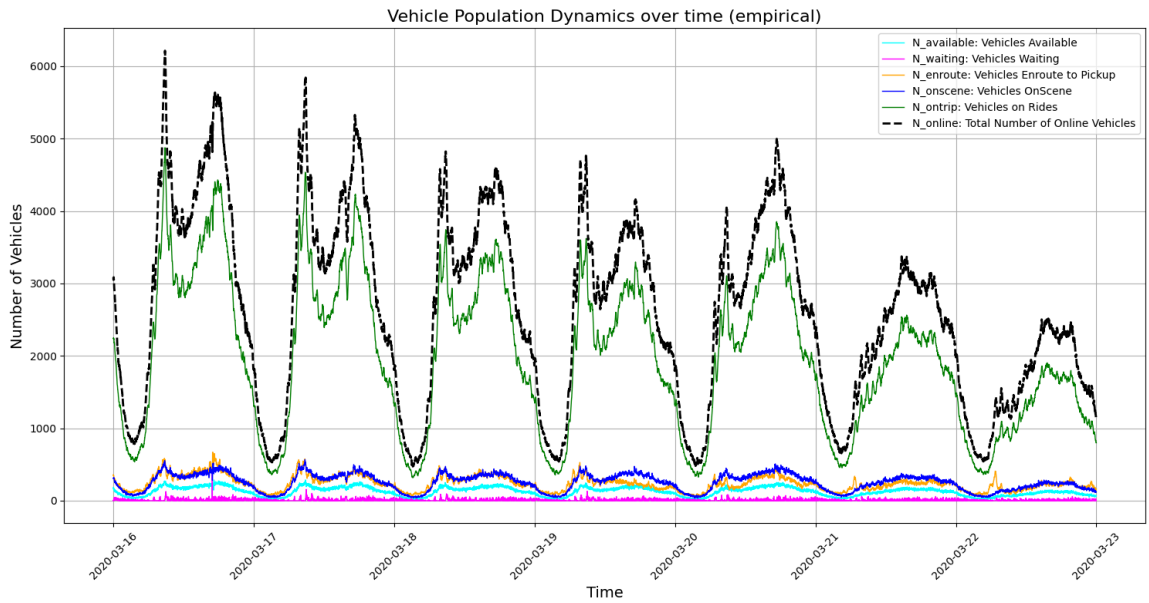

Figure 18. Uber's decline. The bearish period of the first week of covid restrictions in NYC starting March 16 of 2020

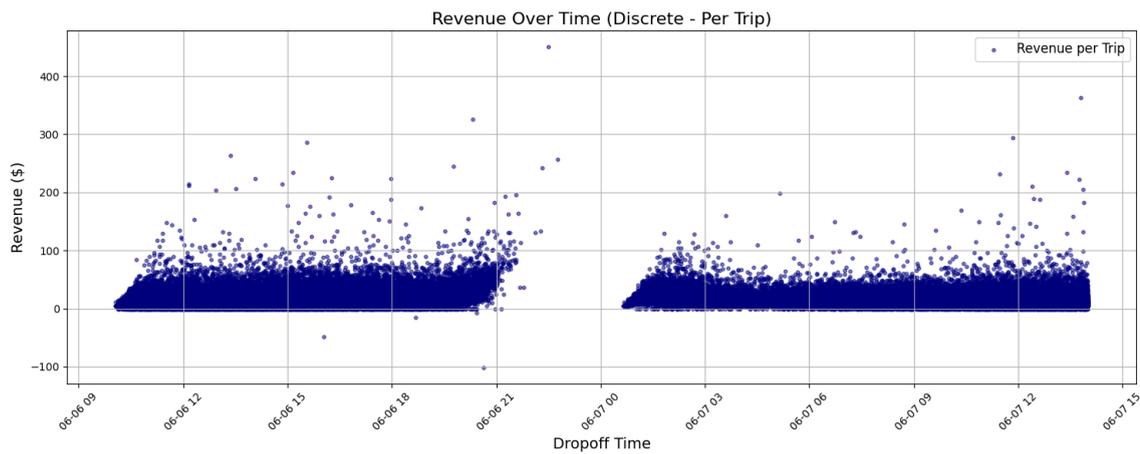

Figure 19. Depiction of the data warm-up and cool-off periods (2 h long each). Also of an extreme extreme shock (shut down) that Uber experienced during many COVID-19 nights.

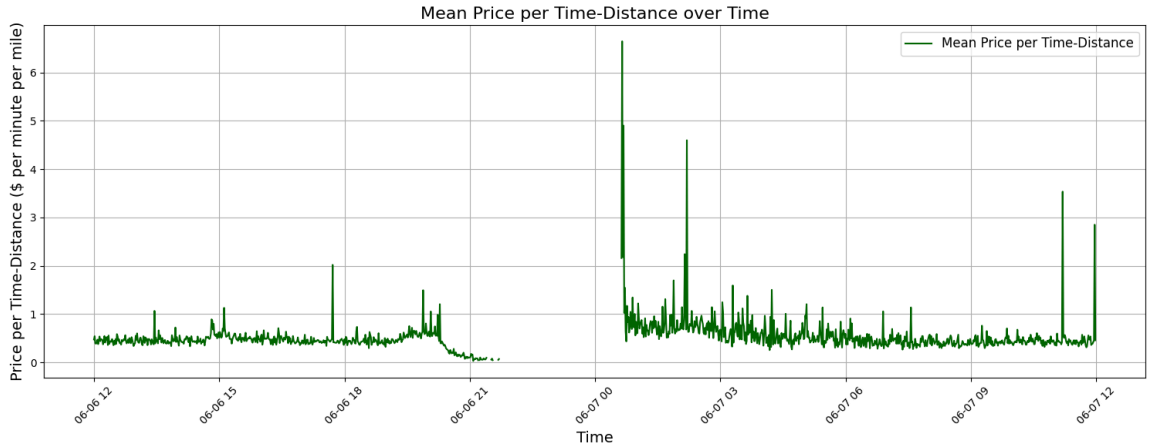

Figure 20. Uber experienced shut downs for hours during many COVID-19 nights.

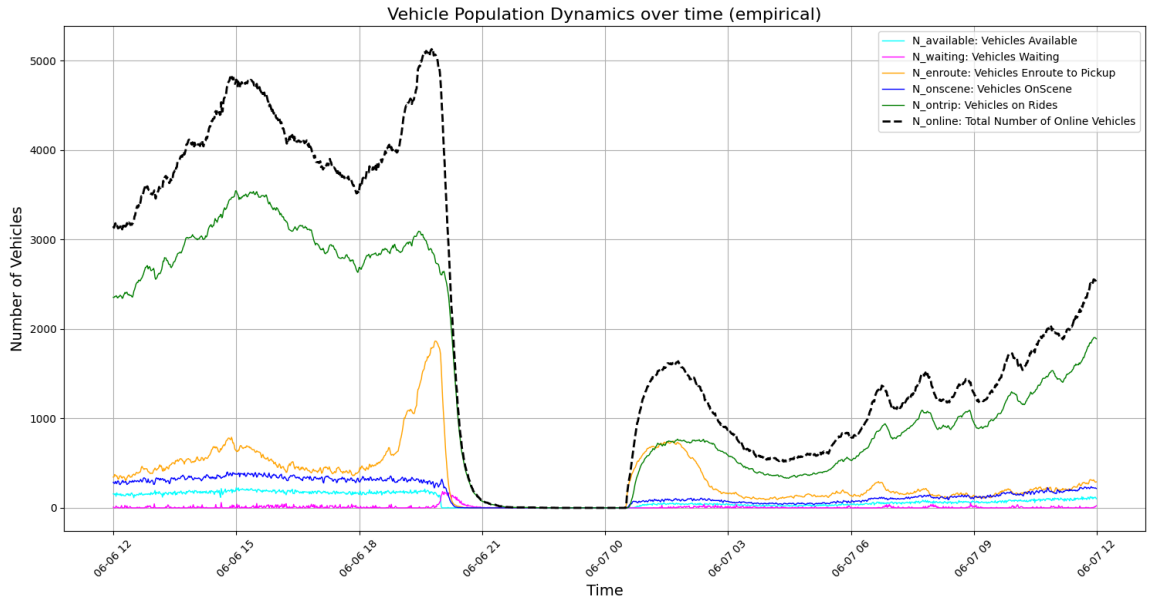

Figure 21. An apparent four and a half hours (20:00-00:30) curfew period in the night.

### 3.1.7 Generalization: Extending the Theory to Other Platforms

Adaptive parameter tuning and structural modifications facilitated simulation of Lyft (removing on-scene states) and Airbnb (coarser time granularity). Tesla's potential CyberCab scenarios could replace driver-centered transitions with vehicle-centered ones. The model's flexible architecture accommodates new modes, validated through quantitative insights into critical platform differences.

### 3.1.8 Synthetic Primary Data Generation: Scenario-Based Simulation and Scenario Analysis

Finally, the validated model produced synthetic data under various hypothetical conditions like autonomous vehicles, and modified business models. This scenario-based analysis further demonstrated generalizability and practical utility for optimizing and scaling on-demand platforms.

### 3.2 Summary

Section 3 outlines the concentrically layered methodology that governs this paper (ranging from big data mining and EDA to analogical thinking to inductive theory development and rigorous validation). Multi-method quantitative was the strategy of choice due to the high-complexity of a ride hailing platform and the high-frequency of its operations. This methodological rigor paves the way for the next section. Section 4 presents the core findings from the data analysis and begins to reveal the emergent physical analogies and system properties characterizing ride-hailing platforms.

## 4.    FINDINGS

This section begins with an exploration of data patterns in ride-hailing operations. It then proceeds with presenting the empirical findings that emerge. Real-time surges in pricing also receive attention, as well as the driver transitions, revenue generation fluctuations, and disruptive externalities like pandemic-induced shutdowns. These findings point toward a deeper, physics-like interpretation of platform dynamics which becomes the emphasis of the next section.

## 4.1 EXPLORATORY PHASE: OBSERVING EMERGENT PATTERNS IN DATA INDUCTIVELY

Initially three periods were data mined in reverse chronological order: the latest dataset from 2024, the COVID period in 2020, and the earliest detailed trip data in 2019. Figure 22 depicts typical ride duration fluctuations in a midweek day (Wednesday, 4 September 2024). This period's data is retrieved from the September 2024 high-volume dataset publicly provided by the TLC Trip Record Data (NYC TLC, 2025). The plot depicts two distinct rush hours and emerging from 400,000 rides were retrieved for that day with positive ride duration via the methodology described in Section 3.

Figure 23 compares the cumulative population of trips that can be used for data validation (non-anomalous) versus the total trips logged in the database in that period. Hence, 485,006 original entries resulted in approximately 400,000 valid trip records, ensuring reliable data for theory development.

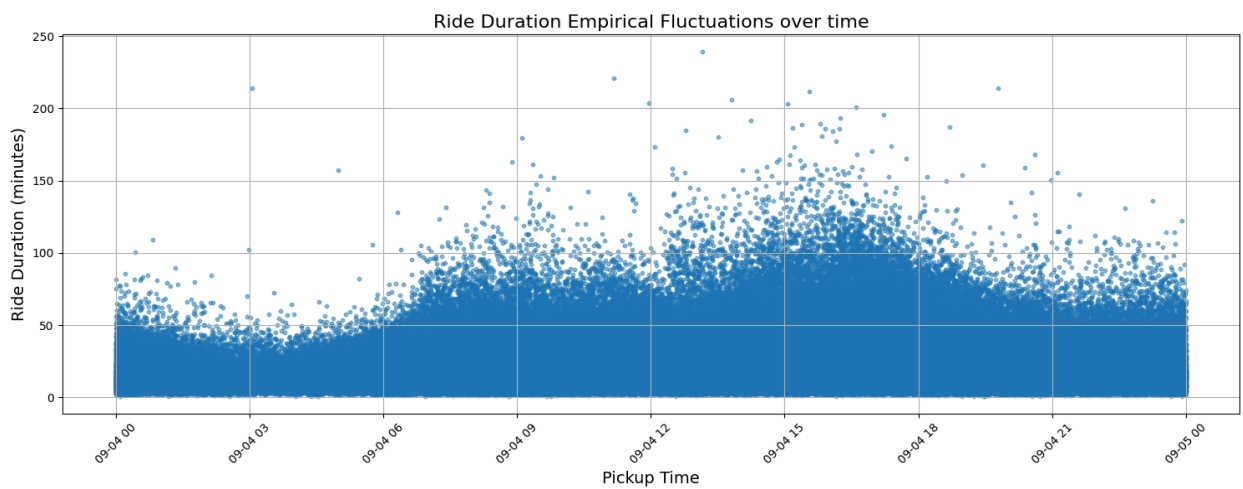

Figure 22 - Wed 4, Sep 2024 - 485,006 trips with positive ride durations (30% increase from pre-COVID 2019 in figure 25).

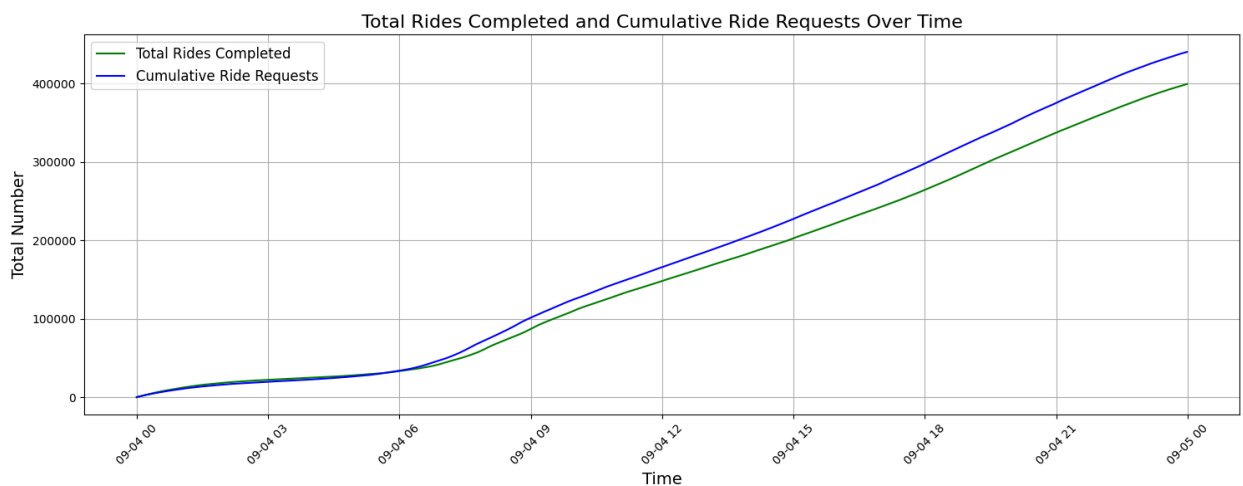

Figure 23 - Wed 4, Sep 2024 - 485,006 trips with positive ride durations.

These ride duration data points came from the TLC dataset's 'driver_pay' field. A small sample of five non-shared trips is shown in Table 1, where each row corresponds to a unique vehicle's trip. The table reveals that per-second timestamps (e.g., 'request_datetime,'

'on_scene_datetime,' and 'dropoff_datetime') were logged, indicating cyclical phases in which drivers proceed from enroute to on-scene to on-trip, thus generating revenue in the on-trip interval. Although Figure 22 contains outlier trips lasting over an hour, the majority of trips were shorter. While trip durations can be linked to time-series or cumulative revenue, these figures alone do not illuminate the interdependencies between riders (C-side) and drivers (B-side).

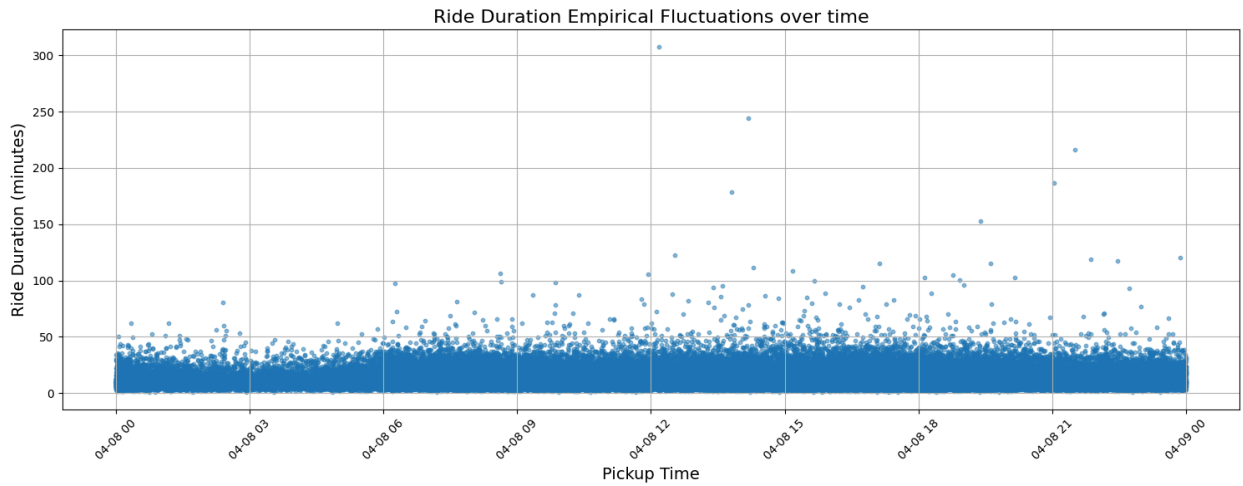

Figure 24. Wed 8, Apr 2020 (2nd month of pandemic restrictions in NYC) - 102,946 trips with positive ride durations. The two rush hour peaks have disappeared (70% drop of rides count from the pre-COVID 2019 in figure 25).

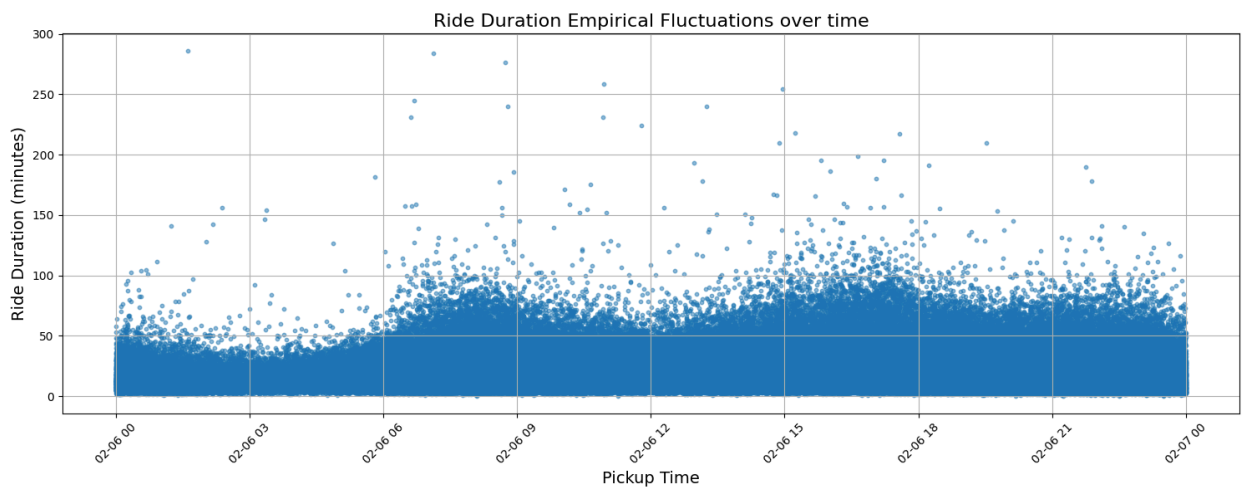

Figure 25 - Wed 6, Feb 2019 - 343,013 trips with positive ride durations.

**Table 1. Five trips extract from the May 2024 dataset of TLC in NYC**

| hvfhs_license _num | request_dat etime | on_scene_da tetime | pickup_dat etime | dropoff_dat etime | trip_ti me | driver_ pay |
|---|---|---|---|---|---|---|
| HV0003 | 2024-05-01 07:04:01 | 2024-05-01 07:05:55 | 2024-05-01 07:07:04 | 2024-05-01 07:16:24 | 560 | 8.45 |
| HV0003 | 2024-05-01 07:53:12 | 2024-05-01 07:57:31 | 2024-05-01 07:58:05 | 2024-05-01 08:02:08 | 243 | 5.39 |
| HV0005 | 2024-05-01 07:17:04 | NaT | 2024-05-01 07:24:40 | 2024-05-01 07:52:26 | 1666 | 25.26 |
| HV0003 | 2024-05-01 07:44:19 | 2024-05-01 07:47:40 | 2024-05-01 07:47:40 | 2024-05-01 07:51:56 | 256 | 5.39 |
| HV0003 | 2024-05-01 07:53:47 | 2024-05-01 07:56:11 | 2024-05-01 07:57:46 | 2024-05-01 08:06:56 | 550 | 7.13 |

Nonetheless, an important causal relationship emerges from the raw data. The log entry for a ride request triggers the driver's subsequent on-trip state, strengthening driver responsiveness and platform trust. The database does not show all ride requests. Only those accepted by drivers were logged. Yet the data imply that plotting per-minute aggregated matched requests (demand side) against per-minute aggregated driver states (supply side) may reveal the transients of dynamic cross-platform network effects pivotal to B2C digital platforms.

Focusing on the same Wednesday in 2024 (Figures 22 and 23), Figure 26 plots that day's empirical population of rider-to-driver matching events, representing valid ride requests that successfully led to on-trip status. Anomalies were filtered out as described in Section 3 and illustrated in Figure 23. Slightly earlier timestamps on each request initiated a matching event, mobilizing a driver who had either just come online or just finished serving another passenger. A notable surge in request acceptance occurs after 6 am, lasting for more than an hour. The data thus suggest that driver populations respond dynamically to rising demand, forming a feedback mechanism indicative of cross-platform network effects.

Figure 27 shows driver population dynamics over time. A minute-level aggregation is used for interpretability, although second-level data were available. When Figures 26 and 27 are viewed in tandem, the exponential-like rise in requests (4 am to 9 pm) appears to coincide with a parallel growth in the number of online drivers. This synchronization reflects a virtuous feedback loop. Rising demand attracts a larger B-side presence, which in turn draws more riders by improving service quality.

Negative feedback loops have also been observed during periods of observed decline. Figure 18 has already depicted a supply mass decline that was observed during the first week of COVID restrictions. When fewer ride requests (demand side - figure 28) corresponded with fewer online drivers.

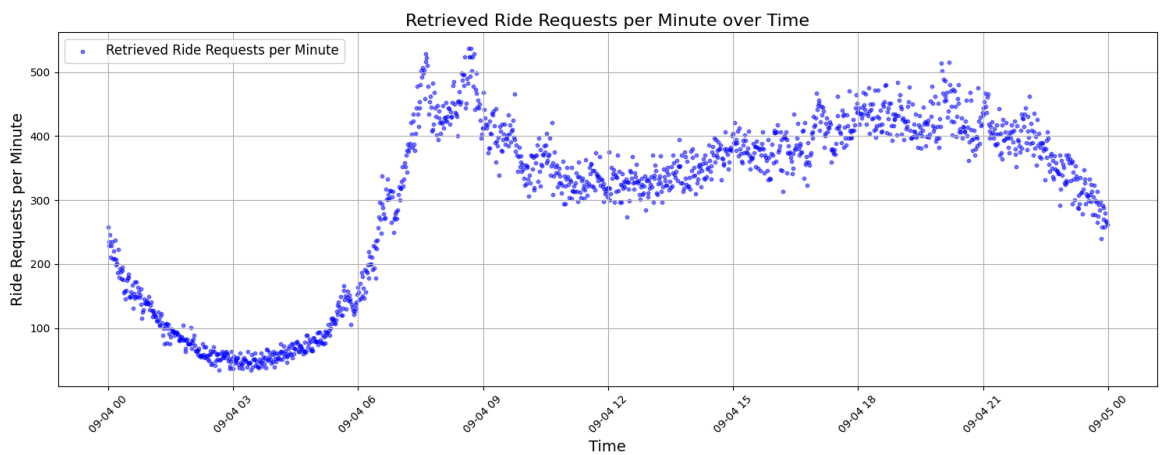

Figure 26. Ride requests that initiated the retrieved trips on Wednesday, 4 September 2024

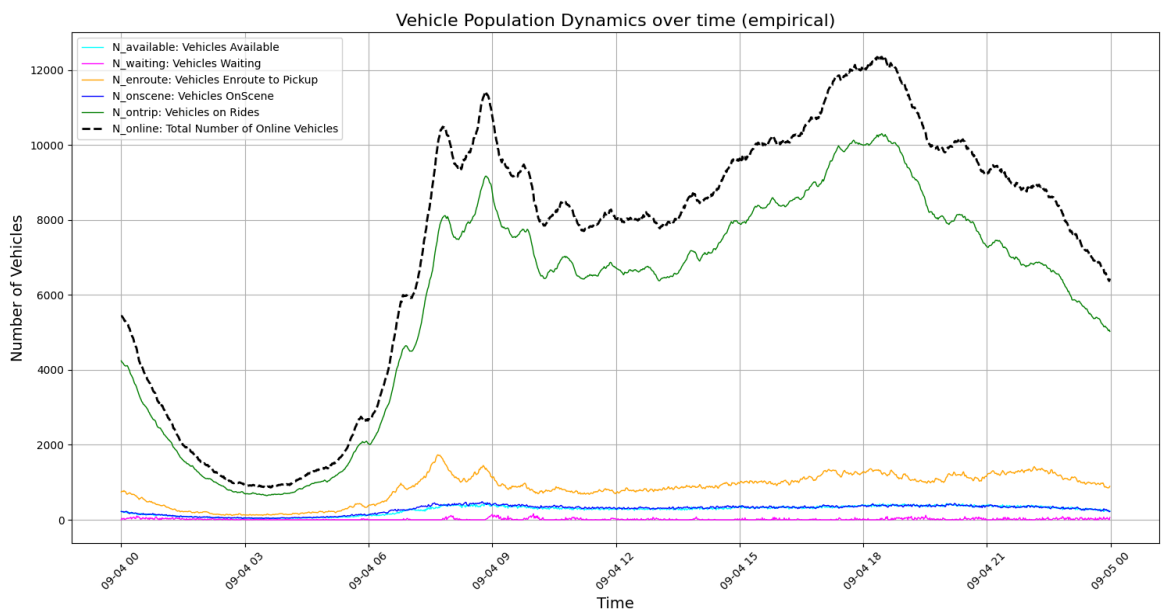

Figure 27. Per second retrieved, per minute aggregated dynamic driver/vehicle state population transient behavior.

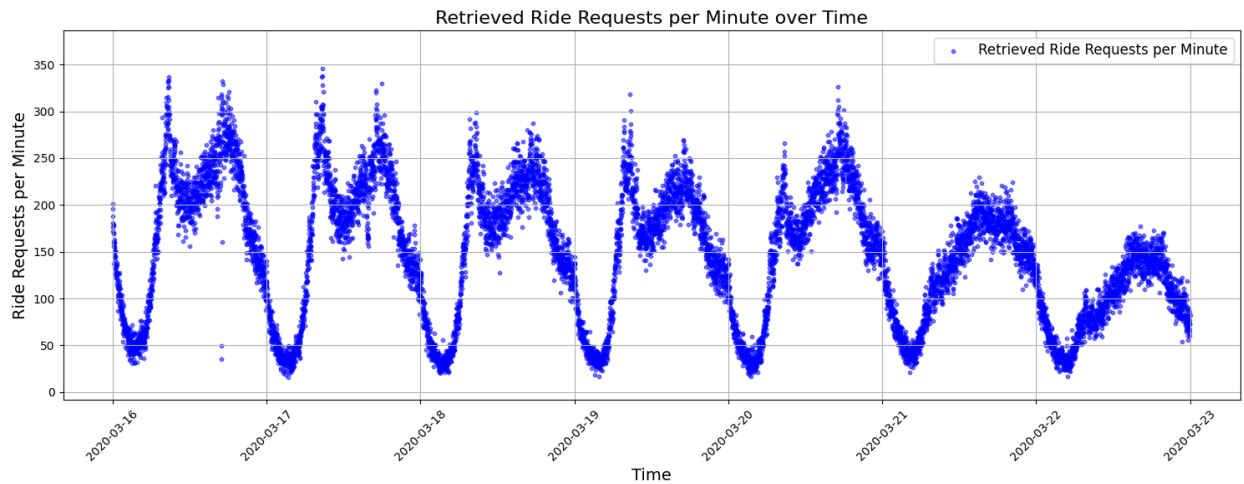

Figure 28. Uber ride requests decline during COVID-19.

Figures 29 and 30 highlight a disruptive 24-hour span starting at noon on 6 June 2020, where apparent curfew conditions lasted four and a half hours. Requests vanished first, prompting drivers to go offline. Once restrictions eased, requests reappeared, mobilizing drivers again. This shock underscores how each side depends on the other for platform continuity, but also how resilience manifests when demand resumes.

The TLC dataset reveals further insights emerging from the historic revenue data of 4 September 2024. Figure 31 highlights the high-frequency, minute-to-minute real-time revenue, while Figure 32 shows a smoother cumulative total. Figure 33 displays mean trip cost per minute, which spikes in early morning hours rather than at rush-hour peaks.

Although this might imply higher potential driver income at night, Figures 34–36 reveal that average revenue per ride actually peaks around 4 pm, even as per-trip cost is lowest during that period. These findings point to complex underlying dynamics: trip cost, volume, and revenue per ride do not always align with the day's busiest commuting windows.

Figures 34–36 reveal that average revenue per ride actually peaks around 4 pm even as per-trip cost is lowest during that period. These findings point to complex underlying dynamics: per minute trip cost, trip volume, and revenue per ride do not always align with the day's busiest commuting windows.

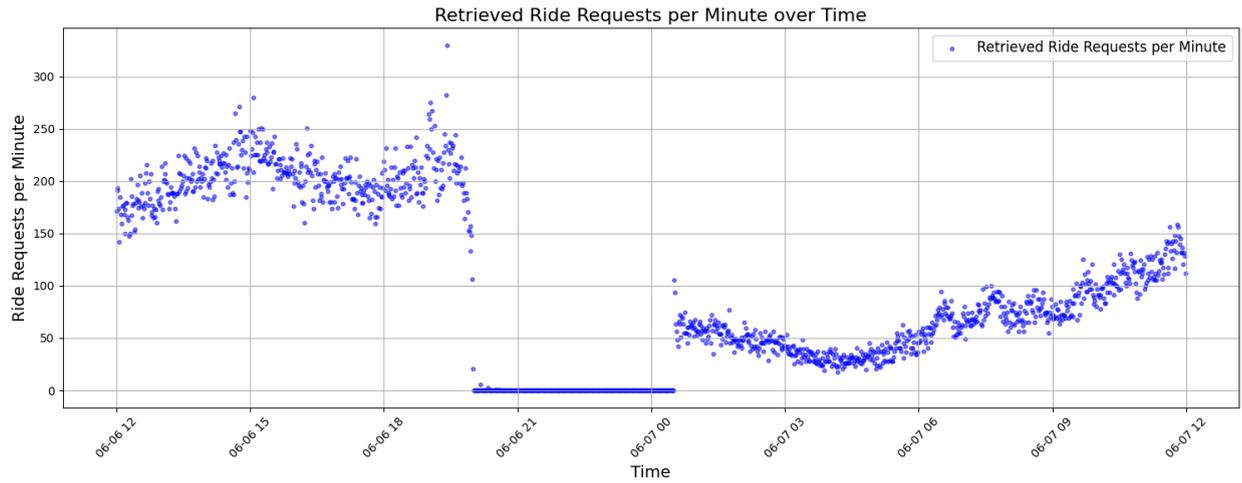

Figure 29. Uber ride requests depletion during a curfew night in 2020.

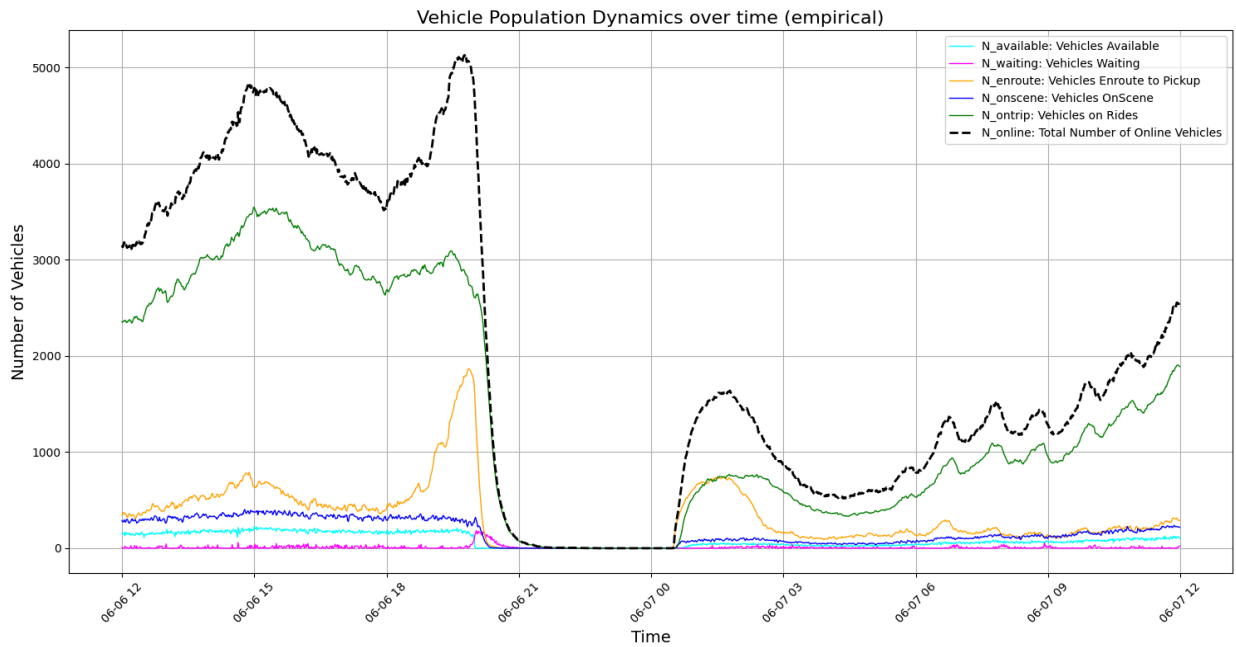

Figure 30. An apparent 20:00 to 00:30 four and a half hours curfew period in the night.

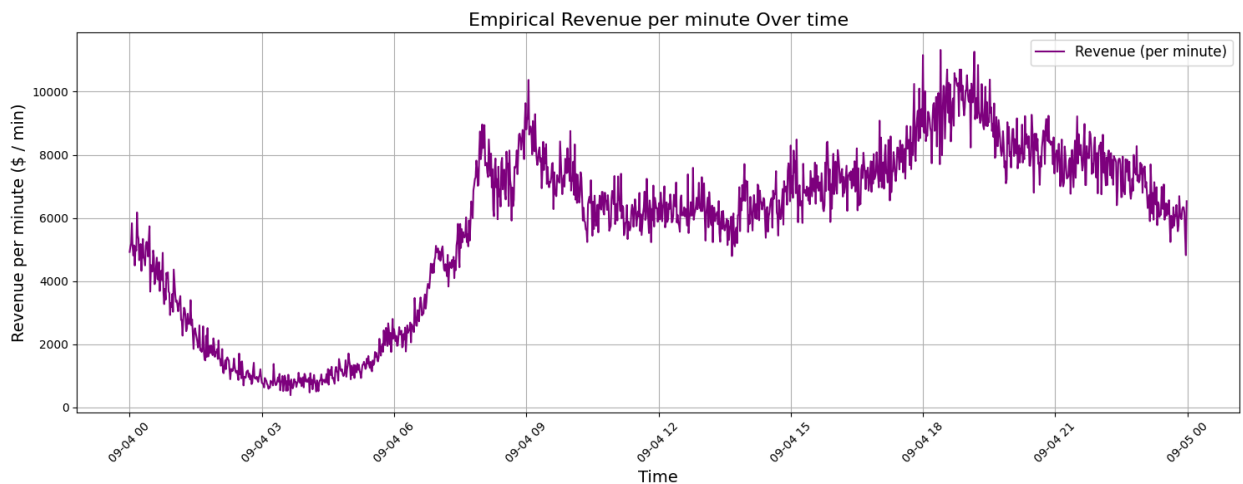

Figure 31. Uber September 2024. High-frequency, minute-to-minute historic revenue.

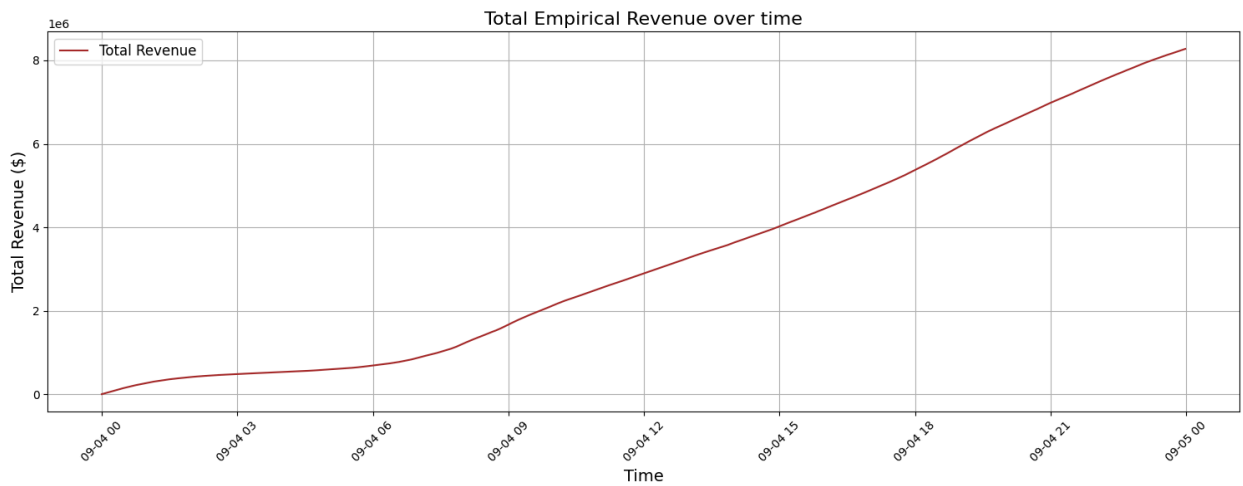

Figure 32. The cumulative total of the revenue in Figure 31.

The data suggest that per-minute revenue (Figure 31) rises in lockstep with heightened rider demand and driver availability around 8 am and 5 pm, reflecting the high number of daily trips in Figure 35. Yet the cost per trip is not at its minimum then. There appears to be coupled behaviours emerging from the data which indicates a need for deeper understanding.

It is becoming evident that network effects dynamics emerge from diverse periods. From typical midweek operations in 2024 (Figures 22–27, 31–37) to the shock events of 2020 (Figures 29–30). The surges and declines observed in both C-side and B-side populations demonstrate that the platform's operation depletes without a sufficient mass on each side. Driver incentives can encourage driver mobilization, but negative forces (like curfews or pandemic restrictions) may induce simultaneous declines in demand and supply.

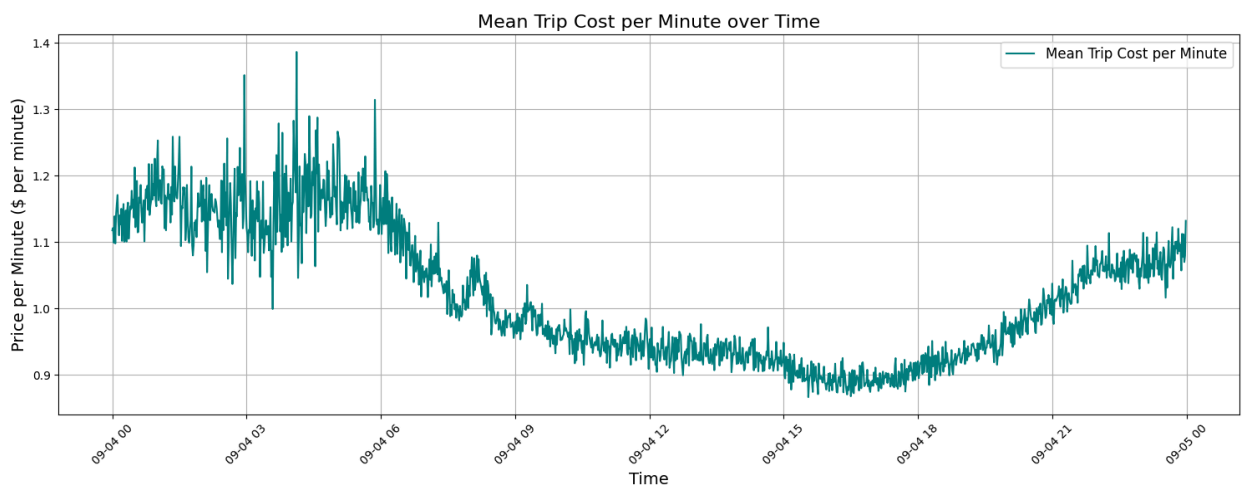

Figure 33. Uber in September 2024

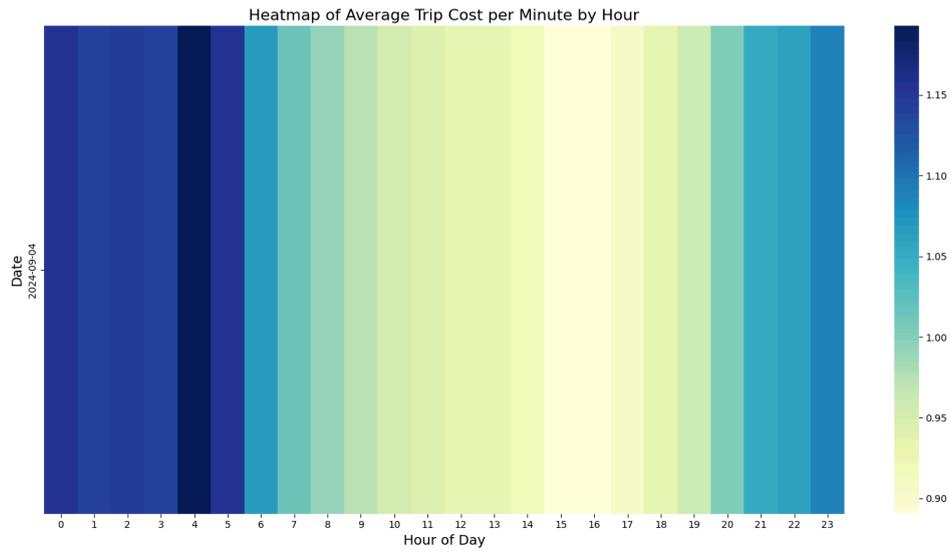

Figure 34. Uber in September 2024

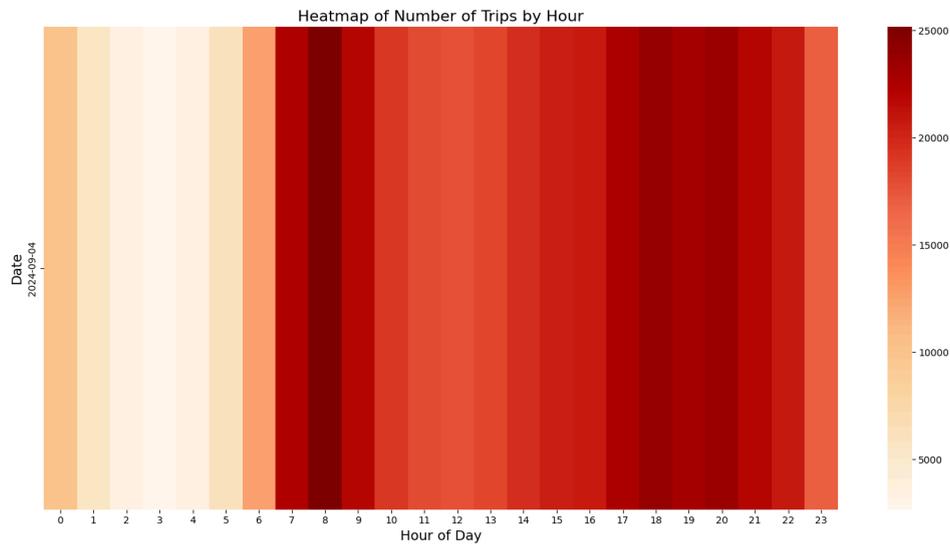

Figure 35. Uber in September 2024

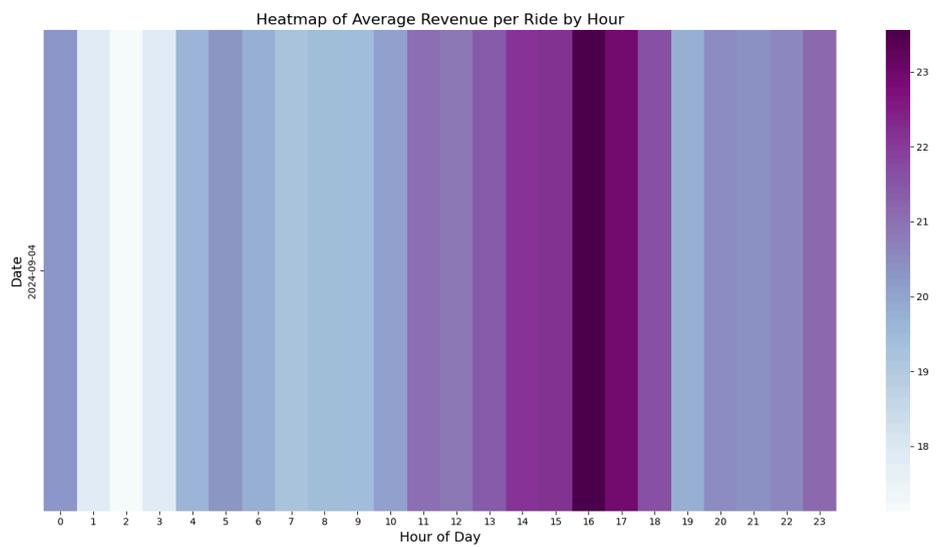

Figure 36. Uber in September 2024

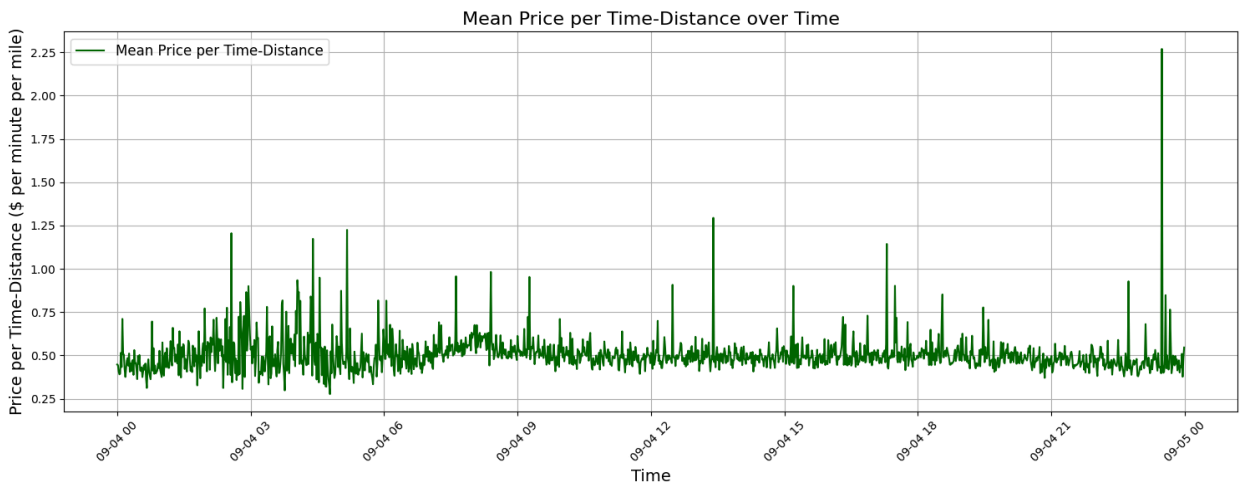

Figure 37. Uber in September 2024

Finally, these exploratory findings underscore the need for a deterministic, transparent quantification of cross-platform dynamics. A scientifically grounded approach, potentially rooted in physics-inspired formulations, may allow real-time optimization and deeper insight into the socio-technical factors driving the platform. Such a method promises a complementary alternative to machine learning alone. Deterministic models are computationally faster and more interpretable, benefiting strategic decisions on surge pricing, resource management, and scenario-based what-if analyses. They also yield realistic synthetic primary data to train AI tools while mitigating the risks of black-box "hallucinations" in neural networks. Altogether, the emergent patterns suggest that platform behavior is governed by feedback loops, efficiency parameters, and stability thresholds that can be systematically captured and leveraged for operational and strategic advantages.

In summary, the observed day-to-day fluctuations in ride requests, driver populations, trip durations, and revenue highlight how cross-platform network effects dynamically shape this B2C digital ecosystem (this addresses O2). Peak periods bring high ride volumes but do not always generate the highest individual trip cost, whereas less-busy times may have elevated average prices yet fewer rides overall . The synergy between empirical observation and theoretical modeling stands poised to advance both the practical implementation and scientific demystification of on-demand mobility platforms.

## 4.2 SYSTEMIC QUALITIES EMERGENT FROM THE DATA

The TLC quantitative data (NYC TLC, 2025) presented in the previous section revealed objective time-domain and frequency-domain patterns arising from the spatio-temporal real-time operations of Uber in NYC. Examining this data more closely may uncover socio-technical and socioeconomic indicators that suggest physical qualities similar to those in certain physical systems, thus motivating a physics-influenced theoretical framework.

### 4.2.1 Coherence

Focusing on driver transitions (supply side) that generate revenue (on-trip transitions) reveals noteworthy statistical properties. The hourly distribution of ride durations displays consistent behavior, as Figure 38 shows: median lines across the boxes nearly align along a straight horizontal axis, indicating that average trip duration remains largely unchanged throughout the day. Rush hours, work hours, nighttime conditions, and fluctuating demand or supply seem to have minimal effect on this core duration. The middle 50% of trip durations (box heights) also vary only slightly, while outliers appear statistically negligible.

Figure 39, which plots the per-minute mean trip duration over a 24-hour window. Most rides last about 18 minutes, with a tolerance of ±12 minutes. Excluding the busy afternoon rush and post-midnight hours refines this to roughly 19 minutes ±3 minutes. Figure 40 reveals that Sunday, 8 September 2024, demonstrates even tighter clustering, presumably due to lower traffic. Five years earlier (6 February 2019, Figure 41) a similarly stable middle 50% and median are observed despite frequent outliers.

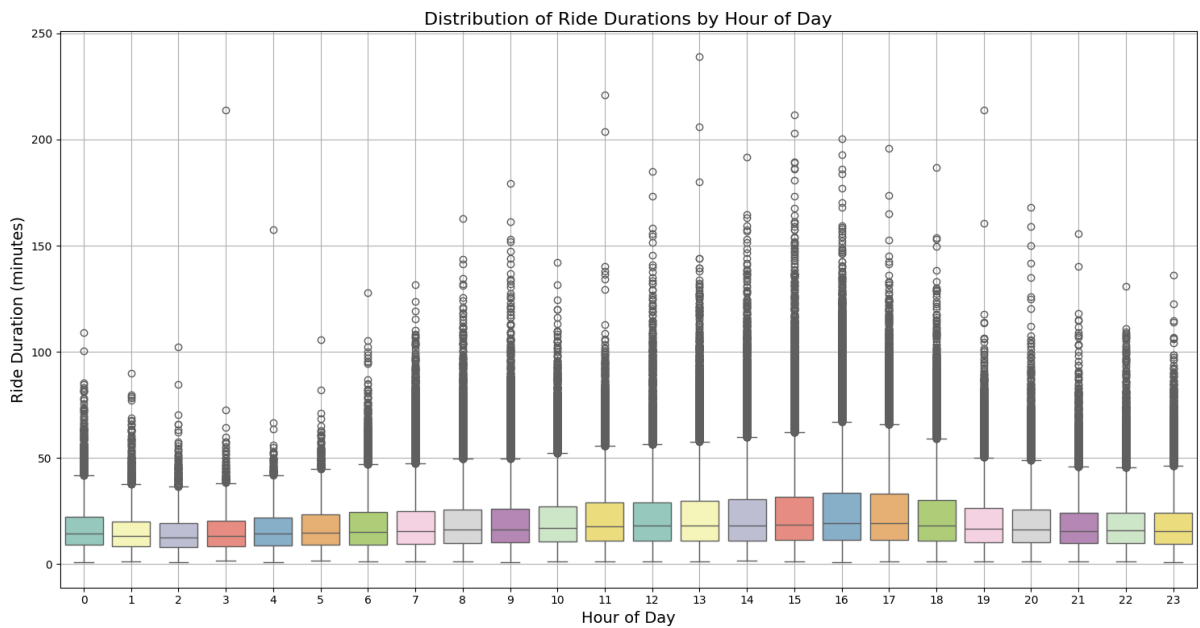

Figure 38. Wed Sep 4, 2024. Median lines across the boxes are nearly aligned

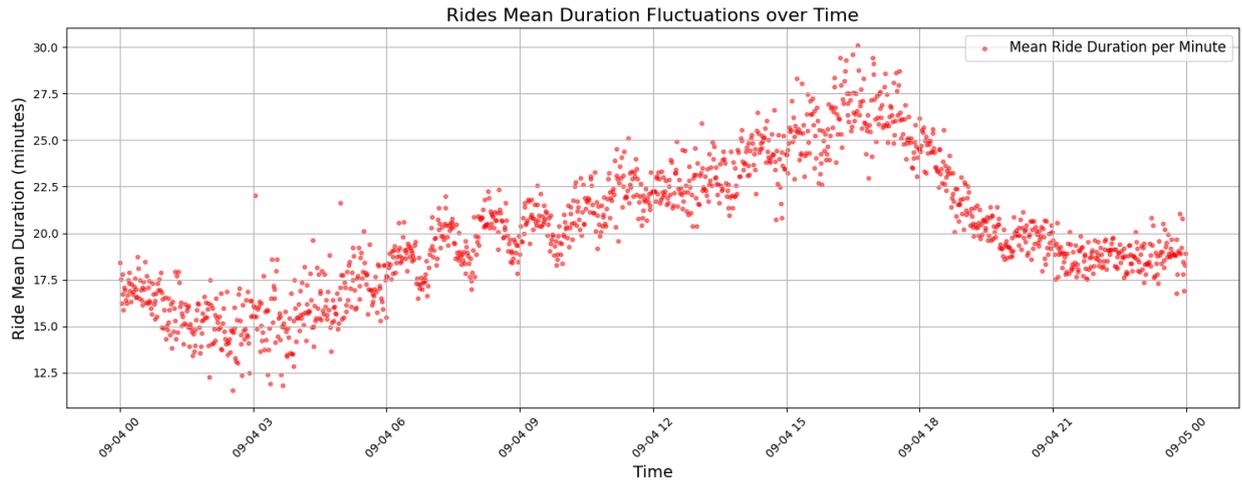

Figure 39. Wed Sep 4, 2024. Durations of the trips that ended during one-minute time slot over a 24-hour period.

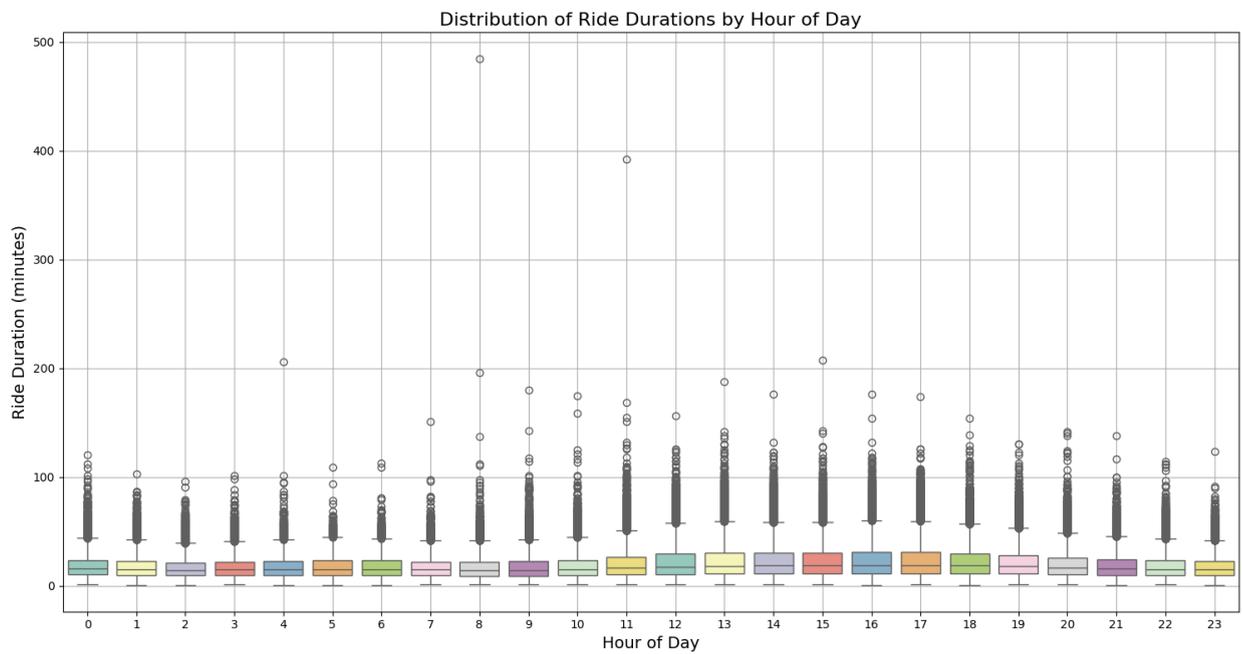

Figure 40 - Sunday, Sep 8 2024. The core of the distribution is even more consistent on Sunday

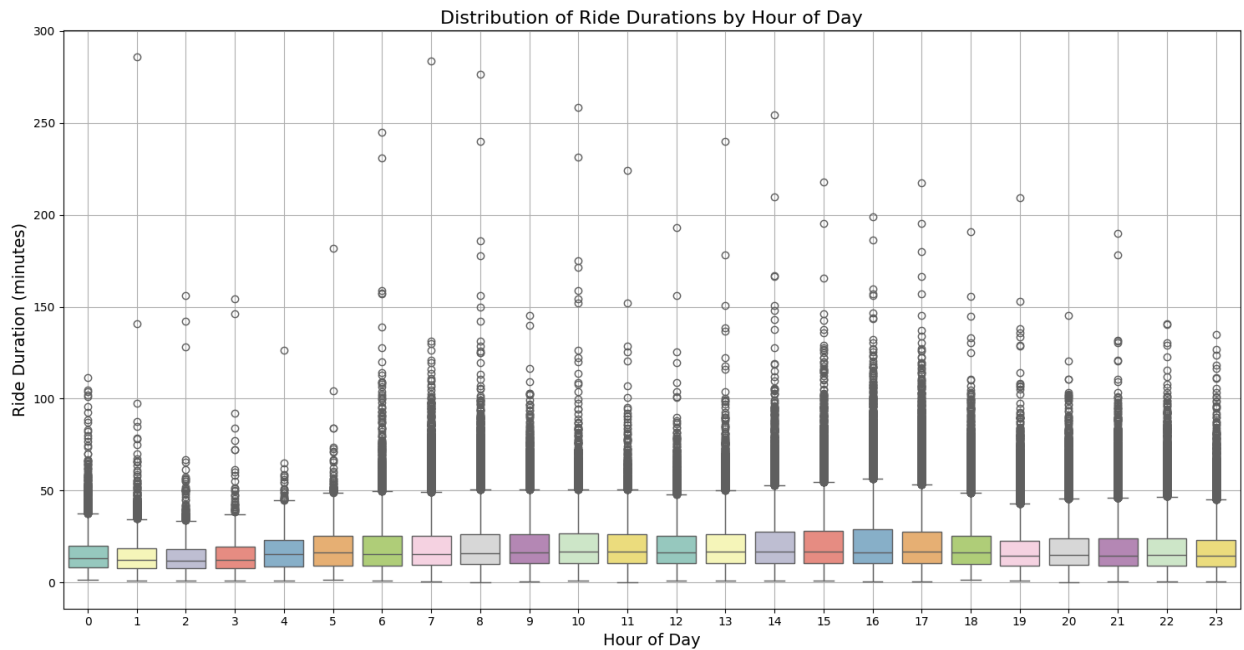

Figure 41. - Wed. 6 in Feb. of 2019. Remarkable alignment of the medians is also observed five years earlier.

Monetary stability complements these durable ride durations. In Figures 42–45, dashed lines depict stable median revenue (even during peak hours), suggesting consistency in average revenue per ride. The box-and-violin shapes reinforce the impression that the majority of rides remain statistically cohesive. Uber's operational mechanisms (e.g., surge pricing in Figures 33 and 37 for 4 September 2024) likely promote this longitudinal uniformity. Non-revenue transitions (e.g., enroute, onscene) also exhibit relatively stable distributions (Figures 46 and 47), hinting at efficient driver positioning and a smooth boarding process.

A deeper systemic quality emerges: the entire socio-technical and socio-economic system of Uber appears organized so that coherence in revenue generation supports profit growth (this addresses O2). The main mode of rides, lasting about 18 minutes at around USD 22 each, seems integral to this profitability. Positive cross-platform network effects align with real-time optimization and socio-technical synchronization, resulting in a coherent, revenue-focused platform design.

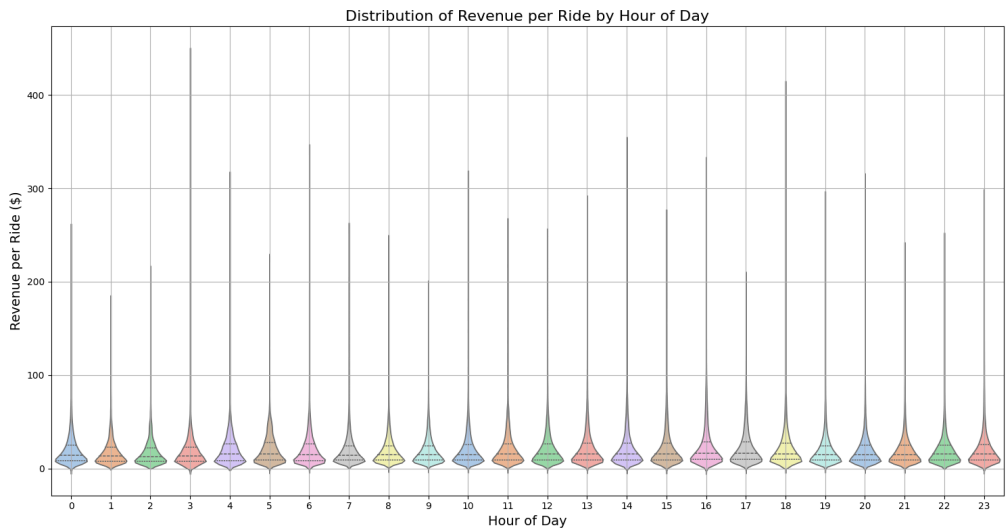

Figure 42. Wed 4, Sep 2024. The median revenue (dashed lines) is particularly stable.

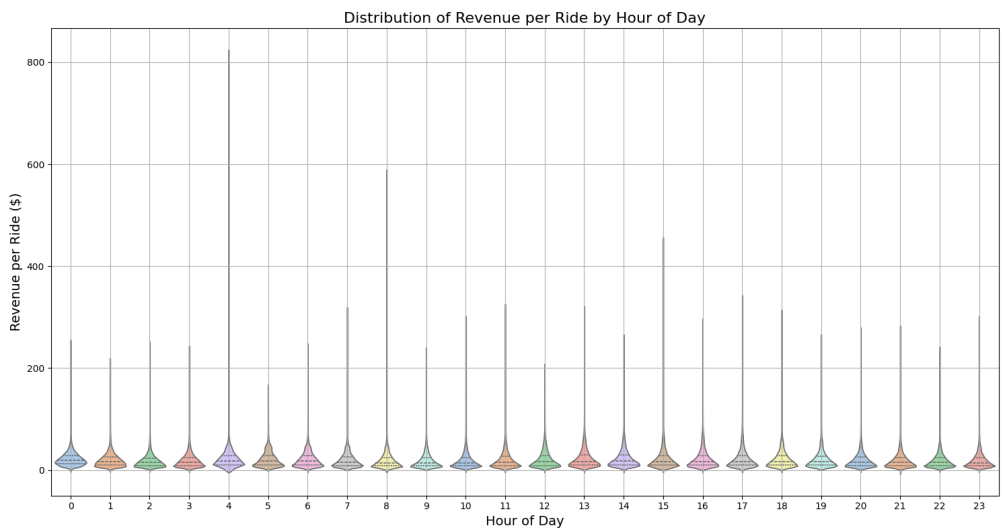

Figure 43. Sunday, Sep 8 2024. The median revenue (dashed lines) is particularly stable.

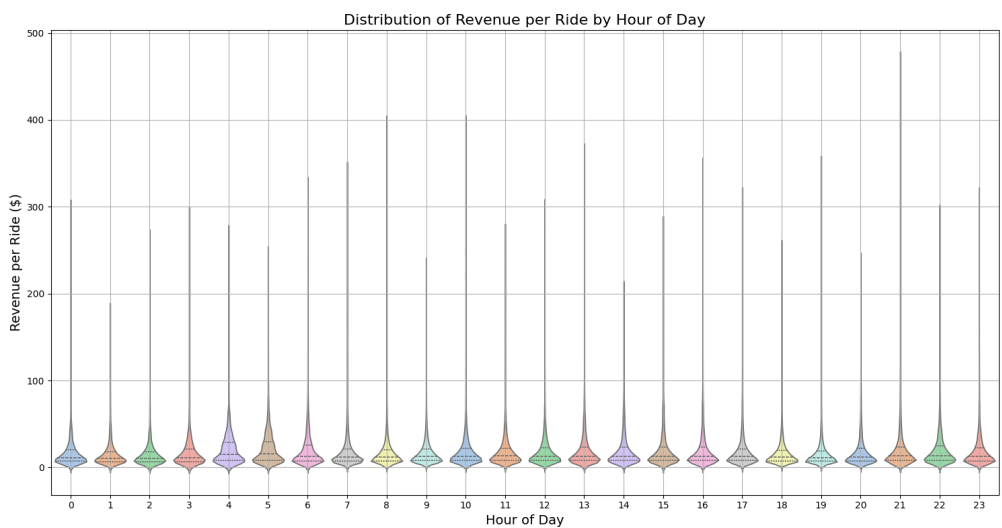

Figure 45. Wed. 6 in Feb. of 2019. The median revenue (dashed lines) is particularly stable.

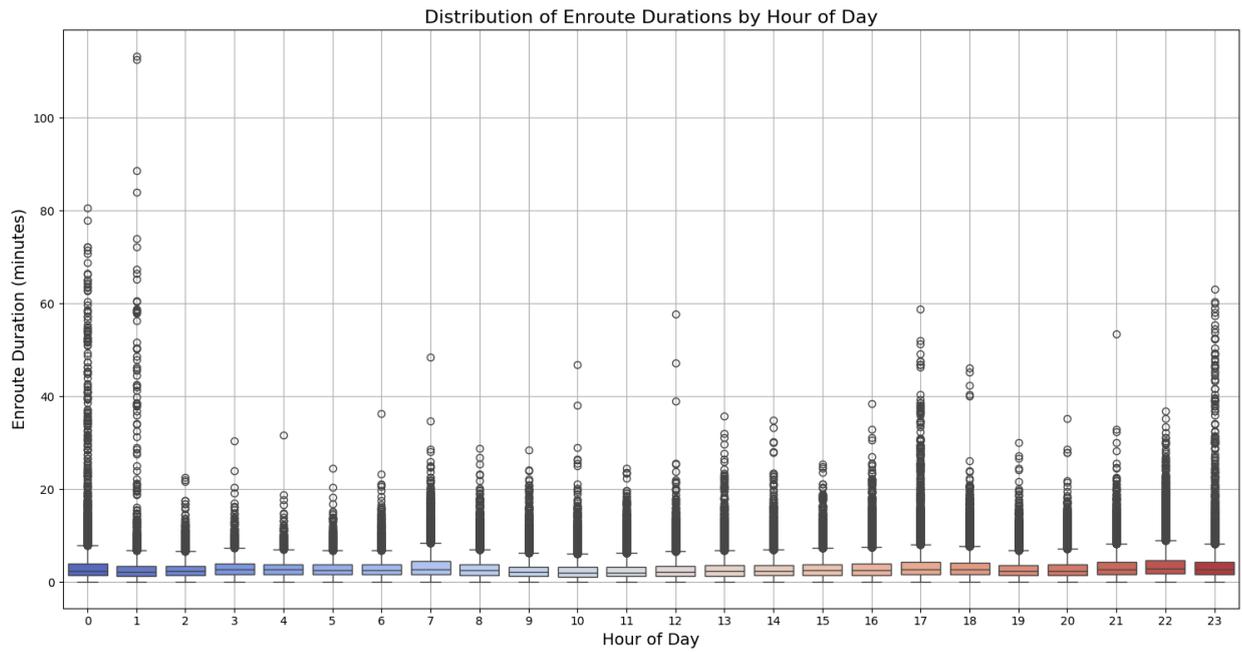

Figure 46. Wed 4, Sep 2024. Uber positions its drivers close to customer locations

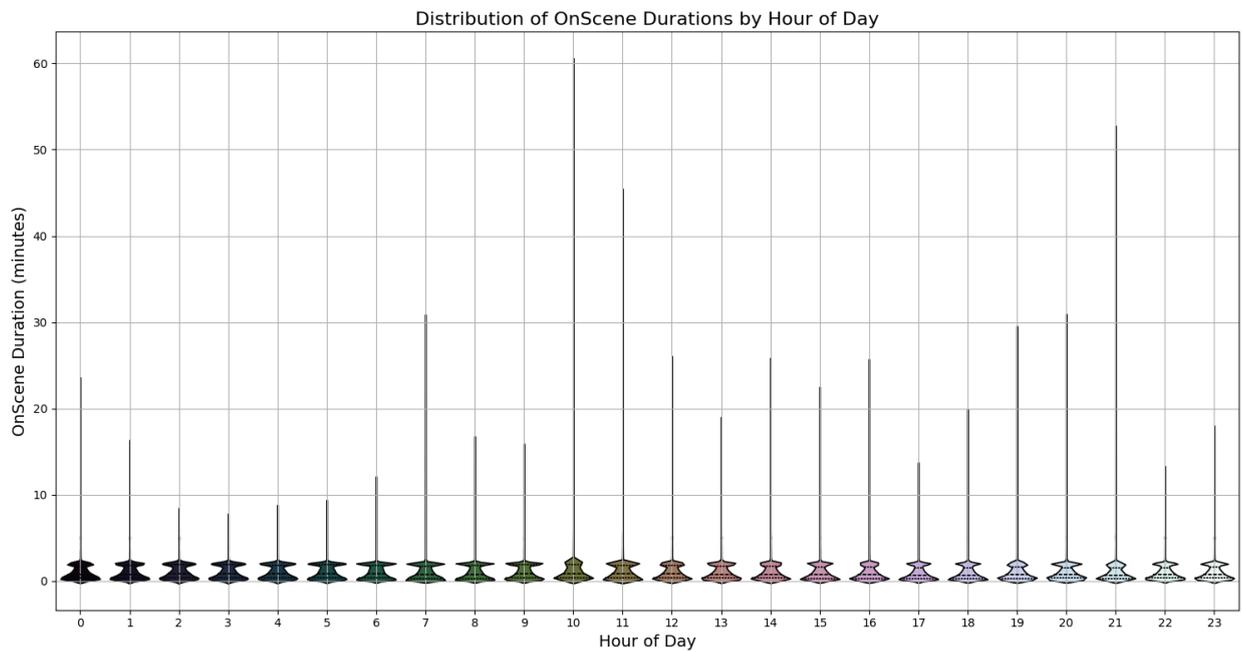

Figure 47. Wed 4, Sep 2024. Uber runs the boarding process consistently and predictably.

### 4.2.2 Population inversion

Data from Figures 17, 18, 21, 27, and 30 suggest that, at virtually any minute in the observed windows, the on-trip driver population ($N\_ontrip$) substantially exceeds the sum of all other states combined ($N\_ontrip \gg N\_available + N\_waiting + N\_enroute + N\_onscene$). This ratio (about 4±1 over the past five years) implies that the platform allocates the bulk of its drivers to revenue-generating tasks at all times, thereby achieving significant asset-light efficiency.

### 4.2.3 Critical mass

The night of 6 June 2020, during the pandemic (shown in Figure 30), illustrates how Uber recovered in real time after several hours of zero activity. Figure 48 focuses on this recovery interval, when the on-trip population inversion was initially absent. Figure 49 shows that about 80 ride requests per minute occurred during the first half-hour, prompting driver mobilization. Because curfews precluded drivers from positioning themselves early, Figure 50 records longer enroute durations before a stable operational state reemerged. After two hours, the dynamic on-trip population once again surpassed all other driver states, restoring population inversion.

Under these conditions, approximately 25 online drivers per minute per ride request coincided with stable network effects and reliable platform operations. Five years of NYC TLC high-volume data reinforce the observation that a stable supply-demand ratio of about 25±5 was routinely maintained by Uber. The minimal sustained median ride requests per minute during stable conditions also hovered around 25.

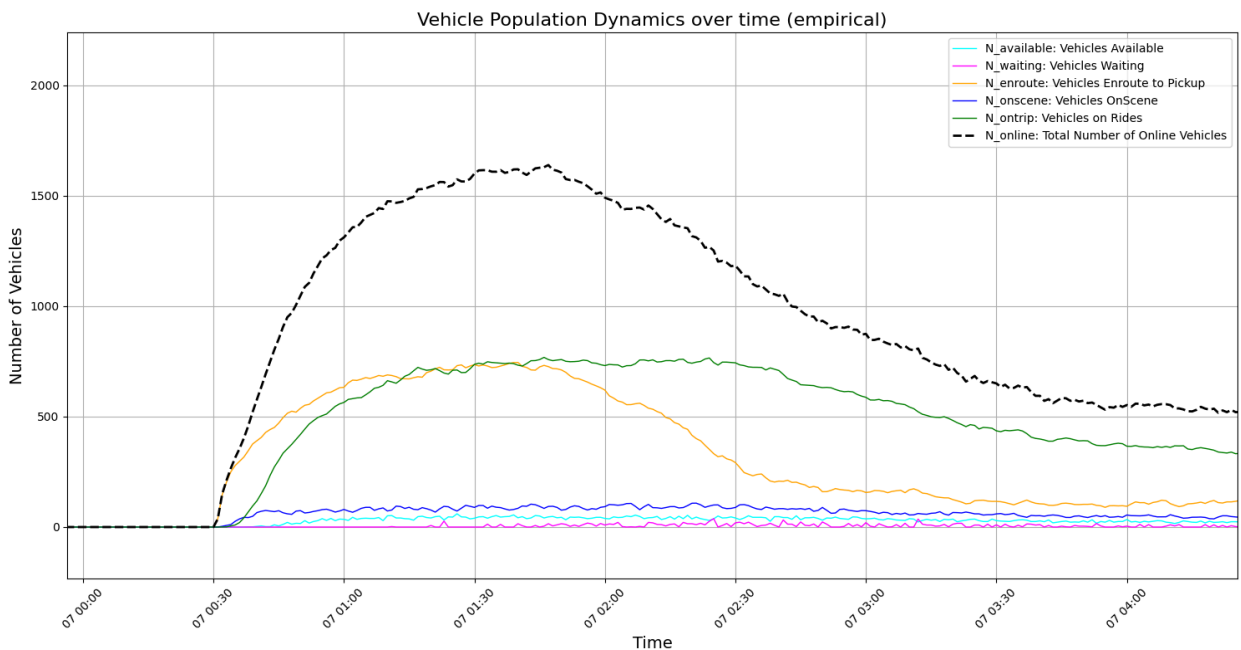

Figure 48 The recovery dynamics of drivers or their vehicles populations

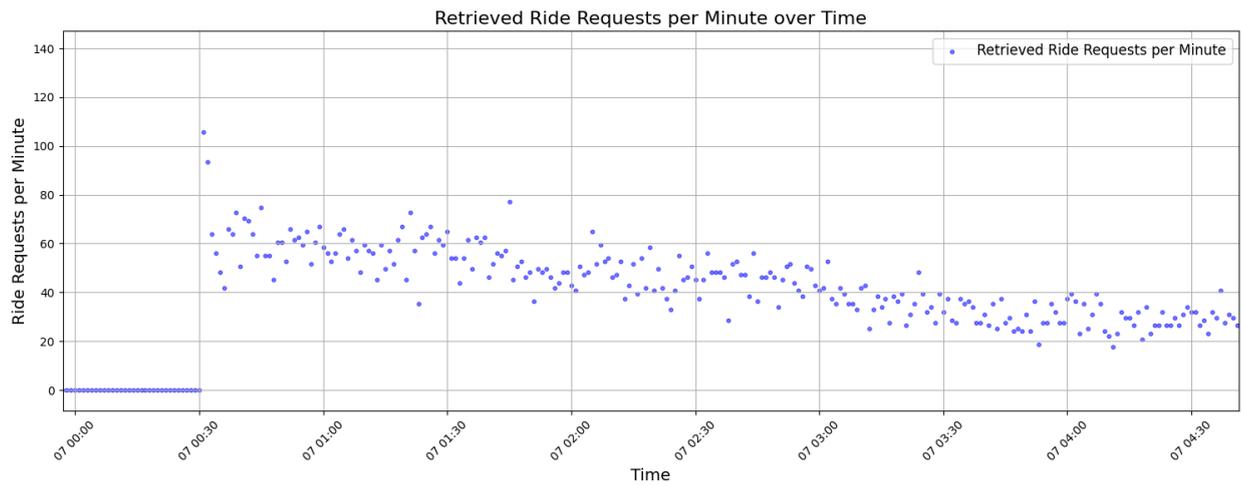

Figure 49. The per minute aggregated ride requests during the recovery process

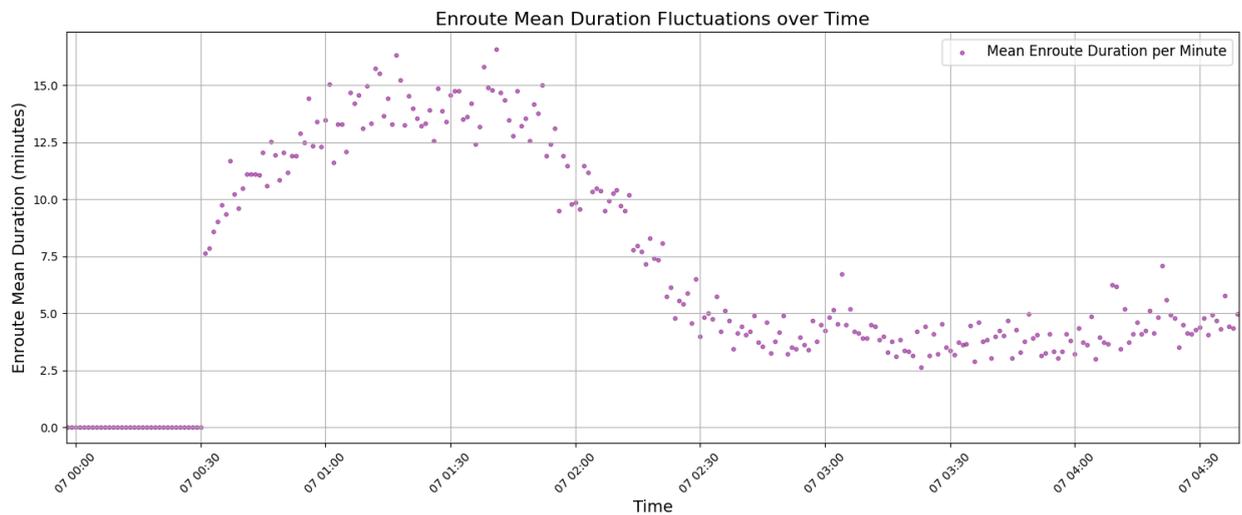

Figure 50. Prolonged enroute durations are observed during the recovery period.

### 4.2.4 Stimulation and the platform's virtual cavity

Once critical mass and stable operations are achieved, a sufficient threshold of on-trip drivers ("main mode" rides) appears to stimulate additional similar trips in both time and space. This "stimulating event" stems from a core group of profitable rides that exhibit temporal-spatial coherence. Through network effects, these main mode trips enhance platform attractiveness, spurring more customers to request comparable rides.

The mobile applications function like virtual reflectors, confining these feedback loops inside a resonant "cavity" and amplifying trust among both sides of the platform. Dynamic pricing and user interfaces are repeatedly reflected in the applications, reinforcing profitable system states. In essence, one main mode ride triggers another, fostering self-sustaining "profit emission."

### 4.2.5 Drivers mobilization

Incoming ride requests mobilize drivers by transitioning them from "available" (ground state) to "waiting" (highest operational state), depicted in Figure 51. They then descend through successive states (enroute, onscene, on-trip) before dropping to a post-trip state that eventually returns them to "available."

If the total driver population remains constant (offline + online) over a day, this mobilization process maintains a stable group of on-trip drivers at any point, thereby preserving population inversion. As a result, the system can offer coherent and efficient trip fulfillment, ensuring that real-time demand meets real-time supply.

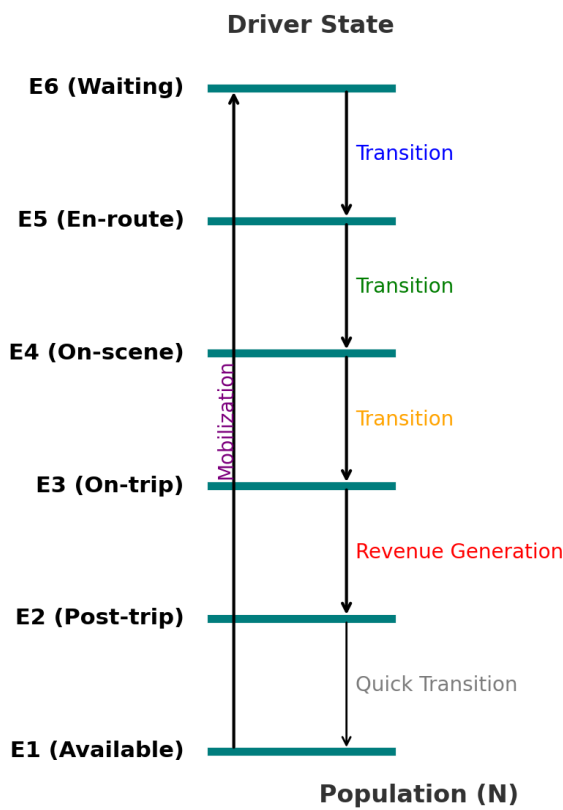

Figure 51. The drivers' mobilization from ground state to most energetic state is the only upward transition.

### 4.2.6 Systemic behavior

Collectively, coherence, population inversion, critical mass, stimulation in a virtual cavity, and drivers mobilization are tightly coupled phenomena. They appear to function as parts of a unified, complex system. The next section introduces a framework to quantify these interdependencies in a coherent manner.

### 4.3 Section summary

The findings in this section confirm the existence of coherent operational metrics and threshold-based behaviors in on-demand ride-hailing. Coherence, critical mass, driver mobilization, and stimulated network effects all emerge as integral to stable, profitable operations. These patterns contextualize the subsequent discussion in Section 5, where the theory and its validation are articulated in a physics-based framework, culminating in real-world optimization applications and scenario testing.

## 5.    DISCUSSION

This section reinterprets the empirical evidence in light of the physics-based perspective. It introduces the model, mapping the emergent patterns observed in Section 4 to laser-physics analogies like coherence, population inversion, and stimulated emission. Through a suite of validation exercises (including extreme shock events) the section demonstrates how the model accurately captures both steady-state and transient behaviors. Also a real-time operations optimization framework is examined to show how the theory translates data-driven insights into actionable strategies.

### 5.1 ANALOGICAL THINKING - Lasers as analogous to on-demand Platforms

The data underscore operational similarities between Uber's real-time dynamics and the well-known physical device called a laser (Light Amplification by Stimulated Emission of Radiation). Laser physics are mathematically robust, extensively tested in laboratories, and embedded in everyday technologies.

Traditional businesses can be linked to incandescent lamps, which emit diffuse light across broad spectra at low efficiency of up to 10%. By contrast, semiconductor lasers concentrate energy into a focused, monochromatic beam, converting 30–70% of input energy into light with minimal waste. Table 2 in appendix A provides an abstract comparison of traditional "lamp-like" business operations vs. "laser-like" asset-light platform models.

By mapping Uber rides to photons in a laser system, a physics-inspired platform theory emerges. The subsequent sections detail these parallels and formulate the corresponding laser-based PASER mathematical model (Section 4.4). The intention is to illustrate how proven laser equations can be adapted to clarify network effects, trust formation, and population inversion in on-demand services like ride-hailing.

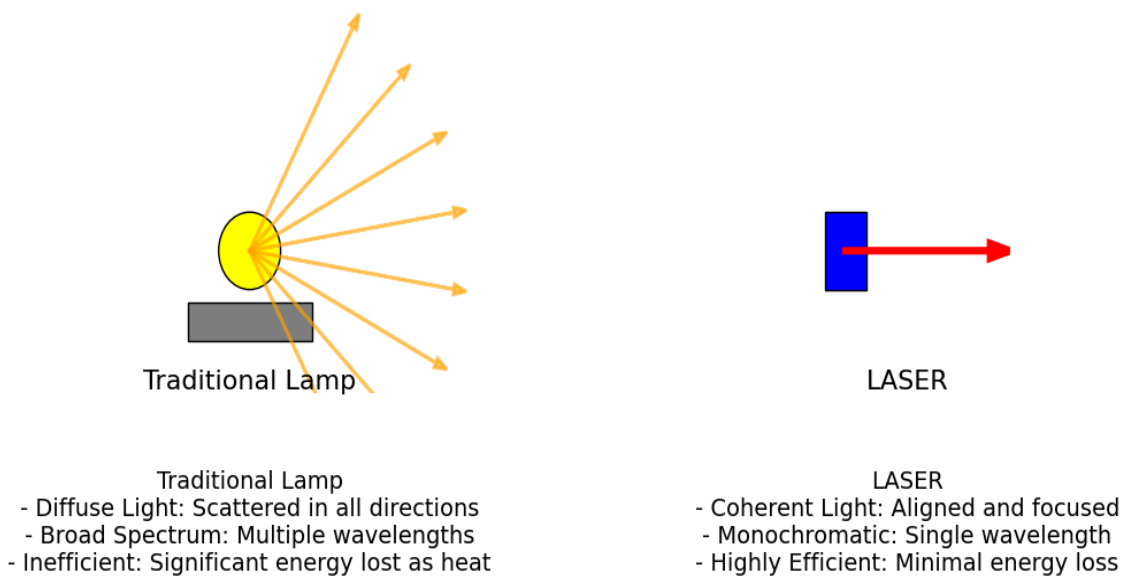

Traditional Lamp
- Diffuse Light: Scattered in all directions
- Broad Spectrum: Multiple wavelengths
- Inefficient: Significant energy lost as heat

LASER
- Coherent Light: Aligned and focused
- Monochromatic: Single wavelength
- Highly Efficient: Minimal energy loss

*Business Parallel:*

**Diffuse and scattered revenue generation through broad strategies**.

*Business Parallel:*

**Coherent and focused revenue generation through optimized, scalable strategies.**

Figure 52. Lamps versus lasers. Traditional businesses versus platforms.

## 5.2 ANALOGICAL THINKING IN DEPTH

In a laser, photons represent fundamental output units of light, just as completed rides on a digital platform constitute fundamental revenue-generating events. Photons collectively form a laser beam, while completed rides, taken together, generate a platform's profit stream. This analogy extends further when considering the laser's gain medium, which amplifies light by maintaining electrons in excited states until they release photons. Analogously, a platform's pool of drivers operates as the "gain medium," since these supply agents are ready to convert ride requests into completed trips, effectively amplifying demand into profitable outcomes.

The concept of beam coherence in a laser resonates with the idea of "profit coherence" in a platform. A laser beam remains highly coherent (photons vibrate in sync with stable frequency) whereas a platform experiences coherent profit flow when the majority of rides align around similar durations, distances, and revenues. Figures 53 and 54 highlight that, when observed statistically, both a laser beam's output and a platform's real-time profit (e.g., in Uber or Lyft) share analogous properties in terms of overall consistency.

A laser's threshold must be exceeded before it emits a steady, coherent beam: only when excited electrons surpass a critical density does lasing begin. In a platform, a similar "critical mass" of supply and demand must be reached for network effects to become self-sustaining. Data plots previously shown in figures 48 and 49 illustrate how demand and supply can build up to initiate strong revenue emissions, whereas figures 55 and 56 reveal how both demand and supply can deplete during a curfew, causing the "lasing" (that is, active revenue generation) shown in figure 57 to shut down. Without adequate driver-rider synergy, operations stagnate and profit "emission" collapses.

The analogy further includes the concept of quenching in a laser: when there is too high an electron density or an inefficient environment, the emission process can be suppressed rather than amplified. Likewise, an oversupply of drivers relative to demand can dilute overall earnings, prolong waiting times, and diminish the per-driver revenue (essentially quenching profitability). In lasers, a single photon can stimulate an excited electron to release another identical photon, a process known as stimulated emission. In platforms, one successful trip tends to stimulate additional trips by fortifying trust, reducing wait times, and making the platform more attractive overall..

Electron population inversion is another key idea in laser physics, where the population of electrons in an excited state must greatly exceed that in lower energy states to sustain continuous lasing. On a platform, a "driver population inversion" arises when a much larger proportion of drivers are actively engaged (on-trip) than in the post-ride state, thereby maximizing the revenue that on-trip drivers produce. Electrons in atoms transition through discrete energy levels. Only certain transitions generate light (photons), while others do not. Drivers similarly transition through states (available, waiting, enroute, on-scene, on-trip) with only the on-trip interval generating revenue. Figure 58 visually presents a six-level laser system mapped to these ride-hailing driver states, reflecting how modern platforms keep drivers efficiently cycling through profitable transitions. For instance, Uber's system encourages drivers to drop quickly to the lowest state (available) before being "pumped" or mobilized back to the waiting state, where they can accept a new request.

This notion of pumping parallels driver mobilization. In lasers, external energy lifts electrons from the ground state to a higher energy band. In ride-hailing, new trip requests and incentives mobilize drivers from idle to waiting states, preparing them to accept matches that yield new revenue. However, not every request effectively mobilizes a driver. Thus, pumping efficiency translates to how many idle drivers become fully engaged per new request. Homogeneous or inhomogeneous broadening in lasers, represents the uniformity or diversity in how the medium responds. Likewise, ride-hailing supplier mass in platforms may exhibit the same properties depending on the uniformity or diversity of driver habits, conditions, pricing, and training.

A laser that emits a narrow linewidth is prized for spectral purity. Analogously, a platform with stable and focused "ride modes" (e.g., a core type of urban ride) achieves revenue purity (fewer disruptions to hamper profitability). Mirror quality in the laser cavity, ensuring maximum reflection and minimal losses, can be linked to the interface quality of rider and driver apps. Well-designed apps "reflect" positive behaviors by ensuring quick matches, trust-building feedback, and frictionless transactions. Equally important is modal confinement, where lasers restrict light to certain spatial patterns. Platforms do the same by focusing on profitable trip modes instead of diluting their operational space.

In lasers, adjusting cavity length affects mode structure, allowing multimode operations. Similarly, a platform might add new services (like deliveries) to expand operational modes under the same infrastructure. When more than one mode is supported, mode competition arises. Losses in lasers due to photon diffraction or reabsorption represent inefficiencies in platforms, such as slow driver responses or trust erosion. Thermal management, crucial to laser stability, becomes friction management in a platform, preventing wait times or driver frustration from "overheating" the system.

The density of photons inside the laser (intracavity photon density) parallels the momentary density of rides, which drives near-real-time revenue. Only a fraction of these photons exit through the partially reflective mirror, forming the actual output beam. This is akin to the fraction of revenue a platform "takes" as a commission. By fine-tuning the degree of output coupling or commission rates, the platform balances profitability with stability.

A platform's infrastructure provides an operational "cavity," confining interactions among supply and demand. Figures 59 and 60 depict how, in a laser's actual cavity, an external energy source excites electrons in the gain medium, generating an avalanche of photons. In the platform, ride requests pump drivers into excited states that produce revenue. Once critical mass is reached and losses are lower than gains, profit surges. Laser stability depends on environmental conditions, while platforms react to policy changes, economic disruptions, or sudden shifts in user behavior. Altogether, the synergy of laser components and the synergy of platform drivers, riders, infrastructure, and algorithms share multiple parallels.

Finally, each constituent in a laser, from gain medium to mirror alignment, must align properly to generate a stable beam. In a PASER, the stimulation of coherent rides, the drivers mobilization, and the management of losses likewise combine to optimize operations. Coherence, critical mass, stimulation, mobilization, and system optimization each reinforce the others, reflecting how an asset-light platform can emulate the finely tuned dynamics of a high-efficiency laser. See table 4 in appendix A for a comprehensive list of the analogies drawn in this section.

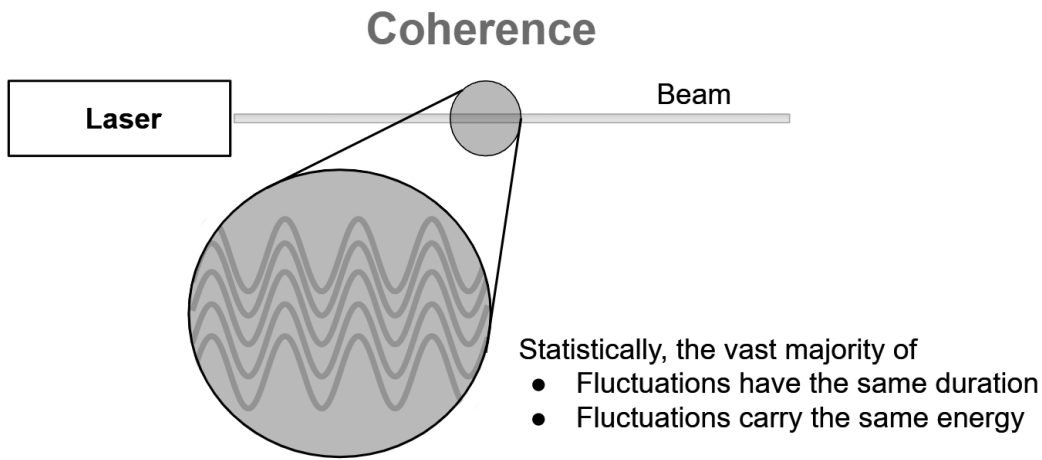

Figure 53. A LASER's beam statistical properties

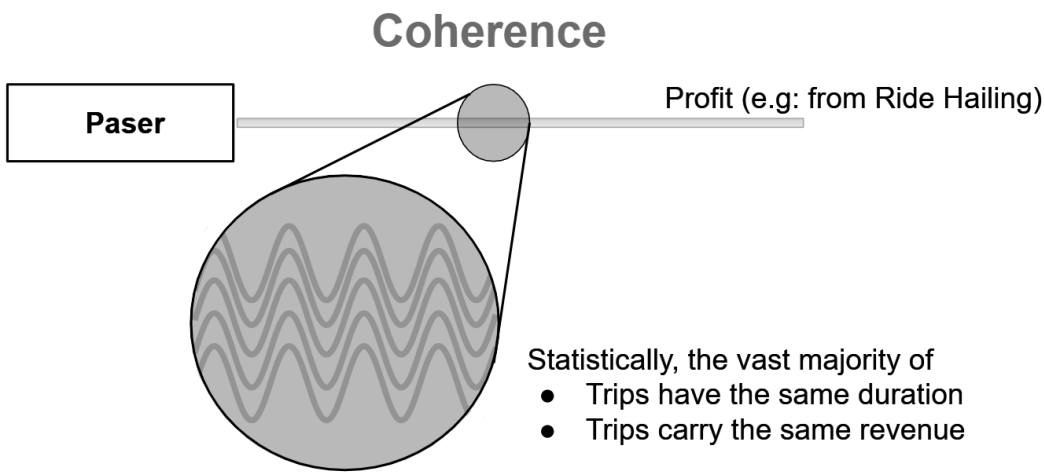

Figure 54. A PASER's revenue and profit statistical properties

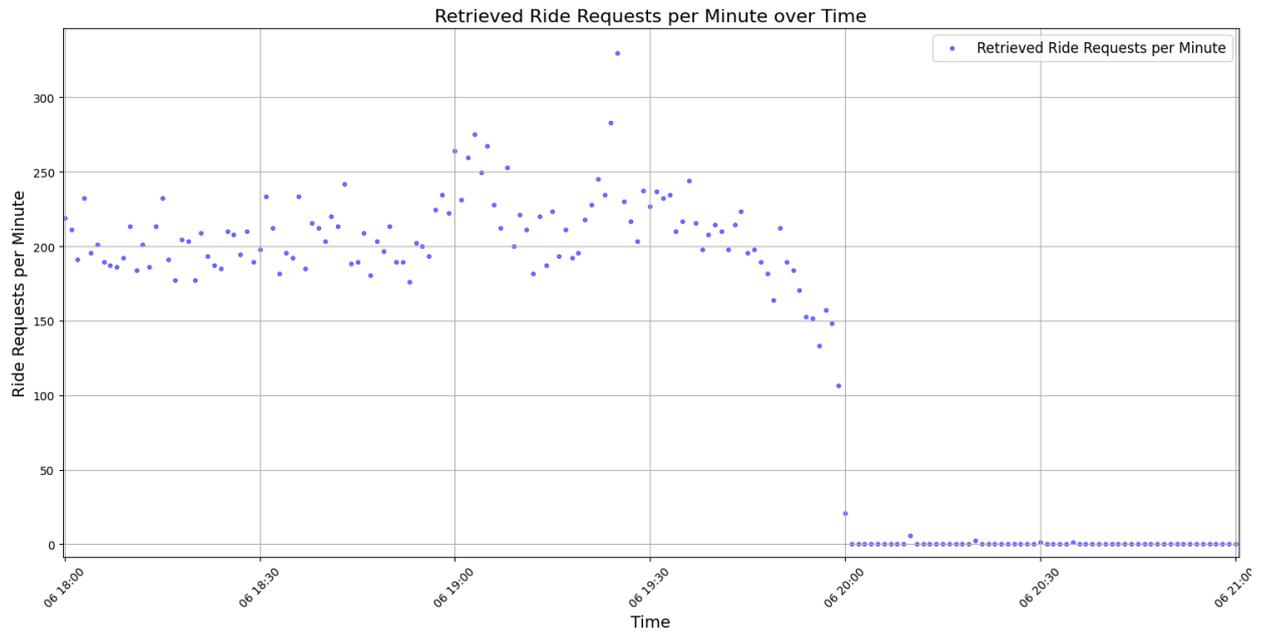

Figure 55. Uber's demand depletion during a curfew period in June 2020

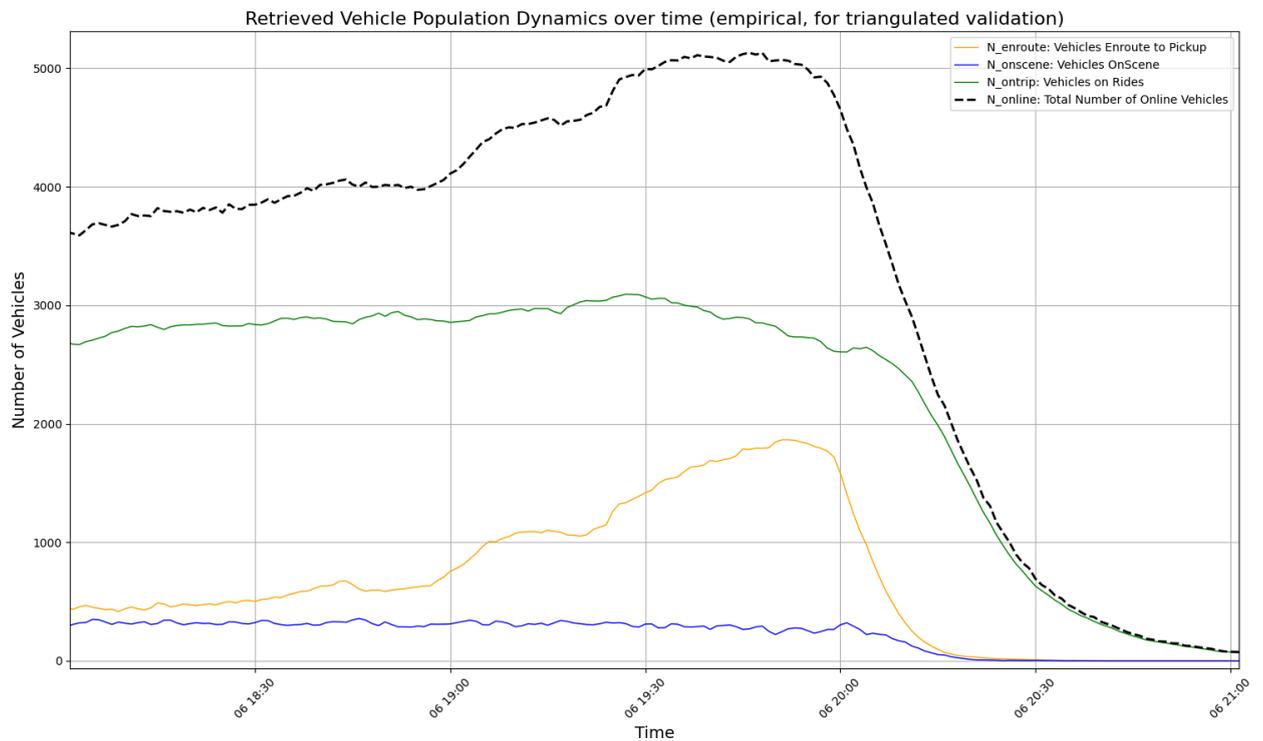

Figure 56. Uber's supplier mass depletion during a curfew period in June 2020

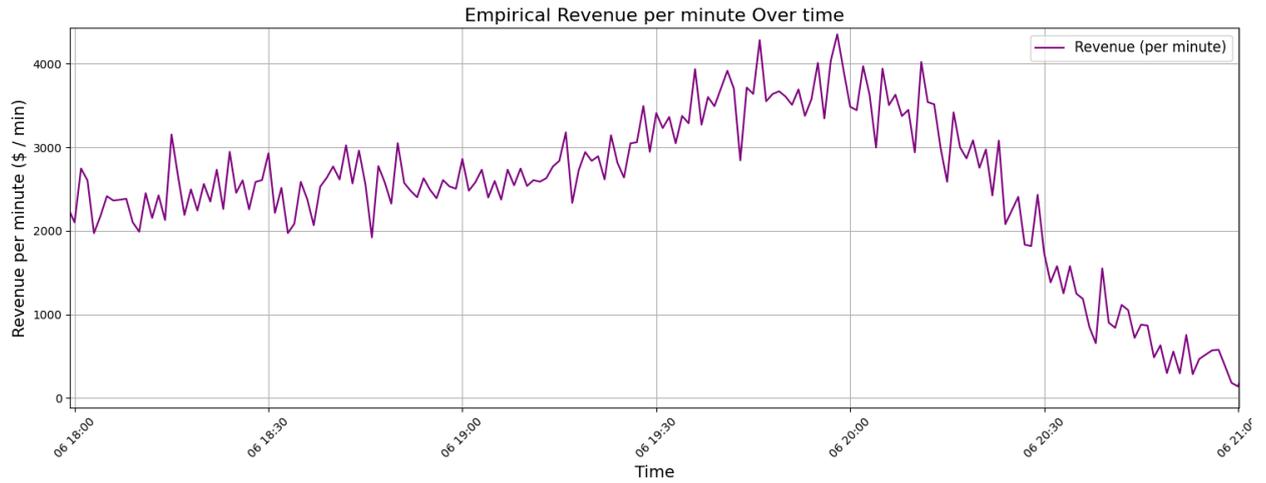

Figure 57. Uber's revenue depletion during a curfew period in June 2020

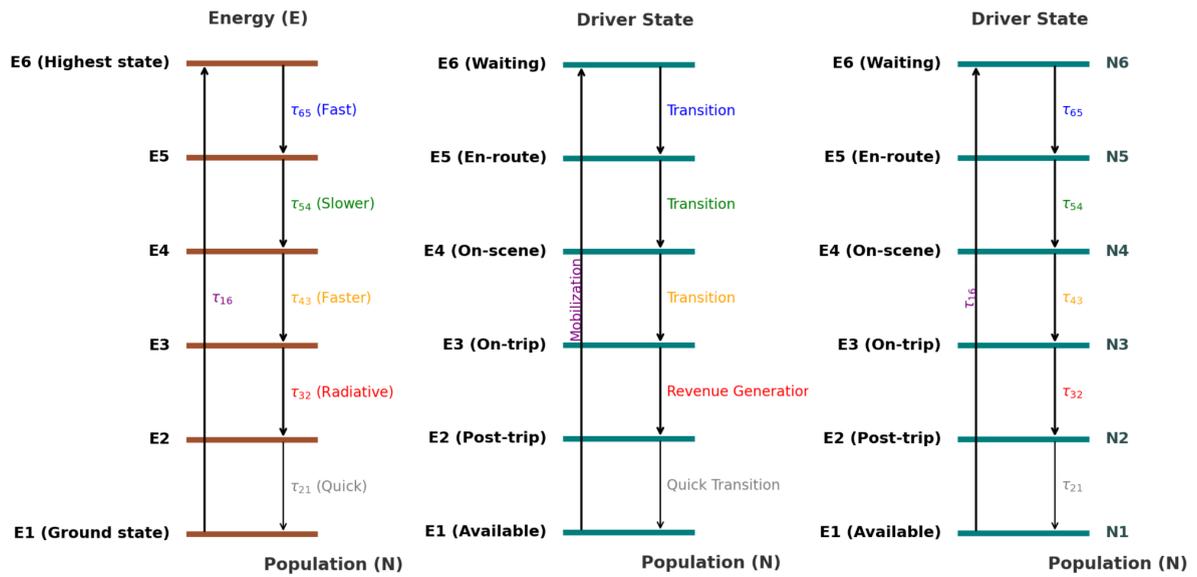

Figure 58. Uber and Lyft driver states mapped to a six level laser system

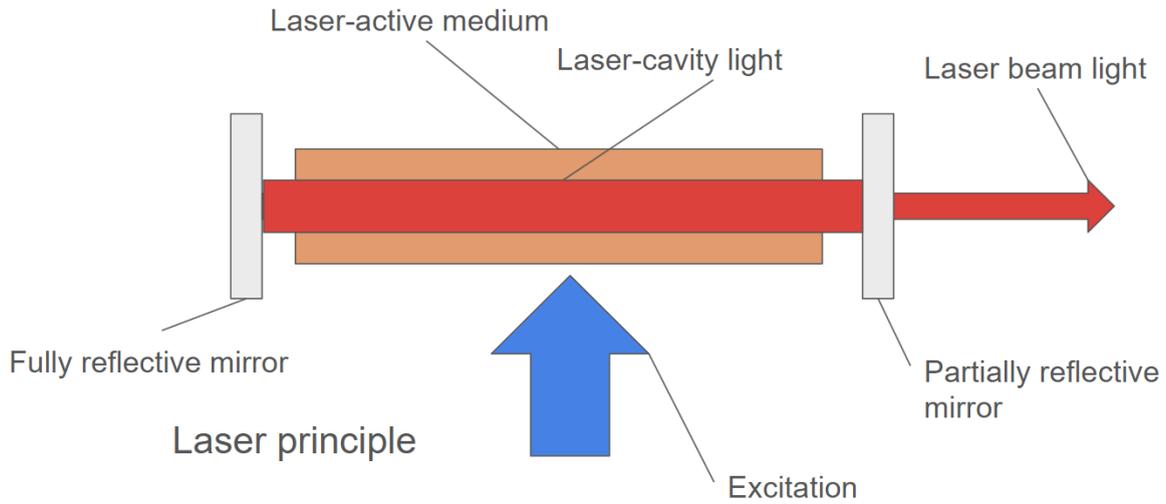

Figure 59. LASER cavity

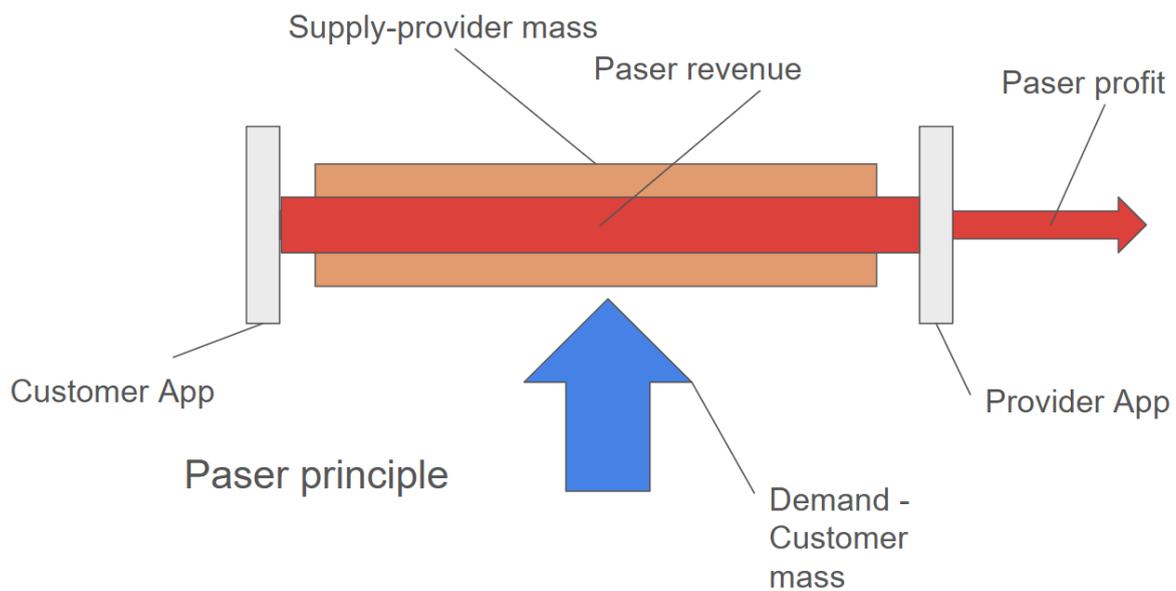

Figure 60. PASER conceptual cavity illustration offering a high-Level view of its principle of operation. - See table 3 in appendix A for a list of analogies with a laser cavity.

## 5.3 QUANTIZATION - PASER MATHEMATICAL MODEL AND EQUATIONS

The PASER framework's mathematical model draws directly from laser physics, offering a structured and novel approach for understanding and optimizing platform dynamics in depth (this addresses O3). The following system of coupled Ordinary Differential Equations (ODEs) encapsulates, in a mathematically elegant manner, the transitions of service providers and the generation of revenue in real time.

Note that the temporal flow of drivers (or vehicles, including autonomous vehicles) through the PASER "states" or "energy levels" follows the closed loop: $1 \to 6 \to 5 \to 4 \to 3 \to 2 \to 1$.

**Model Variables, Parameters, and Notation**

- **Variables:**

  - $N_1$: Number of available vehicles (State $E_1$)

  - $N_2$: Number of post-ride vehicles (State $E_2$)

  - $N_3$: Number of vehicles on rides (State $E_3$)

  - $N_4$: Number of vehicles on scene (State $E_4$)

  - $N_5$: Number of vehicles enroute to pickups (State $E_5$)

  - $N_6$: Number of waiting vehicles (State $E_6$)

  - $\Phi$: Revenue density

- **Parameters:**

  - $W_p$: Vehicle mobilization rate - pumping (vehicles per minute)

  - $\tau_{65}$: Waiting period before dispatch ($E_6 \to E_5$)

  - $\tau_{54}$: Enroute duration ($E_5 \to E_4$)

  - $\tau_{43}$: On-scene duration ($E_4 \to E_3$)

  - $\tau_{32}$: Ride duration ($E_3 \to E_2$)

  - $\tau_{21}$: Post-ride recovery time ($E_2 \to E_1$)

  - $\tau_c$: Revenue lifetime ( tau cavity - the cavity determines its value thus set equal to the mean of the trip durations distribution gaussian fit.

  - $\Gamma$: Efficiency factor of revenue emission

  - $\beta$: Fraction of spontaneous emission of revenue

  - $\sigma$: Stimulated emission cross-section - platform's technical quality

  - $\alpha$: Coefficient of the agent's (supplier's) trust to the platform

**Seven Coupled Simultaneous ODEs**

1. **Available Vehicles ($N_1$):**

$$\frac{dN_1}{dt} = \frac{N_2}{\tau_{21}} - W_p N_1 \tag{1}$$

- **Inflow**: Vehicles transitioning from post-ride state $(N_2)$ back to the available state $(N_1)$ at rate $\frac{N_2}{\tau_{21}}$.

- **Outflow**: Vehicles being mobilized to waiting state $(N_6)$ at rate $W_p N_1$.

2. **Post-trip Vehicles** $(N_2)$:

$$\frac{dN_2}{dt} = \frac{N_3}{\tau_{32}} - \frac{N_2}{\tau_{21}} \tag{2}$$

- **Inflow**: Vehicles completing rides $(N_3)$ an entering post-ride at rate $(\frac{N_3}{\tau_{32}})$.

- **Outflow**: Vehicles transitioning back to available state $(N_1)$ at rate $\frac{N_2}{\tau_{21}}$.

3. **Vehicles on Rides** $(N_3)$

$$\frac{dN_3}{dt} = \frac{N_4}{\tau_{43}} - \frac{N_3}{\tau_{32}} \tag{3}$$

- **Inflow:** Vehicles departing from on-scene state $(N_4)$ to start rides at rate $\frac{N_4}{\tau_{43}}$.

- **Outflow:** Vehicles completing rides and moving to post-ride state at rate $\frac{N_3}{\tau_{32}}$.

4. **Vehicles On Scene** $(N_4)$

$$\frac{dN_4}{dt} = \frac{N_5}{\tau_{54}} - \frac{N_4}{\tau_{43}} \tag{4}$$

- **Inflow:** Vehicles arriving on scene from enroute state $(N_5)$ at rate $\frac{N_5}{\tau_{54}}$.

- **Outflow:** Vehicles starting rides and transitioning to on-ride state $(N_3)$ at rate $\frac{N_4}{\tau_{43}}$.

5. **Vehicles Enroute** $(N_5)$

$$\frac{dN_5}{dt} = \frac{N_6}{\tau_{65}} - \frac{N_5}{\tau_{54}} \tag{5}$$

- **Inflow:** Vehicles dispatched from waiting state ($N_6$) to enroute state at rate $\frac{N_6}{\tau_{65}}$.

- **Outflow:** Vehicles arriving on scene at rate $\frac{N_5}{\tau_{54}}$.

6. **Waiting Vehicles ($N_6$)**

$$\frac{dN_6}{dt} = W_p N_1 - \frac{N_6}{\tau_{65}} \tag{6}$$

- **Inflow:** Vehicles mobilized from available state ($N_1$) to waiting state at rate $W_p N_1$.

- **Outflow:** Vehicles dispatched to enroute state ($N_5$) at rate $\frac{N_6}{\tau_{65}}$.

7. **Rate Equation for the Revenue Density ($\Phi$) - assumes a conceptual volume for the conceptual platform cavity**

$$\frac{d\Phi}{dt} = \Gamma\alpha\sigma\left(N_3 - N_2\right)\Phi + \beta\left(\frac{N_3}{\tau_{32}}\right) - \frac{\Phi}{\tau_c} \tag{7}$$

- **Stimulated Revenue (Gain Term - Coherent Revenue):** $\Gamma\alpha\sigma\left(N_3 - N_2\right)\Phi$ represents stimulated revenue generation, proportional to the supplier population inversion ($N_3 - N_2$). It directly relates to the platform's profit threshold once the critical mass of suppliers (drivers) and customers (riders) is reached. When the gain exceeds losses, the system sustains profitable operations. The coefficient $\Gamma\alpha\sigma$ quantifies both the platform's design quality and the trust it fosters, allowing revenue to function as the optimization target. For instance, if $\Gamma\alpha\sigma = 0.8$, the platform can theoretically convert 80% of requests into rides. If, at the same time, $\alpha = 0.7$, it reflects the drivers' trust in the platform.

- **Spontaneous Revenue (Incoherent Revenue):** $\beta\left(\frac{N_3}{\tau_{32}}\right)$ represents revenue generated spontaneously by rides that are temporarily aligned with the main mode (coherent) but spatially misaligned (incoherent). Though these rides contribute to total revenue, they neither drive nor stimulate more main-mode rides that are both temporally and spatially coherent (positioning drivers outside hotspots).

- **Loss Term (Revenue not stimulating additional revenue):** $-\frac{\Phi}{\tau_c}$ represents the decay of the revenue over time. Setting the mean of the rides duration distribution gaussian fit to e.g.: $\tau_c = 20$ minutes implies that rides lasting more than 20 minutes incur a net "loss" to overall revenue (they do not contribute to the favourable main mode of the rides generating the bulk of the revenue - outlier trips position drivers far from hotspots).

**Explanation of the Terms**

- **Vehicle Mobilization Rate ($W_p N_1$)**

    ○ Represents the rate at which available vehicles ($N_1$) are mobilized to the waiting state ($N_6$) - either on the spot or through repositioning.

- **Transition Rates ($\frac{N_i}{\tau_{ij}}$)**

    ○ Define the rate at which vehicles transition from state $E_i$ to state $E_j$.

    ○ $\tau_{ij}$ is the average time a vehicle spends in state $E_i$ before transitioning to $E_j$.

- **Conservation of mass- Supply side (drivers/vehicles)**

    ○ Analogous to the conservation of mass principle in physics. The total number of online vehicles ($N_{online}$) is conserved:

$$N_{online} = N_1 + N_2 + N_3 + N_4 + N_5 + N_6 \qquad (8)$$

- **Conservation of Energy - Demand side (ride requests)**

    ○ Causality principle - Trips completed can not exceed rides requested (Conservation of energy - there is no free revenue in the platform): 'Total trips completed' < 'Total trips requested'

## 5.4 MODEL VALIDATION

A systematic PASER model validation was conducted to assess accuracy and robustness. Six principles were applied:

1. Data-driven modeling

2. Shock-period testing

3. Sensitivity analysis

4. Simultaneous validation of five high-certainty metrics

5. Testing across multiple historical periods for both Uber and Lyft

6. Iterative tuning of revenue density parameters

Per-minute aggregated raw data at per-second granularity balanced readability with precision. The model's predictive power for dynamic vehicle-state populations and real-time revenue was evaluated, reflecting network effects driven by rider requests, driver incentivization, and revenue generation.

### 5.4.1 Near 100% certainty raw data driving the model

Four near-100%-certainty data streams (retrieved from various historical periods for Uber and Lyft) fed the model: (1) the per-minute ride requests for Uber (Figure 61), (2) the per-minute mean enroute durations (Figure 62), (3) on-scene durations (Figure 63), and (4) on-trip durations (Figure 64). These represent, respectively, the consumer-side demand and the supplier-side operational times in this two-sided platform.

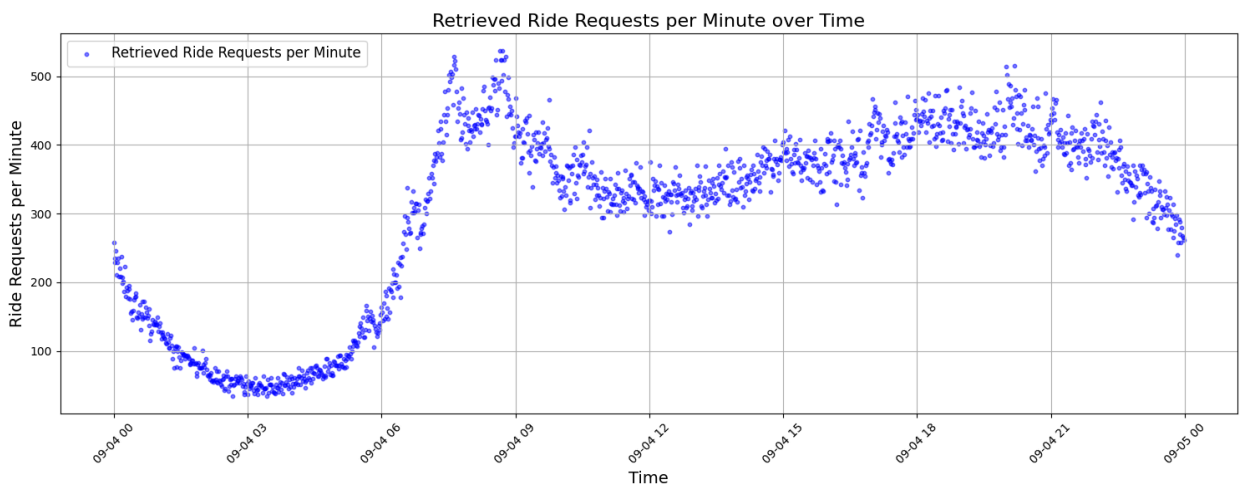

Figure 61. Uber, September 4, 2024. The consumer-side demand for rides

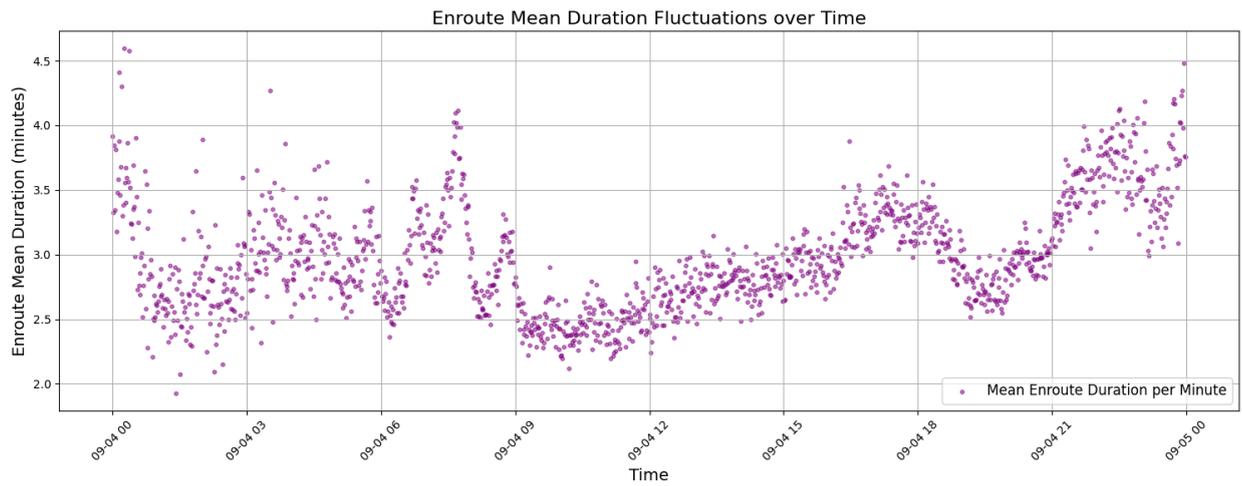

Figure 62. Uber, September 4, 2024. Minute-by-minute mean durations of enroute transitions.

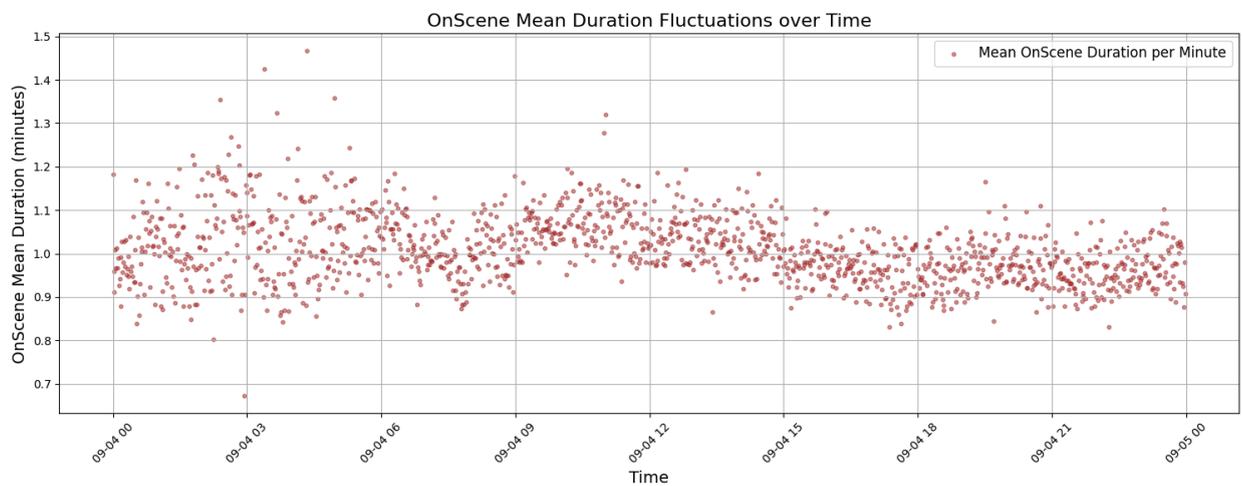

Figure 63. Uber, September 4, 2024. Minute-by-minute mean durations of onscene transitions.

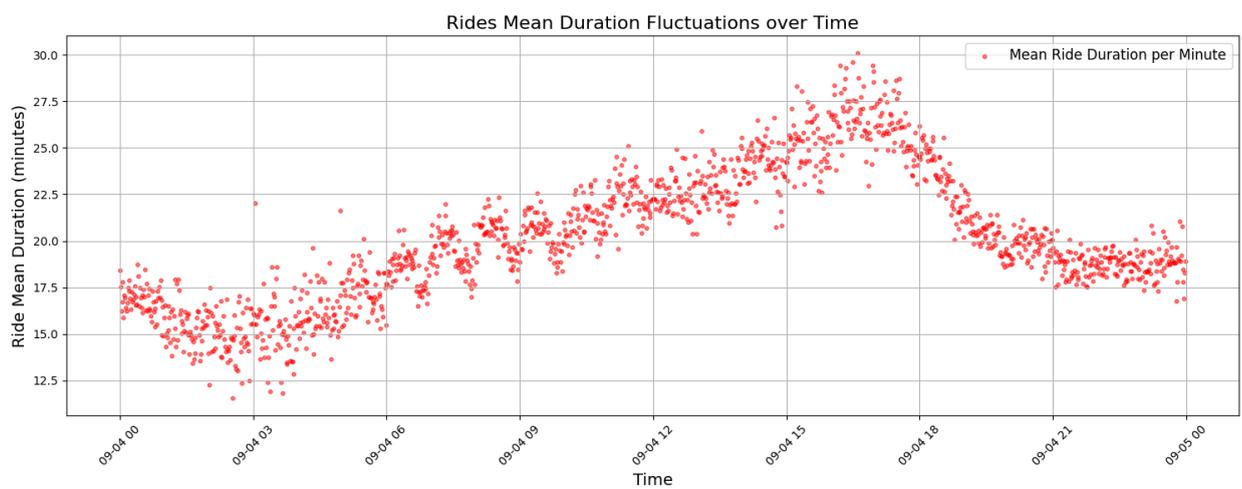

Figure 64. Uber, September 4, 2024. Minute-by-minute mean durations of on-trip (ride) transitions.

### 5.4.2 Data driven simulations accuracy testing - network effects

Data-driven simulations reconstructed Uber's historical real-time dynamics for validation while also calculating unlogged operational elements. In Figure 61, ride requests trigger an incentivization mechanism mobilizing drivers at a "pump rate" (drivers/min), lifting them from the "available" (ground) state to the "waiting" (highest) state. Their acceptance of ride requests (gain term in the revenue density equation - Equation (7) in Section 4.4) shapes platform efficiency. Initial conditions (e.g., per-minute revenue and driver states) were set from historical data at simulation start.

A boundary condition ensures that the total model-predicted online driver population cannot exceed the empirically retrieved upper limit. Conservation of mass and energy similarly cap on-trip drivers by actual ride requests. Figure 65 depicts the driver or vehicle "pumping," showcasing how the platform elevates suppliers to a more active state.

Solving seven coupled ODEs (Section 4.4) via the Radau method at a 1-minute timestep over 24 hours required ~120 seconds in Python on an i7-6700HQ CPU; a C++ version may be up to 100x faster. The simulations produced an overall MAPE below 4%. Figures 66–68 demonstrate a three-point validation of supplier states (enroute, on-scene, on-trip), each predicted with high accuracy. Validation extends into a fourth dimension by forecasting the dynamic PASER intra-cavity revenue in Figure 69.

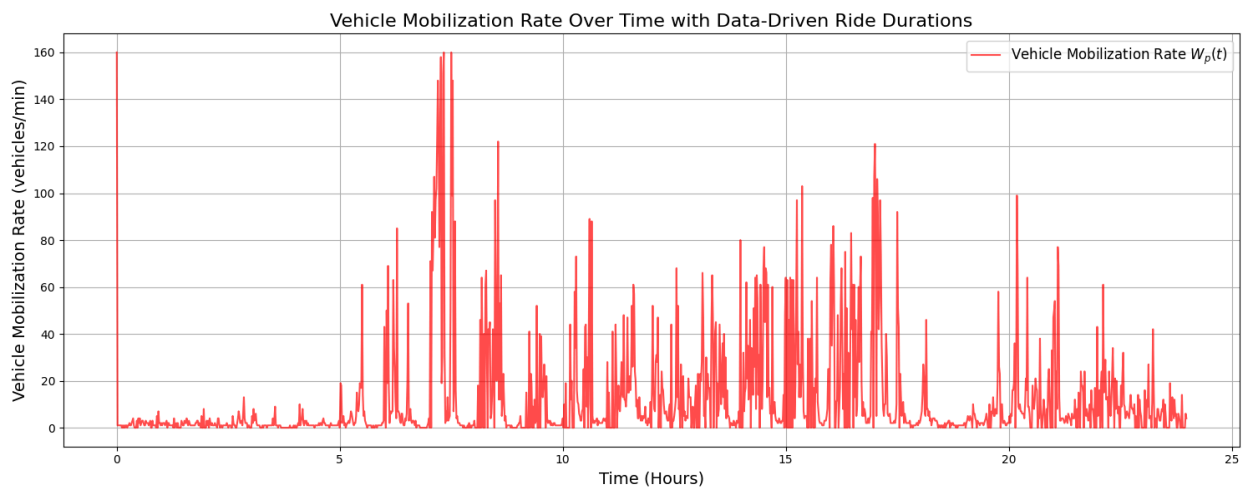

Figure 65. Uber, September 4, 2024. Drivers/vehicles mobilization rate.

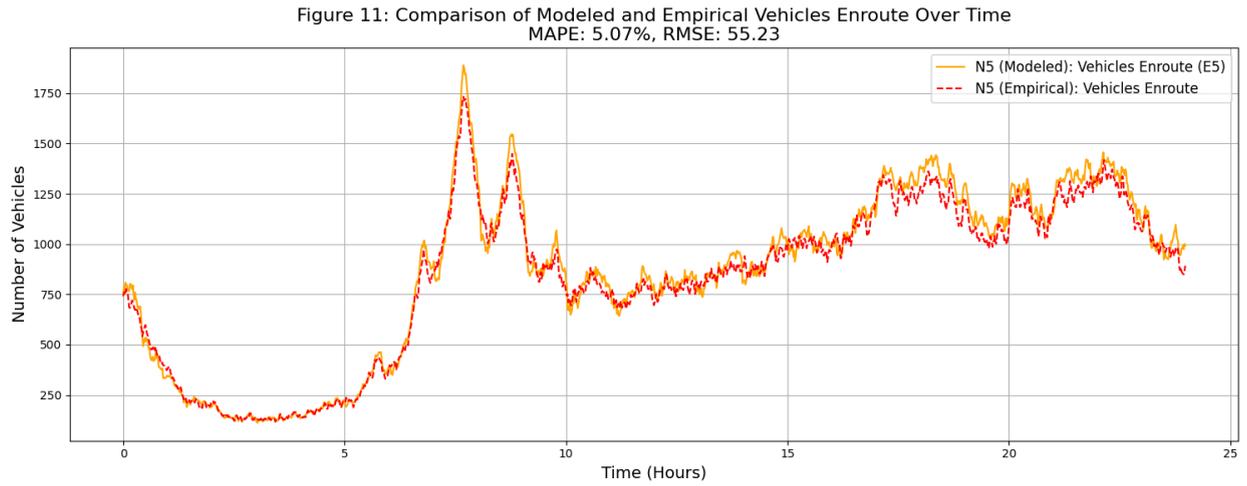

Figure 66. Validation of enroute drivers population prediction

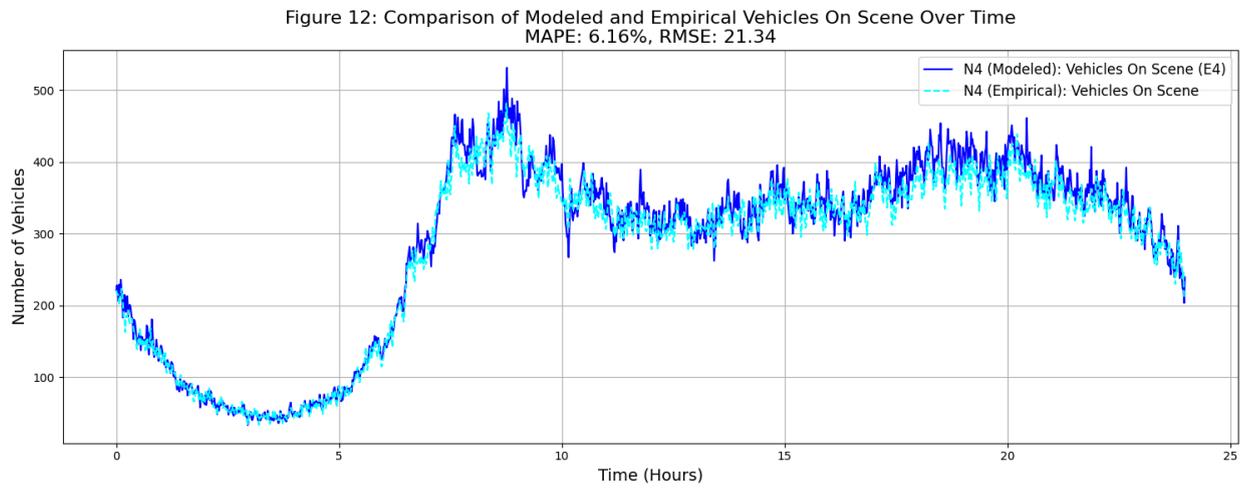

Figure 67. Validation of onscene drivers population prediction

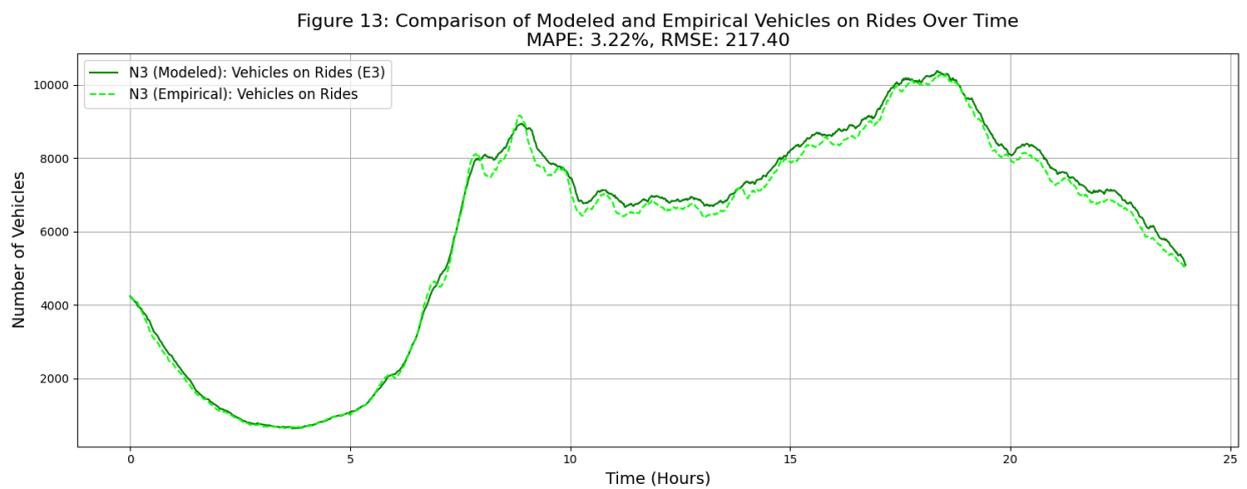

Figure 68. Validation of on-trip drivers population prediction

### 5.4.3 Data driven simulations accuracy testing - revenue density

The Radau solver imparted a smoothing effect on high-frequency fluctuations in the empirical data, yet its prediction of real-time revenue dynamics maintained a 12.44% MAPE error. Cumulative revenue over time (Figure 70) proved even more accurate, showing that per-minute smoothing still captures the aggregate trend robustly. Higher-frequency simulations (per-second) could enhance resolution if needed. Iterative tuning of parameters in the revenue density equation (Eq. (7) in Section 4.4) yielded Figure 69's results. Adjusting the gain term (gamma × alpha × sigma $\Gamma\alpha\sigma$), spontaneous term ($\beta$), and main ride lifetime ( $\tau_c$ ) produced the final parameter set for Uber in September 2024 (Table 5). Different historical conditions require updated parameter values reflecting shifts in business models, apps, IT infrastructure, driver loyalty, and other factors.

Remaining on the same historical day, the model was tested against Lyft's raw data. Figure 71 shows Lyft receiving fewer ride requests per minute and experiencing milder afternoon peaks than Uber. Correspondingly, Figure 72 indicates lower driver mobilization. PASER predicted Lyft's on-trip population (Figure 73) even more accurately, yet revenue dynamics (Figure 74) lagged behind Uber-level accuracy because the parameter set used (Table 5) was calibrated for Uber. Tuning PASER specifically to Lyft reveals the operational differences shaping its distinct business model, algorithms, and platform infrastructure.

When calibrating PASER to Lyft's data (September 2024), the main ride duration ($\tau_c$ = 20 min) was assumed unchanged due to NYC's common traffic context. However, Figure 75 shows that Lyft handles more atypical trips than Uber, implying a higher spontaneous fraction ($\beta$). Competitiveness also suggests that alpha (driver responsiveness, trust) be higher. Iterative tuning supported these insights, indicating that Lyft's apps and algorithms might be 25% superior and that it serves 31% more atypical trips with drivers who are 14% more satisfied. Figure 76 and Table 6 confirm these differences. Tables 7 and 8, alongside Figures 79–81, detail parameter adjustments for Uber and Lyft in July 2019.

Experimental results in Figures 76–81 confirm the model's adaptability to multiple operators across various historical periods. Given the relevance of shock periods (e.g., COVID-19), significant disruptions such as curfews caused substantial operational downtime for both platforms, further underscoring the importance of robust parameter calibration.

**Table 5. Revenue (Φ) density calibrated parameters for Uber's operations in Sep 2024.**

| Uber in September 2024 | | |
|---|---|---|
| Parameter | Symbol | Value |
| Technical efficiency | Γ | 0.8 |
| Social efficiency (drivers responsiveness, trust) | α | 0.7 |
| Overall socio-technical efficiency | Γα | 0.56 |
| Apps and algos quality, platform design quality | σ | $40 \times 10^{-7}$ |
| Trips coherence | β | 0.65 |
| Ride lifetime (main mode) | $\tau_c$ | 20 min |

(For reference: Figures 69 & 70 present the corresponding simulation results.)

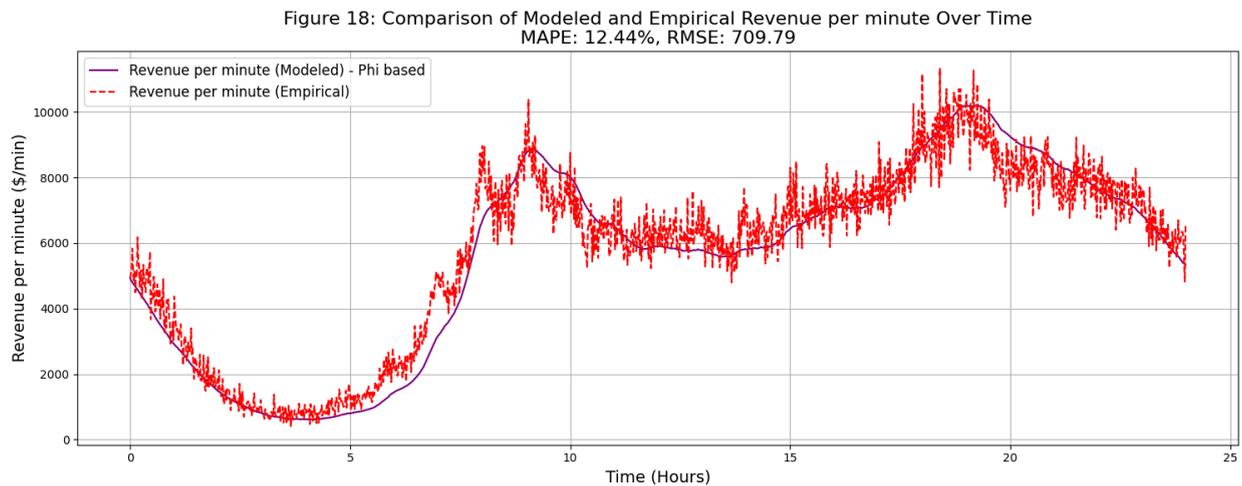

Figure 69. Revenue per minute over time for the for Table 5 parameter values

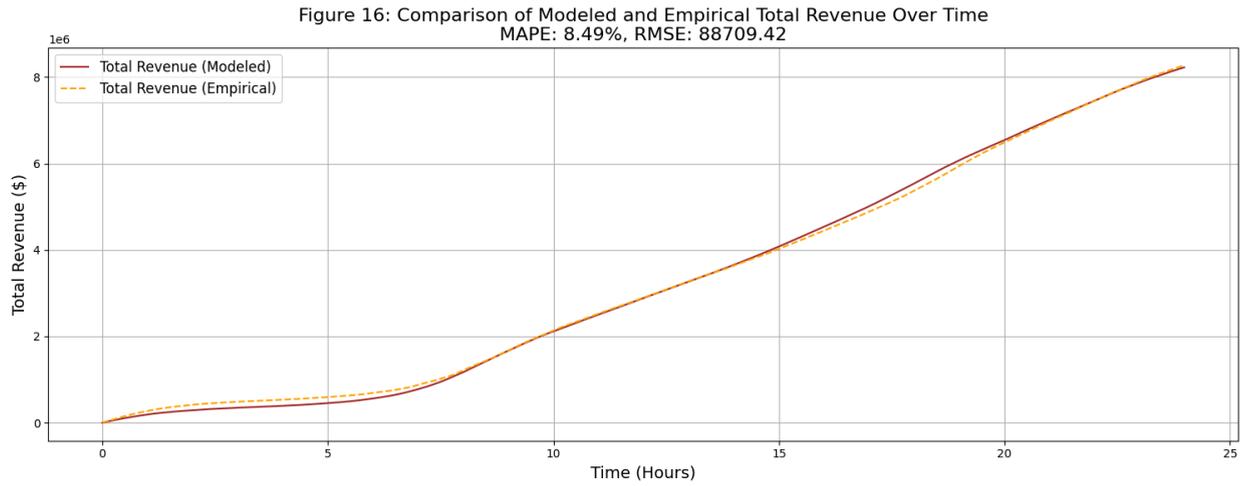

Figure 70. Total revenue over time for the for Table 5 parameter values

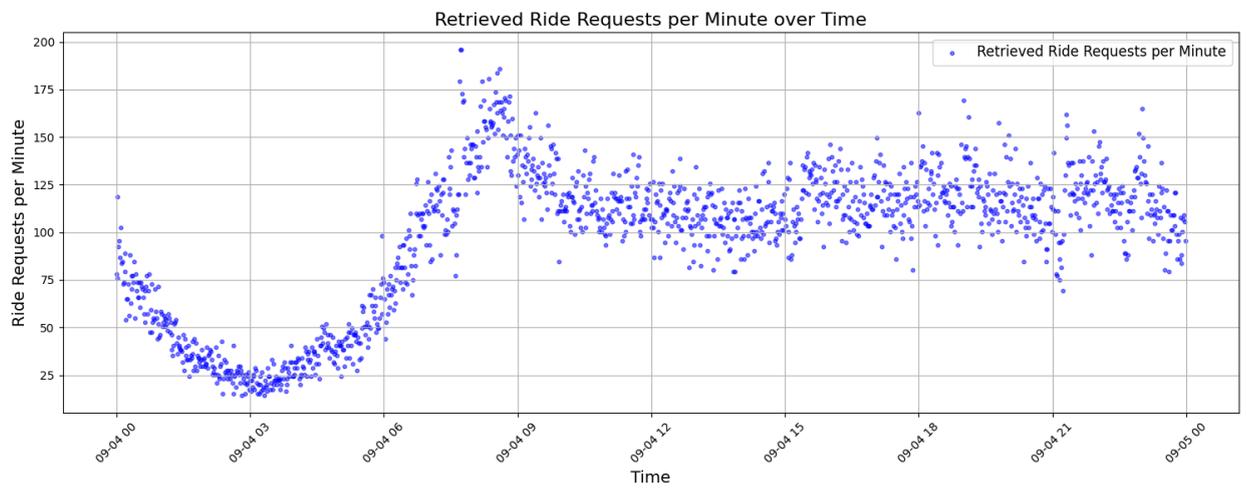

Figure 71. Lyft - requests per minute

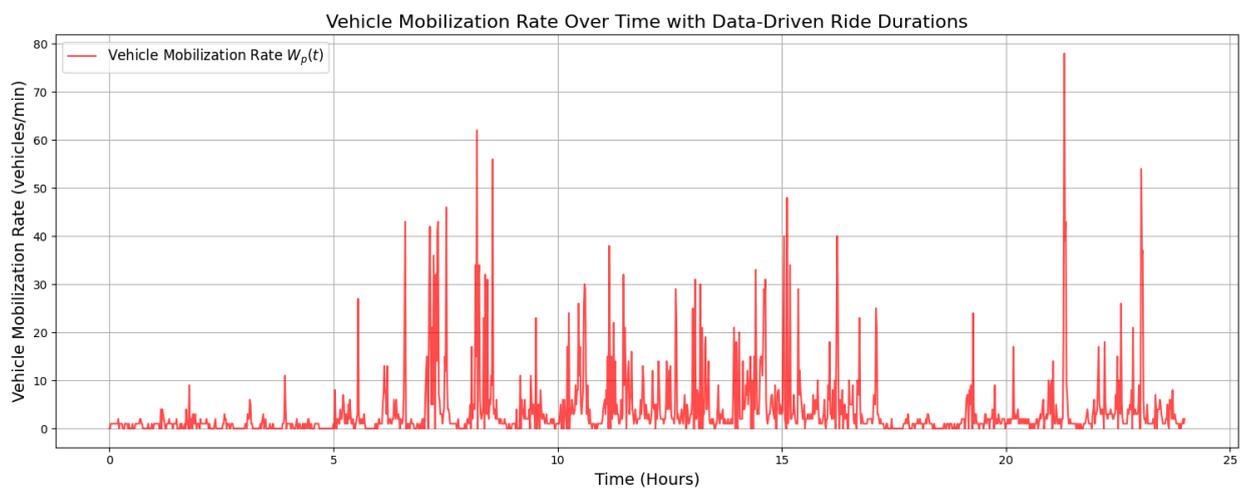

Figure 72. Lyft - supplier mobilization.

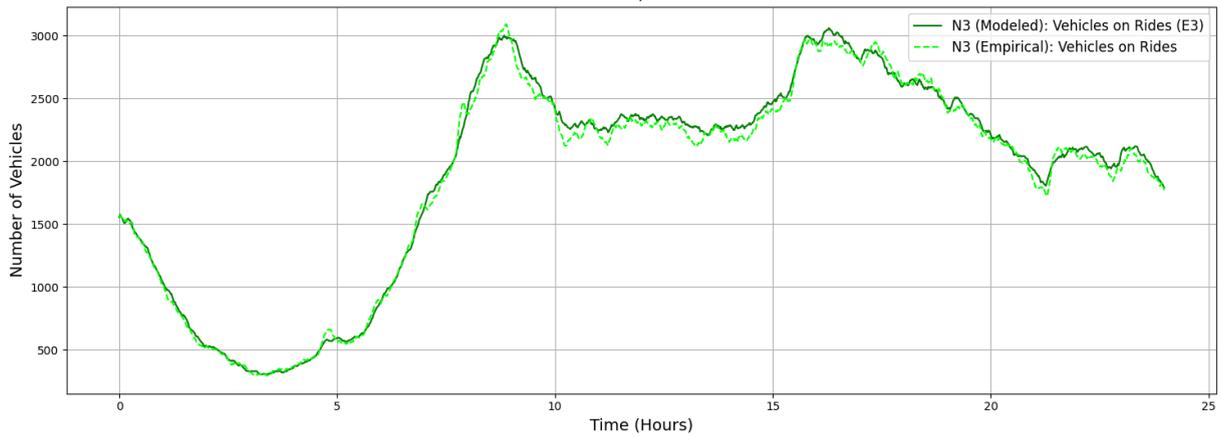

Figure 73. Lyft's on-trip population dynamics prediction accuracy

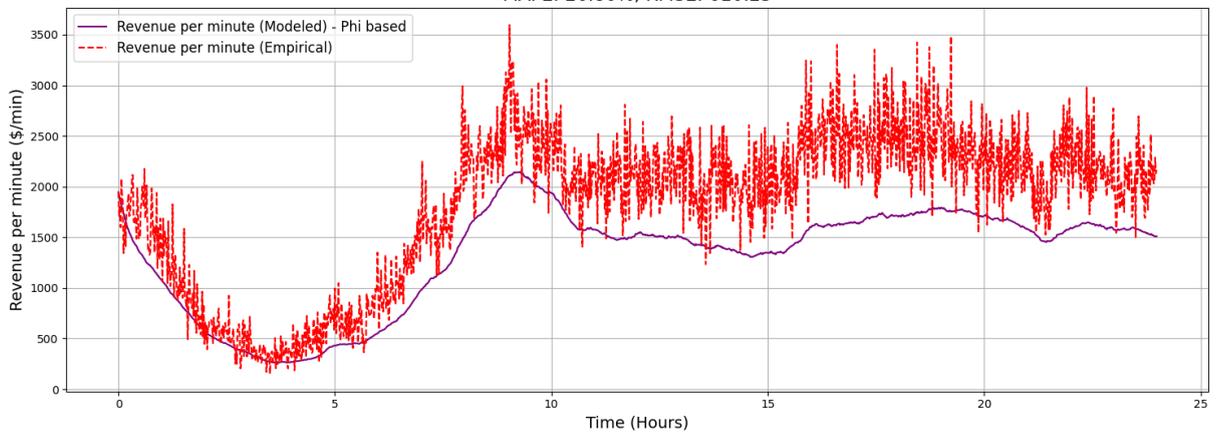

Figure 74. Lyft revenue density before calibrating its revenue density parameters.

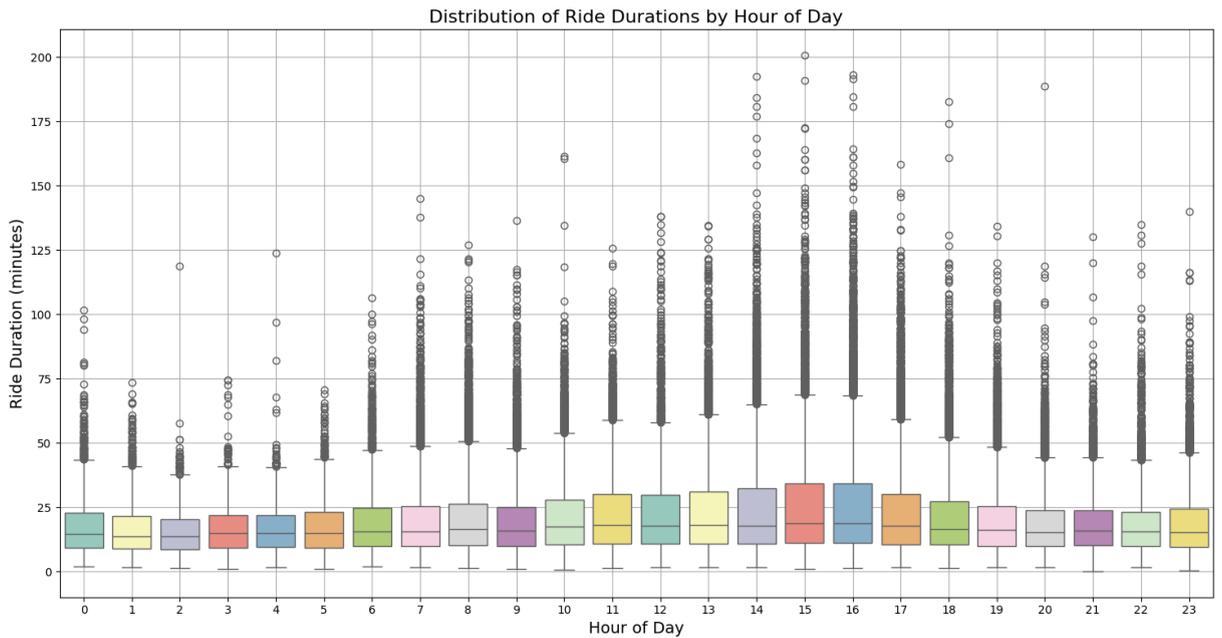

Figure 75. Lyft's distinct trip distribution - September 2024.

**Table 6. Revenue (Φ) Density Parameters for Lyft in September 2024.**

| Lyft - Sep 2024 | | | |
|---|---|---|---|
| **Parameter** | **Symbol** | **Value** | **Difference from Uber Sep 2024** |
| Technical efficiency | Γ | 0.8 | 0% |
| Social efficiency (drivers responsiveness, trust) | α | 0.8 | +14% ↑ |
| Overall socio-technical efficiency | Γα | 0.64 | +14% ↑ |
| Apps and algos quality, platform design quality | σ | $50 \times 10^{-7}$ | +25% ↑ |
| Trips coherence | β | 0.85 | +31% ↑ |
| Ride lifetime (represents the main mode ride duration) | $\tau_c$ | 20 min | Unchanged |

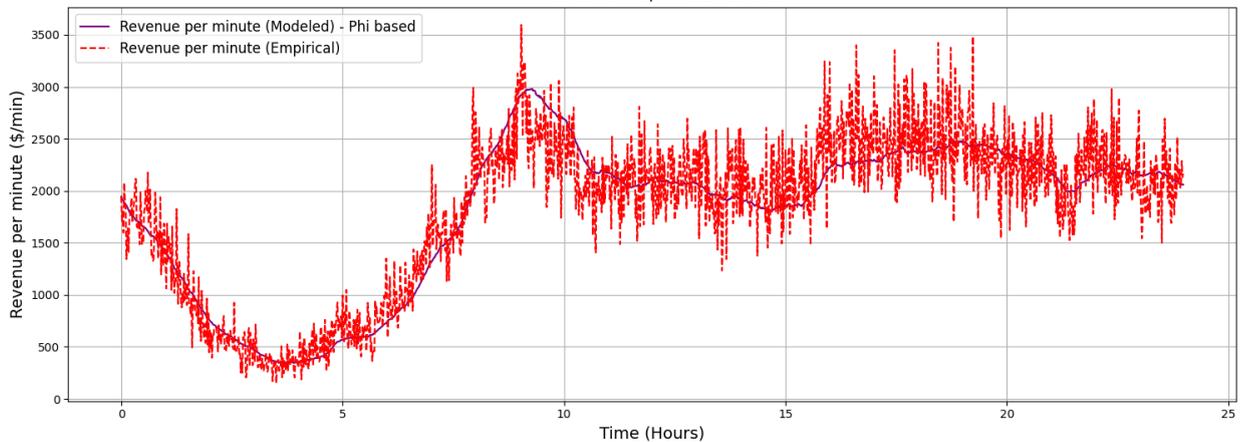

Figure 76.  Lyft (September 2024) after calibrating its revenue density parameters

**Table 7. Revenue (Φ) Density Parameters for Uber in July 2019.**

| Uber – July 2019 | | | |
|---|---|---|---|
| **Parameter** | **Symbol** | **Value** | **Difference from Uber Sep 2024** |
| Technical efficiency | Γ | 0.6 | -25%↓ |
| Social efficiency (drivers responsiveness, trust) | α | 0.8 | +14% ↑ |
| Overall socio-technical efficiency | Γα | 0.48 | -15% ↓ |
| Apps and algos quality, platform design quality | σ | $30 \times 10^{-7}$ | -25% ↓ |
| Trips coherence | β | 0.7 | +7% ↑ |
| Ride lifetime (represents the main mode ride duration) | $\tau_c$ | 20 min | Unchanged |

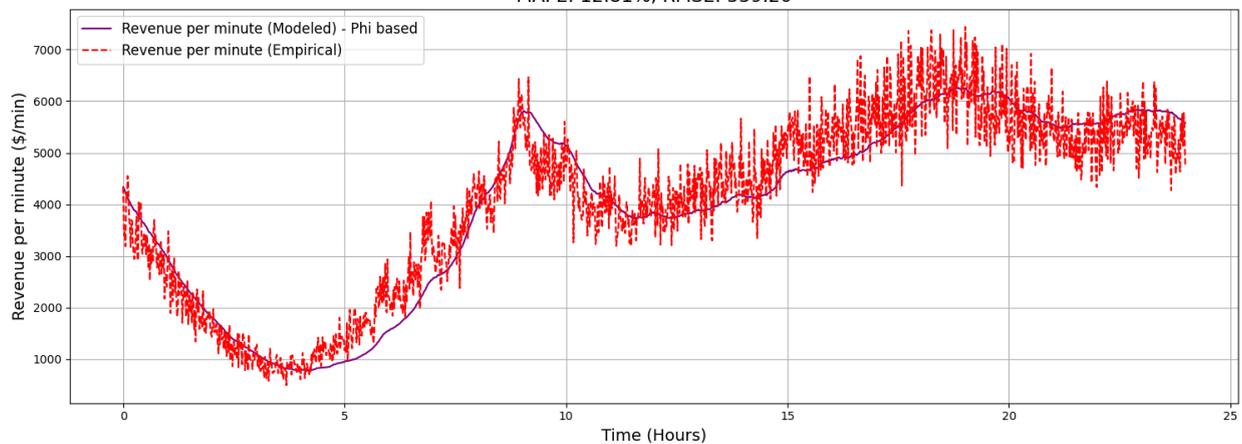

Figure 77. Uber (July 2019) revenue density after parameters calibration (table 7)

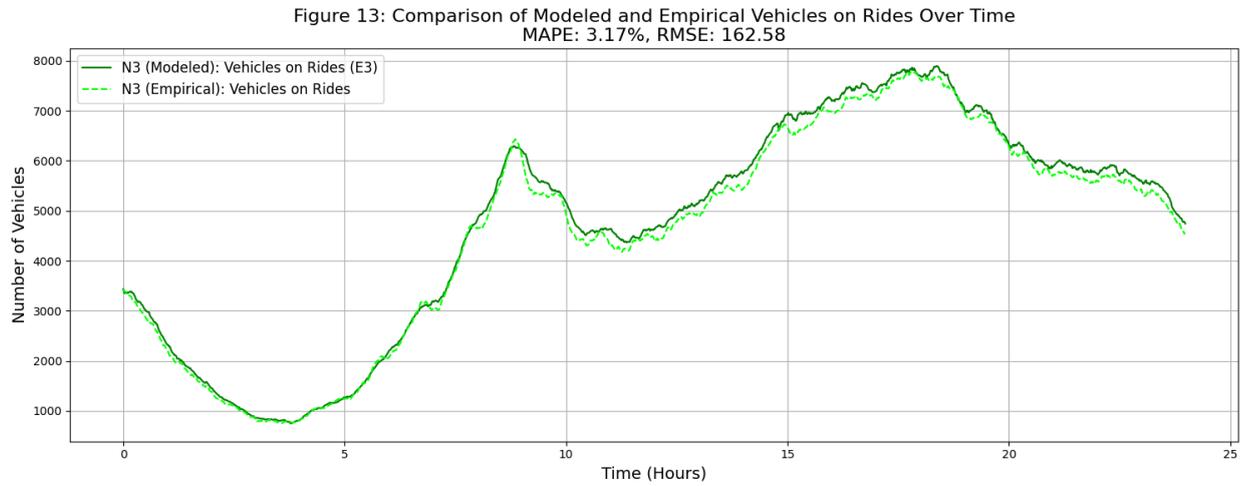

Figure 78. Uber (July 2019) network effects calculations accuracy.

**Table 8. Revenue (Φ) Density Parameters for Lyft in July 2019.**

| Lyft - July 2019 | | | |
|---|---|---|---|
| **Parameter** | **Symbol** | **Value** | **Difference from Lyft Sep 2024** |
| Technical efficiency | Γ | 0.7 | -13%↓ |
| Social efficiency (drivers responsiveness, trust) | α | 0.85 | +6% ↑ |
| Overall socio-technical efficiency | Γα | 0.59 | -8% ↓ |
| Apps and algos quality, platform design quality | σ | $35 \times 10^{-7}$ | -30% ↓ |
| Trips coherence | β | 0.8 | -6%↓ |

| Ride lifetime (represents the main mode ride duration) | $\tau_c$ | 20 min | Unchanged |
|---|---|---|---|

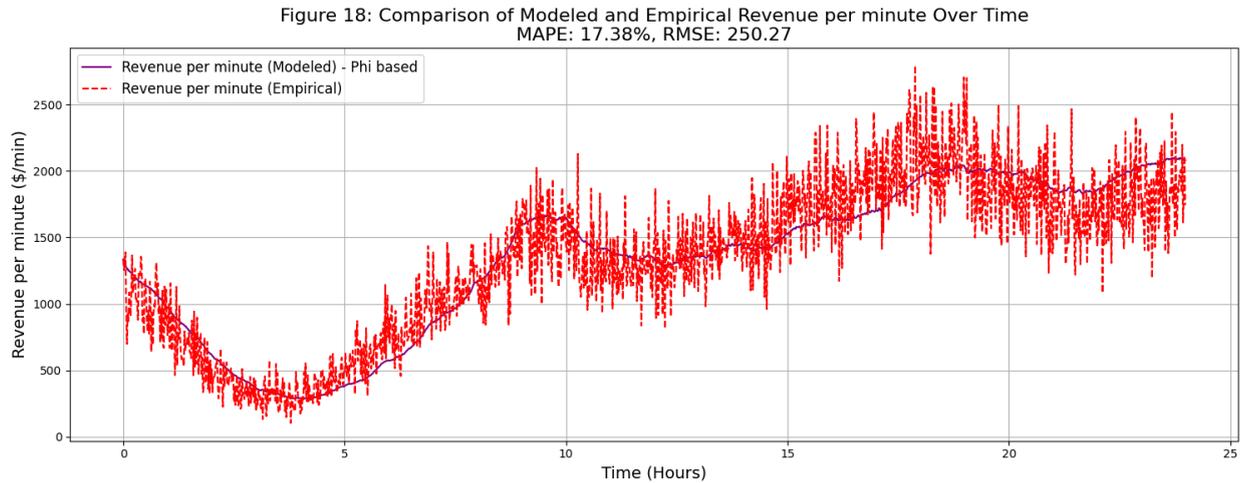

Figure 79. Lyft in July 2019 - revenue density after parameters calibration (table 8)

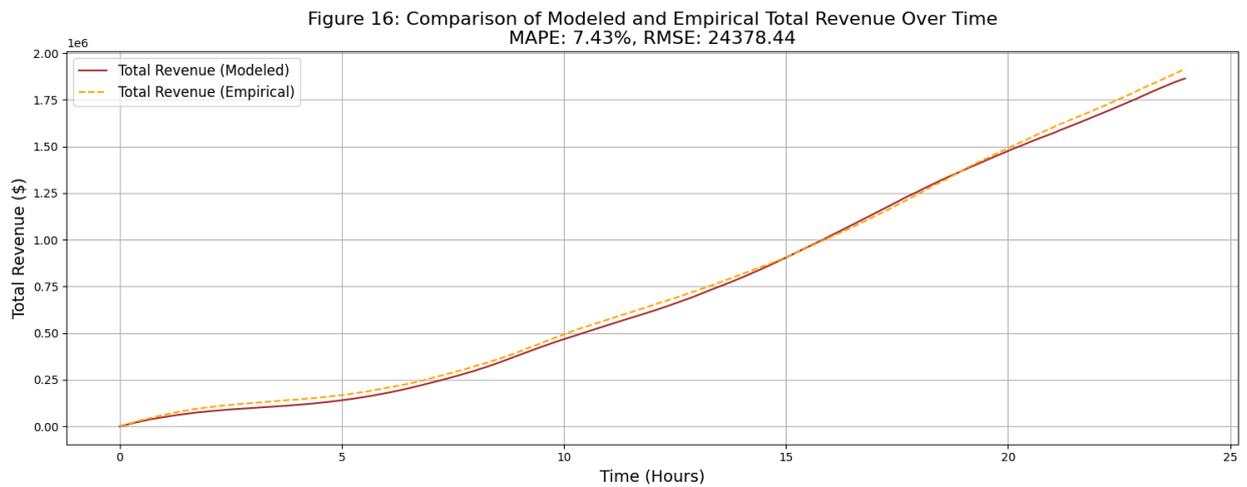

Figure 80. Lyft in July 2019 - Total revenue over time for the for Table 8 parameter values

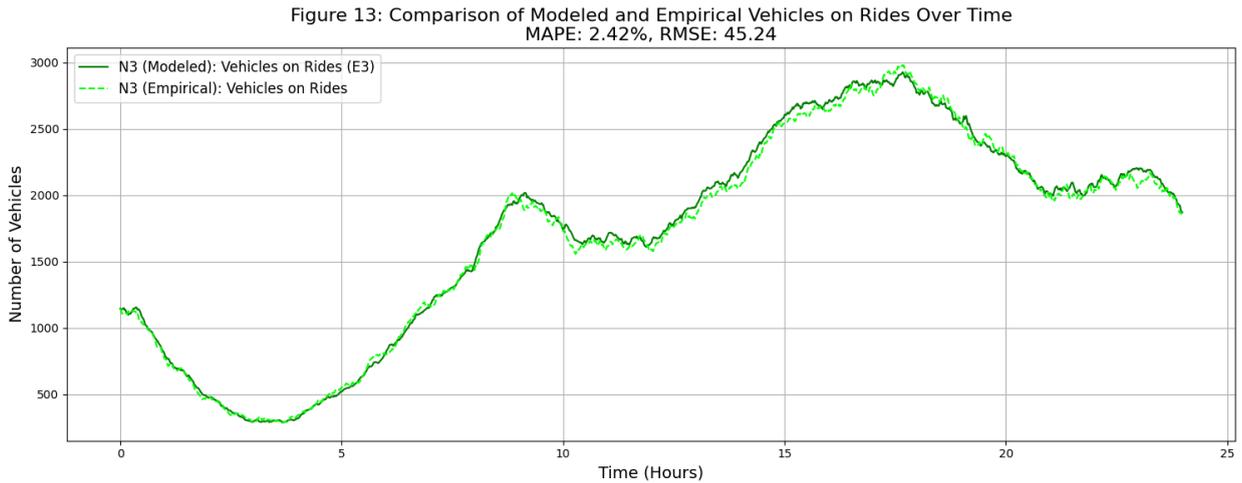

Figure 81. Lyft in July 2019 - Modeled vs. empirical on-ride drivers population for the Table 8 parameter values.

### 5.4.4 Data driven simulations sensitivity and stability testing

The decline week's empirical network effects (Figures 18 and 28) and daily revenue dynamics (Figure 82) revealed that using July 2019 parameter settings (Table 7) nearly doubled the MAPE error, indicating the pandemic's significant change in ride mix. Figure 83 shows that rides became notably shorter, narrowing box plots and whiskers, thereby reducing the median trip duration. By lowering $\tau_c$ from 20 to 17 minutes, predictive accuracy improved again (Figures 84 and 85), demonstrating the model's robustness and real-world grounding.

An extreme shock test during an overnight curfew (Figures 29, 30, 48–50, 55–57) challenged the model even further (this addresses O3). Figures 86 and 87 show that while MAPE in per-minute revenue increased, cumulative revenue improved. ODE system solver tuning could capture abrupt transitions more precisely. Overall the model demonstrated strong fidelity, adaptability, and shock stability.

Figures 88–92 provide proof of this adaptability. The driver mobilization rate (Figure 88) reflects dramatic operational changes under curfew and Figures 89–91 confirm accurate representation of network effects in that shock period. Figure 92 implies how the supplier population inversion drives real-time revenue generation.

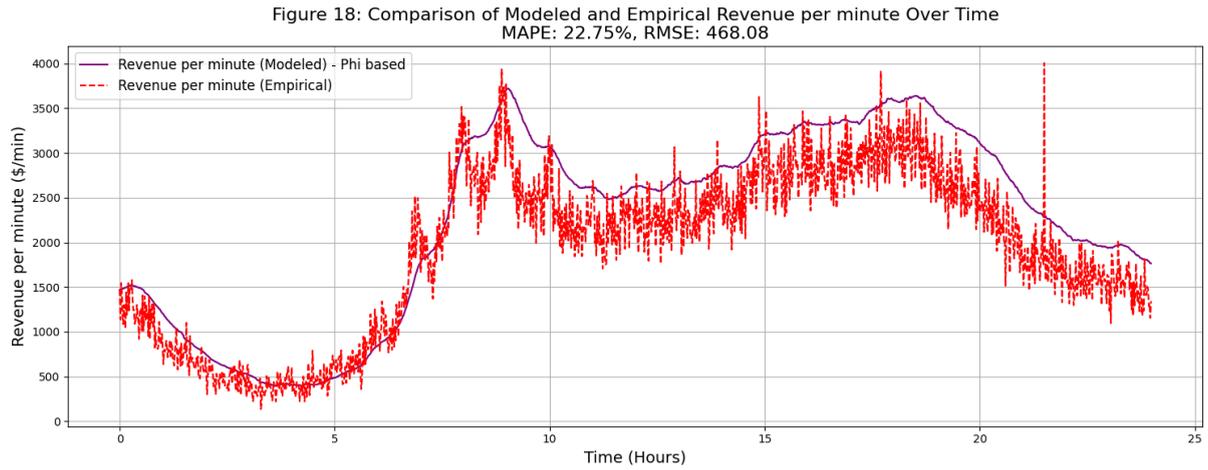

Figure 82. Uber 18 March 2020 ($\tau_c = 20$ min) - 1st week of pandemic restrictions.

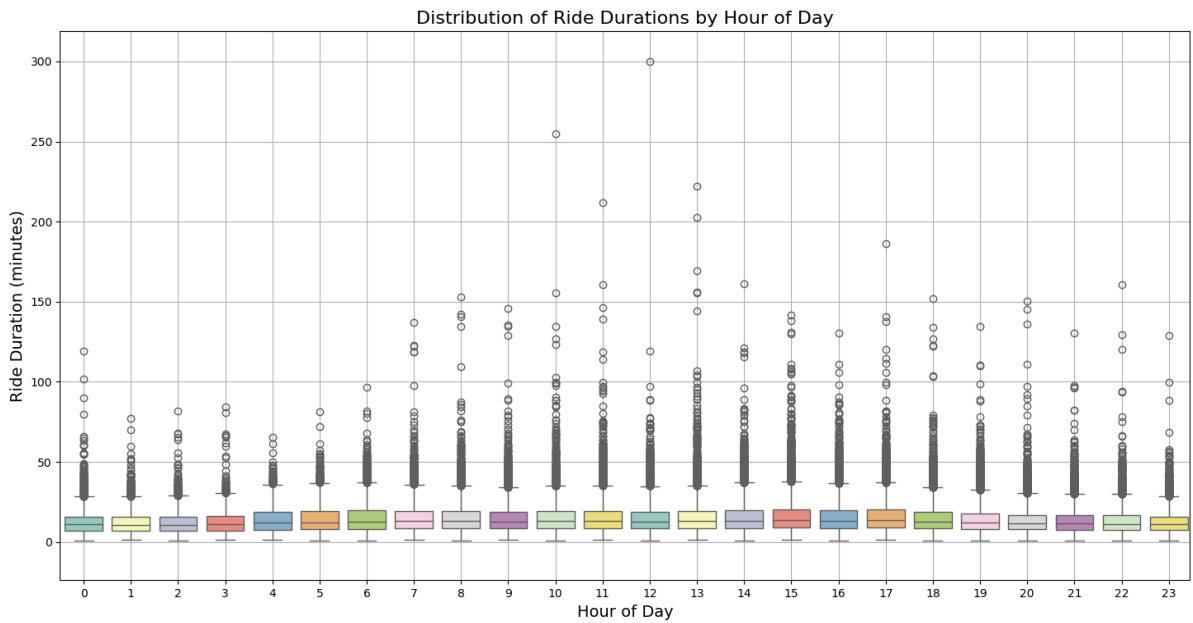

Figure 83. Uber 18 March 2020 (in 1st week of pandemic restrictions)

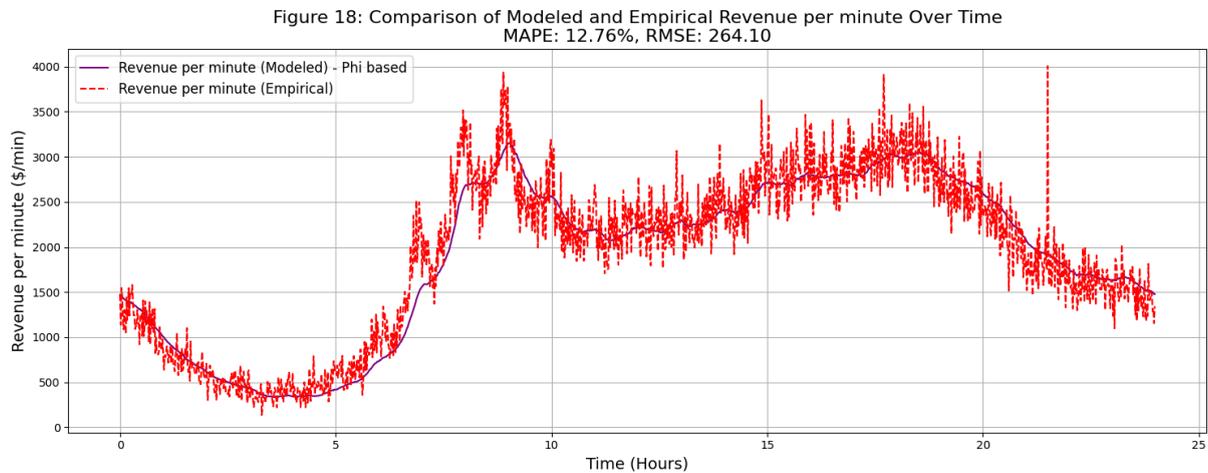

Figure 84. Uber 18 March 2020 ($\tau_c == 17$ min)

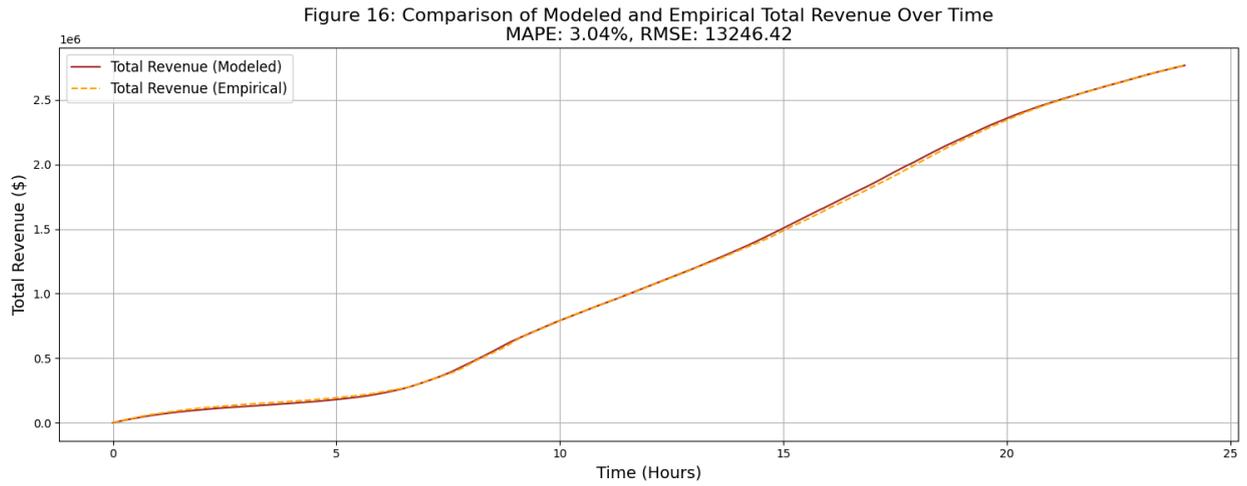

Figure 85. Uber 18 March 2020 ($\tau_c == 17$ min)

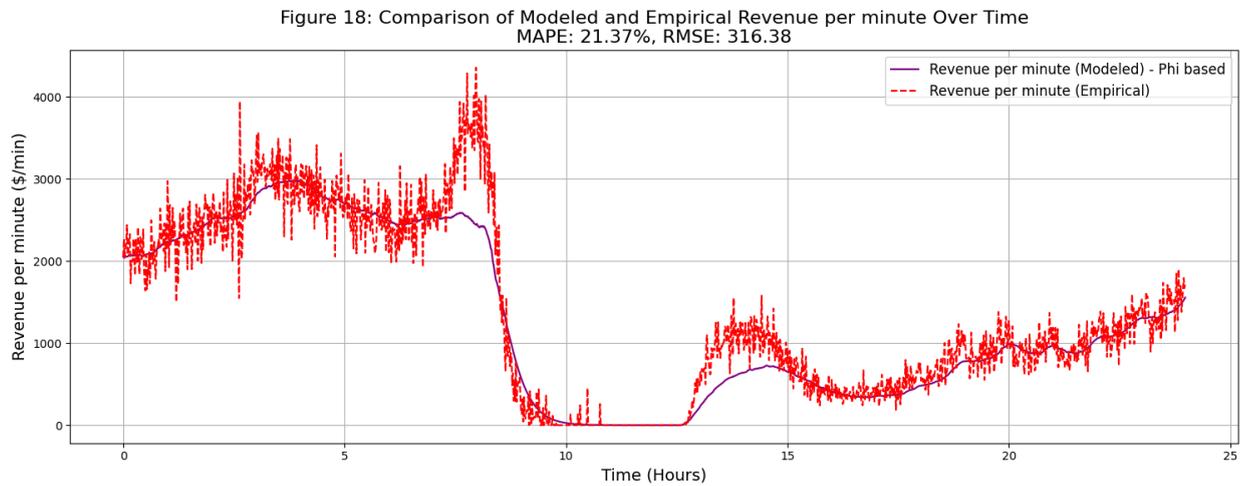

Figure 86. Uber, June 6, 2020 curfew period ($\tau_c = 19$ min).

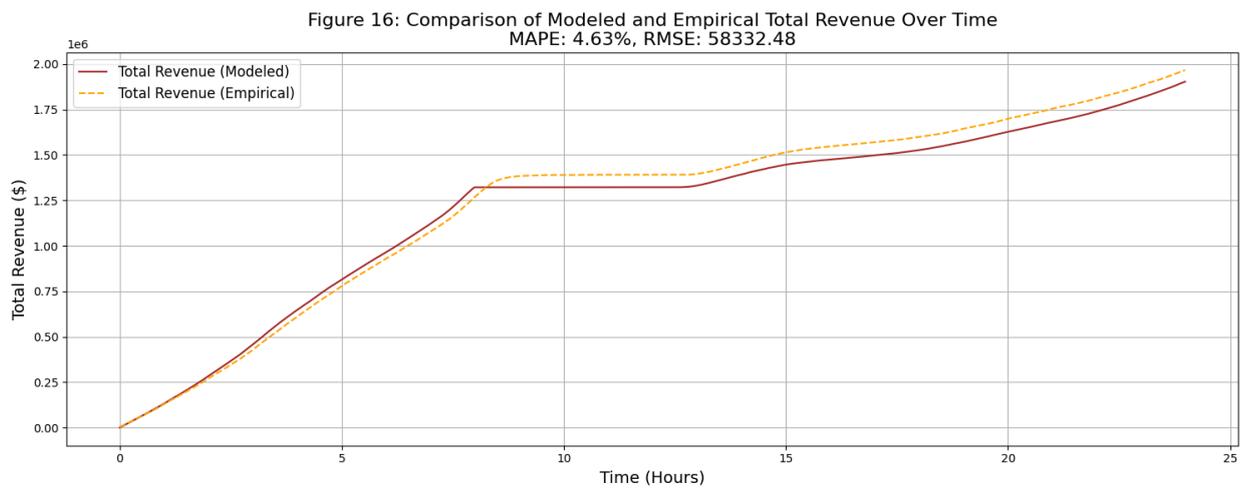

Figure 87. Uber, 6 June 2020, curfew period

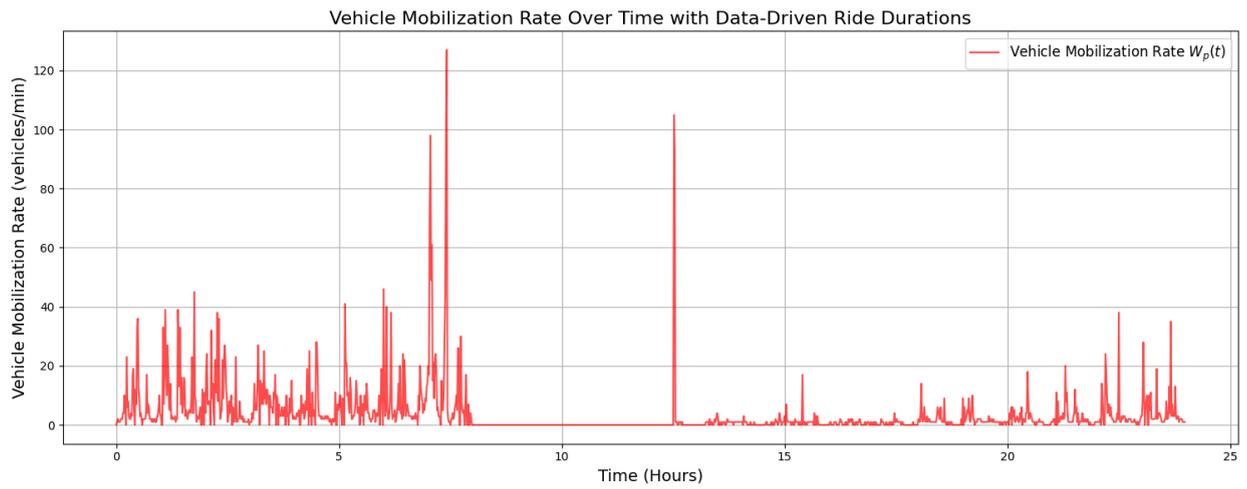

Figure 88. Uber, 6 June 2020 curfew period

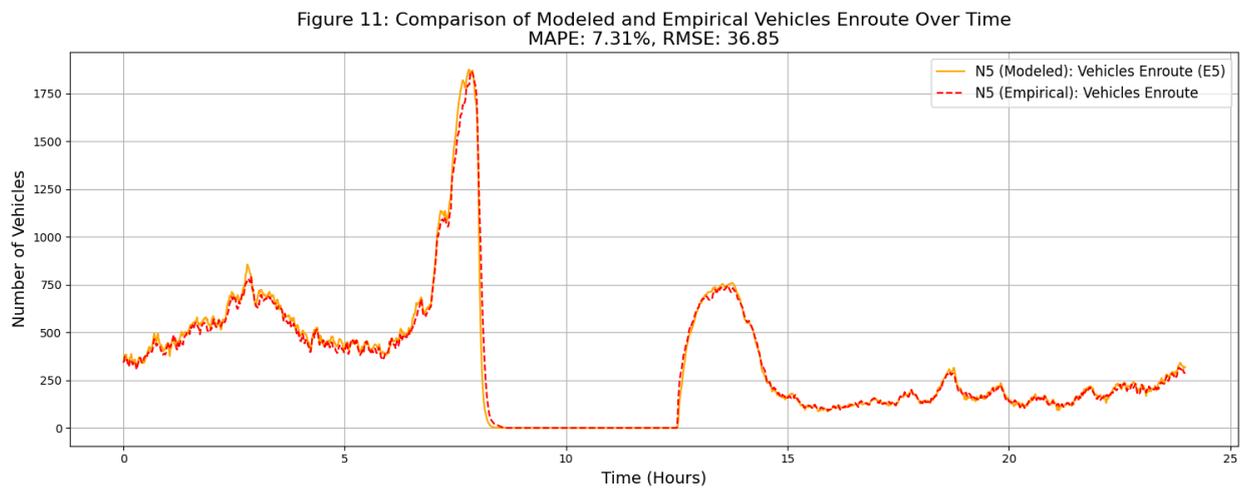

Figure 89. Uber, 6 June 2020 curfew period

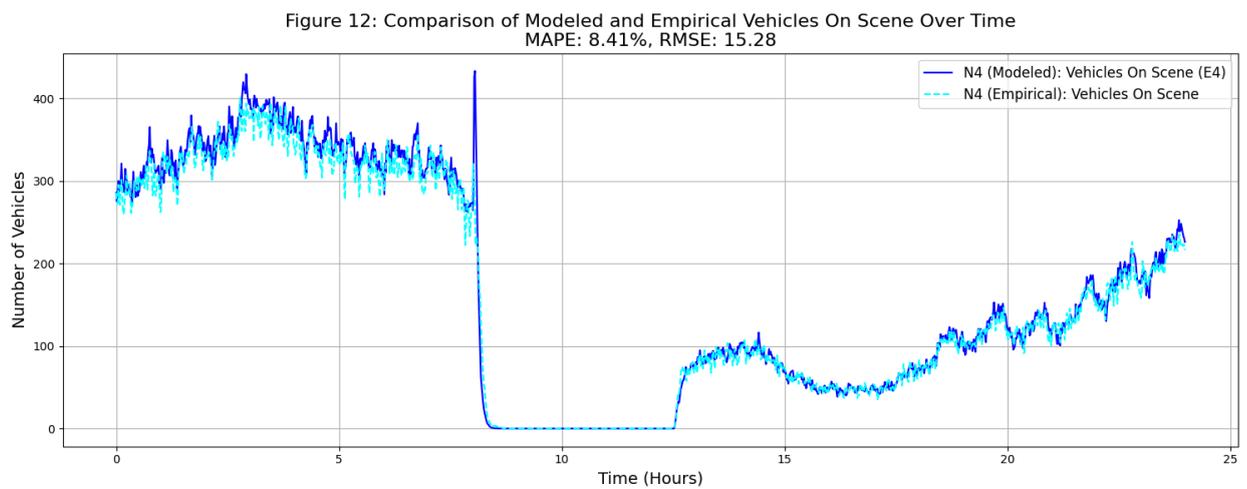

Figure 90. Uber, 6 June 2020 curfew period

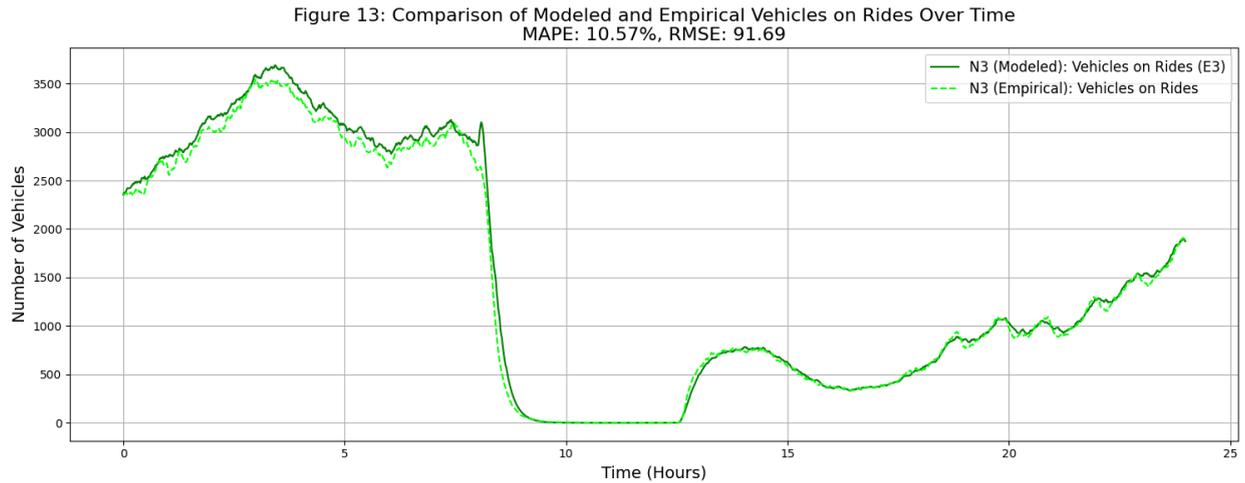

Figure 91. Uber, 6 June 2020 curfew period

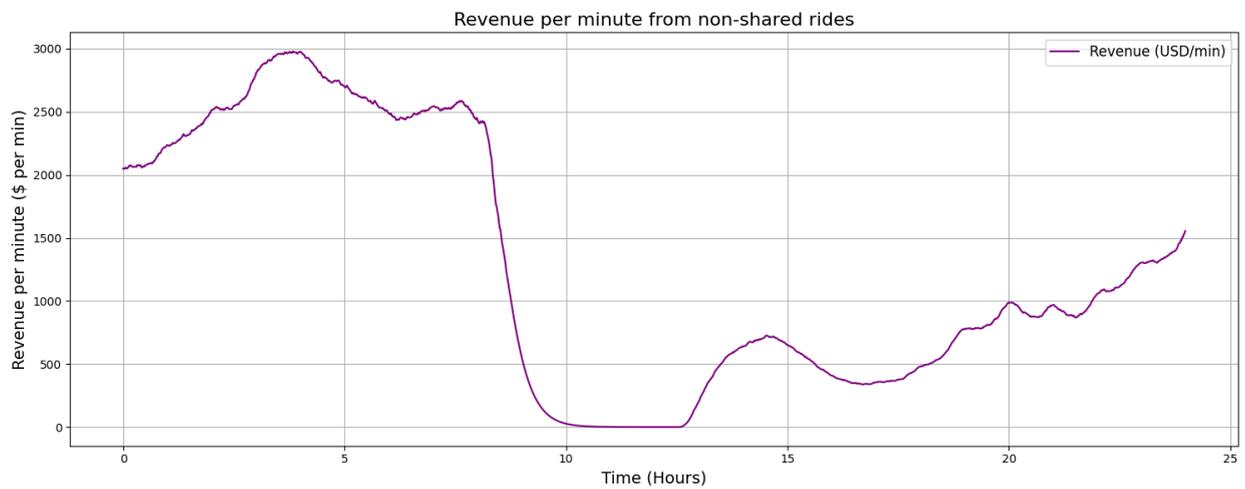

Figure 92. Uber, 6 June 2020 curfew period

## 5.5 SPATIO-TEMPORAL CONSIDERATIONS

Although the PASER mathematical model primarily captures temporal dynamics, it does incorporate macroscopic spatial behavior. In principle, it functions as an adaptive spatio-temporal model for several reasons:

- Parameter β (analogous to spontaneous emission in lasers) quantifies the fraction of trips aligning with the dominant revenue mode in terms of revenue but are lacking spatial coherence. These trips do not reinforce operational feedback loops. This is akin to spontaneously emitted photons that dissipate without contributing to coherent lasing.

- Gain Term Γασ also reflects how the platform focuses on the dominant ride mode. For example, Uber appears to handle fewer outlier rides than Lyft. Figure 93(a) illustrates how the platform's rider-app and driver-app confine the distribution of served rides (red line) that encompasses the most profitable main mode (orange line). It also depicts that the platform does not serve a portion of the requests. Figure 93(b) contrasts the aforementioned conceptual modal operation with the real ride durations distribution retrieved from the historical data of a single business day.

- Parameter $\tau_c$ bears an indirect spatial influence: if $\tau_c$ is set to 20 minutes, it effectively discourages rides significantly longer than 20 minutes (both temporally and spatially). Shorter $\tau_c$ fosters preference for more localized trips further discouraging spatially outlier trips.

Furthermore, PASER is geographically scalable. This paper validated the model for NYC as a whole, but one could simulate a "swarm" of localized PASERs across different city areas, facilitating spatio-temporal simulations at an adjustable geographic scale. A purely spatial approach can be appended by modeling the influx and outflux of drivers per area, governing the local online population.

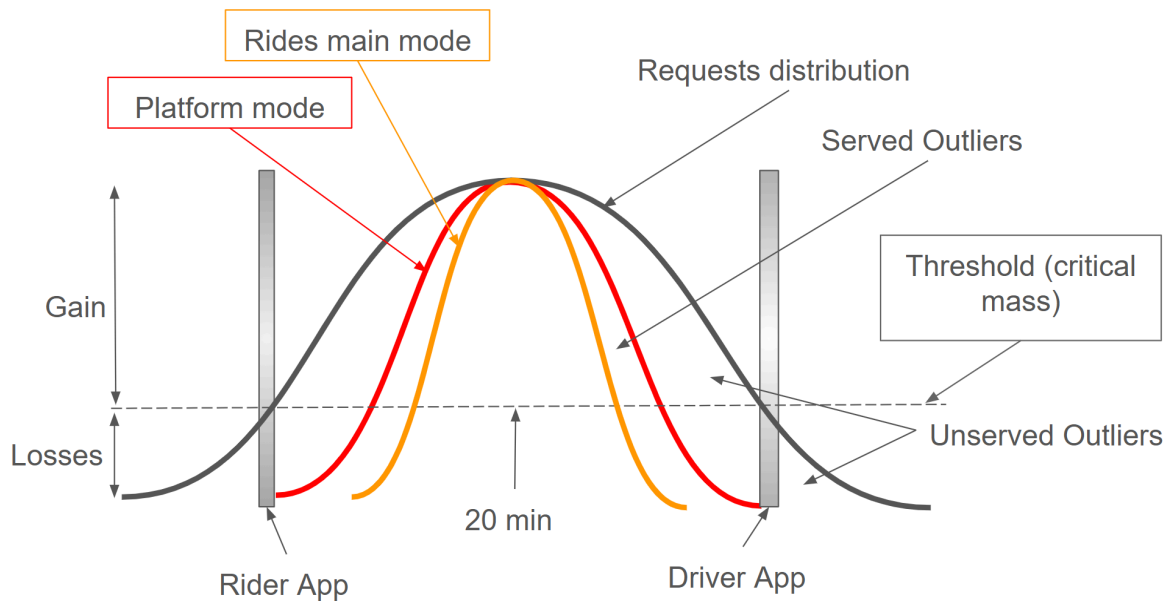

Modal Operation in PASER: Main Mode vs. Outliers
Threshold Illustration: Gains, Losses, and Critical Mass

Figure 93. (a). PASER conceptual depiction of ride modes confinement, critical mass, and gain region.

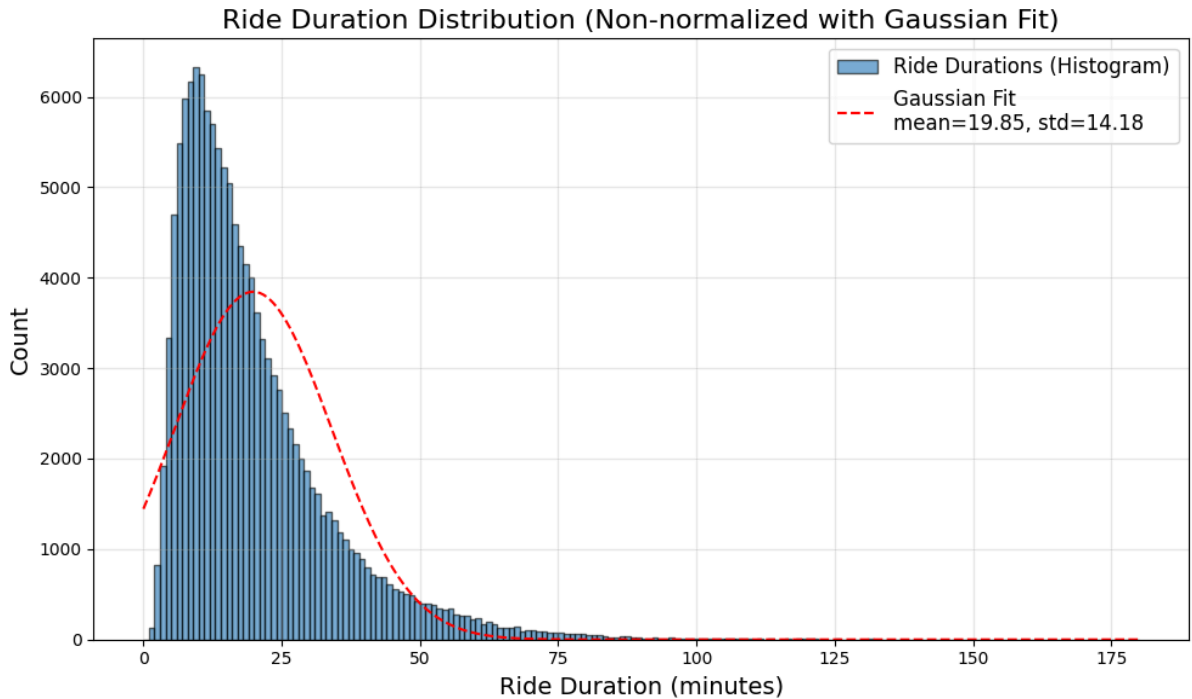

Figure 93. (b). A single day's, real rides duration distribution and its gaussian fit.

## 5.6 SURGE PRICING WITH PASER

The PASER framework is inherently well-suited for strategic surge pricing (this addresses O4). Its real-time, dynamic mathematical model positions it as a valuable tool for physics-based trip cost prediction and optimization. The mean per-minute pricing plotted in Figures 95 to 98, derived from data four years apart, exhibits greater volatility than the empirical pricing. Although the MAPE between the predicted and empirical pricing is not negligible, the corresponding RMSE (Root Mean Square Error) of no more than USD 0.3 indicates that the model closely approximates the empirical values in practical terms.

The increased volatility suggested by the model may imply a more equitable pricing approach for the same total revenue, given that it is informed by physics and conservation principles. The model's volatility also aligns more closely with the variability in time-distance pricing (Figure 94). In addition, it more distinctly captures rush-hour price surges and anticipates higher prices in the early post-midnight hours (patterns that appear more natural). This observation could as well be seen as evidence that the spatial behaviors accommodated by the PASER model, described in Section 5.2, are effectively captured.

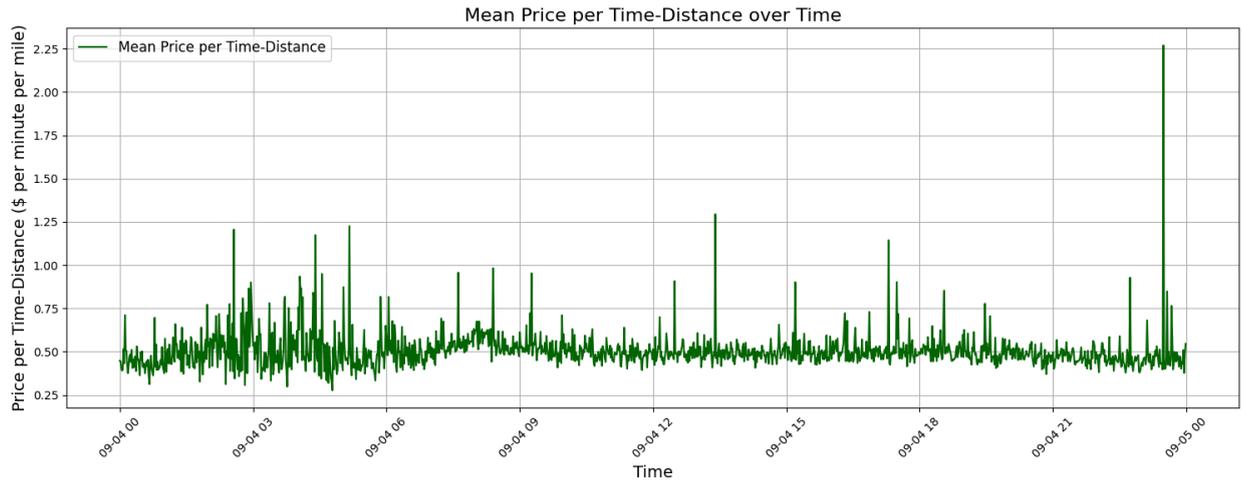

Figure 94. Uber, 4 September 2024, NYC. Per minute aggregated, time-distance empirical surge pricing.

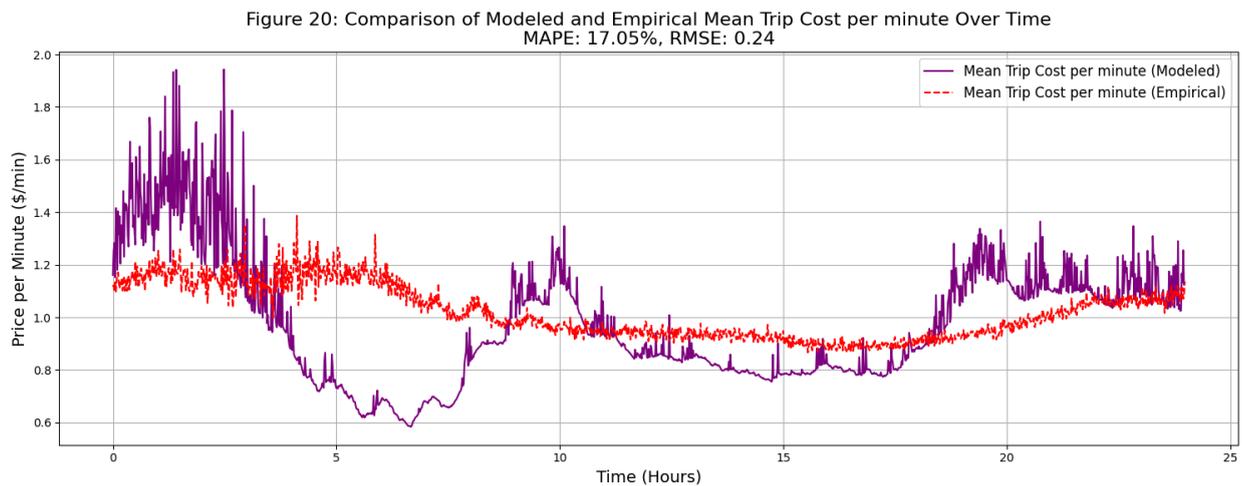

Figure 95. Uber, 4 September 2024, NYC. Per minute aggregated surge pricing.

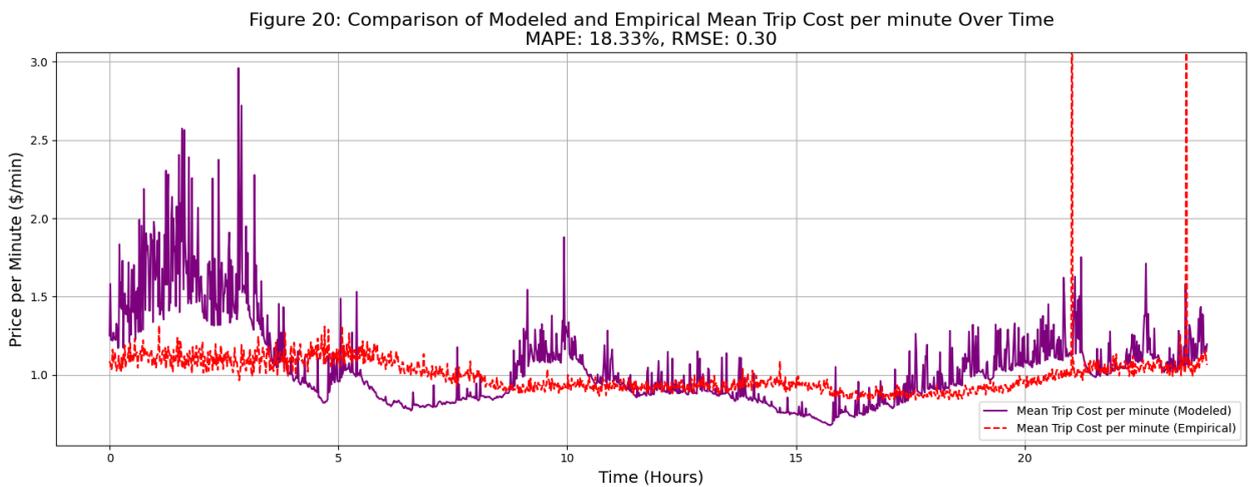

Figure 96. Lyft, 4 September 2024, NYC. Per minute aggregated surge pricing.

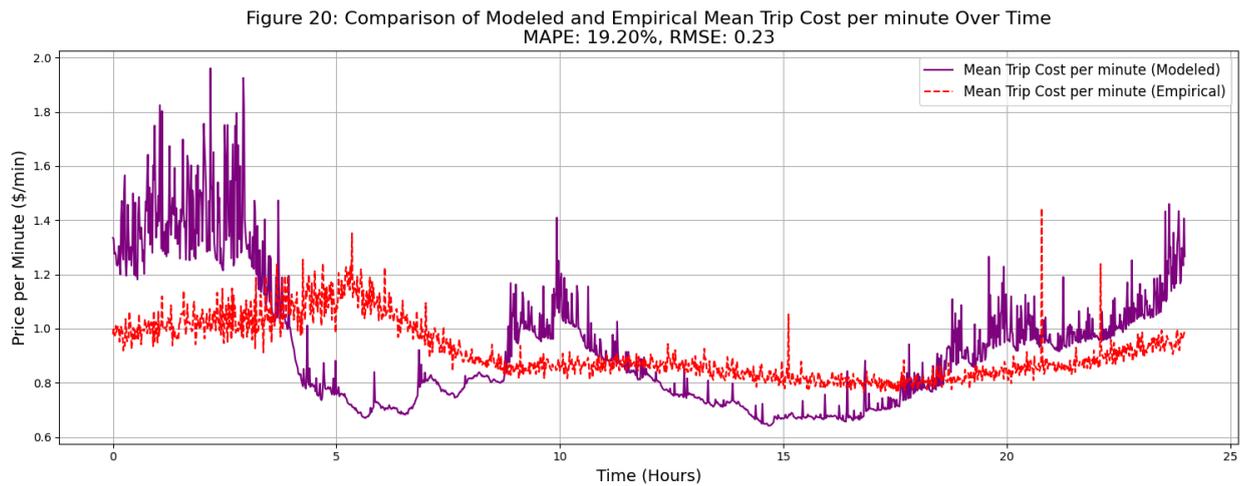

Figure 97. Uber, 3 July 2019, NYC. Per minute aggregated surge pricing.

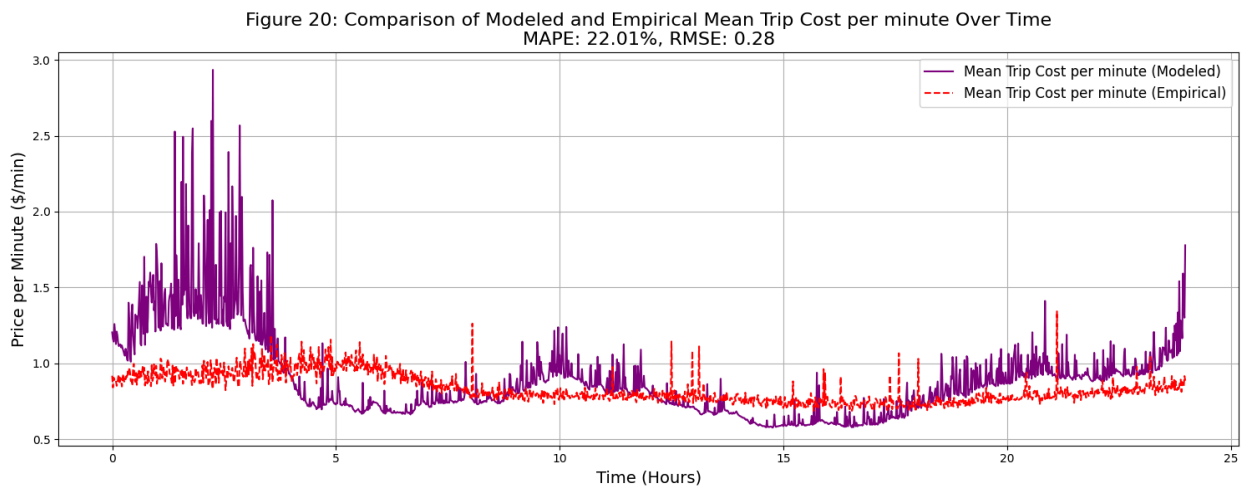

Figure 98. Lyft, 3 July 2019, NYC. Per minute aggregated surge pricing.

## 5.7 PLATFORM OPTIMIZATION FRAMEWORK

All PASER parameters (e.g. $\Gamma$, $\alpha$, $\sigma$, $\beta$, $\tau_c$, $W_p$, $\tau_{ij}$) constitute controllable platform dimensions impacting profitability, stability, and responsiveness. By analogy to laser physics, these "engineering" controls determine whether the system "lases" (i.e., achieves coherent profit) and how sustainably it does so. PASER's physics-based design allows for representing key platform parameters in a unified mathematical model at per-second granularity, capturing network effects, trust-building, and real-time revenue emission. It enables either analytical or numerical optimization of parameters, guiding design choices in surge pricing, driver incentives, and trust-building mechanisms (figure 99).

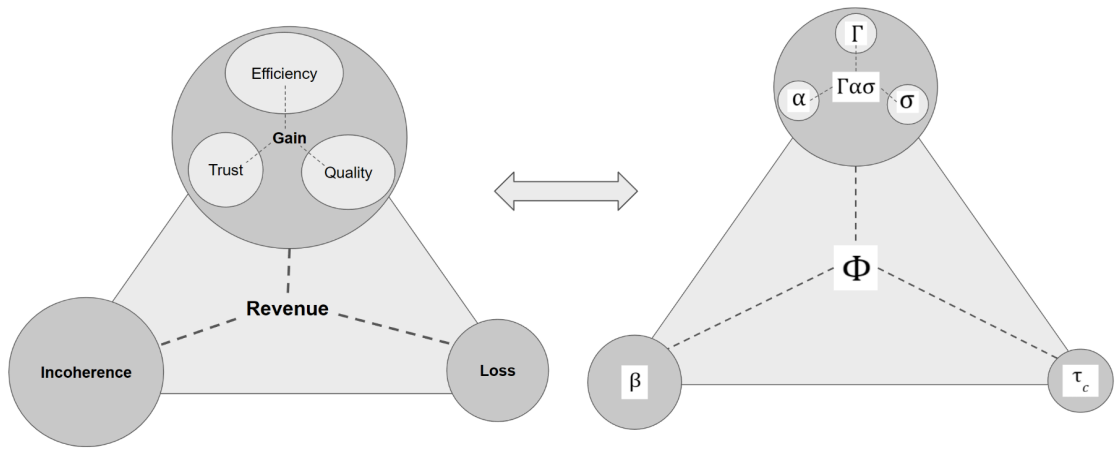

PASER revenue optimization

Figure 99. Depiction of a platform optimization framework offered by the PASER theory

## 5.8 GENERALIZATION OF PASER BEYOND UBER

Building on the concepts presented thus far, this section demonstrates how PASER can be extended from Uber's ride‑hailing operations (a B2C platform) to Uber Eats (a B2B2C platform) and beyond. It is established next that PASER is, by design, multimodal. It is then illustrated how different platforms may be captured by adapting the framework.

### 5.8.1 Uber Eats as a Multimodal PASER

In standard ride-hailing PASER (as discussed in Sections 5.1–5.7), the model focuses on one principal "source" of profit emission: completed rides. Uber Eats, by contrast, involves two main streams of revenue generation within its ecosystem. The revenue from restaurants and the revenue from couriers (drivers). This arrangement naturally lends itself to a multimodal PASER interpretation. In laser terms, one can envision a dual-mode laser that supports two frequencies (colors) simultaneously.

Figure 100 illustrates these additional elements. The on-restaurant transition (analogous to "on-scene" in Uber) now does emit revenue because the restaurant transaction itself generates a fee or commission. Subsequently, the order transitions to delivery (on-ride), producing a second "emission" event. Each revenue mode (restaurant vs. driver) is governed by potentially distinct parameters in the revenue equations because restaurant interactions and courier operations differ:

- $\Gamma_r \alpha_r \sigma_r$ for restaurants, capturing the technical platform efficiency, social efficiency (trust), and cross-section of stimulated revenue specifically in the restaurant segment.

- $\Gamma_d \alpha_d \sigma_d$ for couriers/drivers, reflecting the same conceptual parameters but tied to the delivery side of the transaction.

Accordingly, two revenue equations (one for restaurant revenue, one for delivery revenue) emerge, each with its own gain term, spontaneous term, and decay term:

$$\frac{d\Phi_r}{dt} = \Gamma_r \alpha_r \sigma_r \left(N_4 - N_3\right) \Phi_r + \beta_r \left(\frac{N_4}{\tau_{43}}\right) - \frac{\Phi_r}{\tau_{c_r}} \tag{9}$$

$$\frac{d\Phi_d}{dt} = \Gamma_d \alpha_d \sigma_d \left(N_3 - N_2\right) \Phi_d + \beta_d \left(\frac{N_3}{\tau_{32}}\right) - \frac{\Phi_d}{\tau_{c_d}} \tag{10}$$

The model thus seamlessly expands to eight coupled ODEs (the original six vehicle‑state transitions plus these two new revenue Equations (9) and (10)). Conceptually, the two mirrored apps remain (driver and consumer) but an additional "mirror" (the restaurant app) is introduced.

Hence, the perception that Uber Eats is a multimodal PASER is conceptually correct: it enables two coherent revenue channels to be emitted in parallel, each with distinct properties.

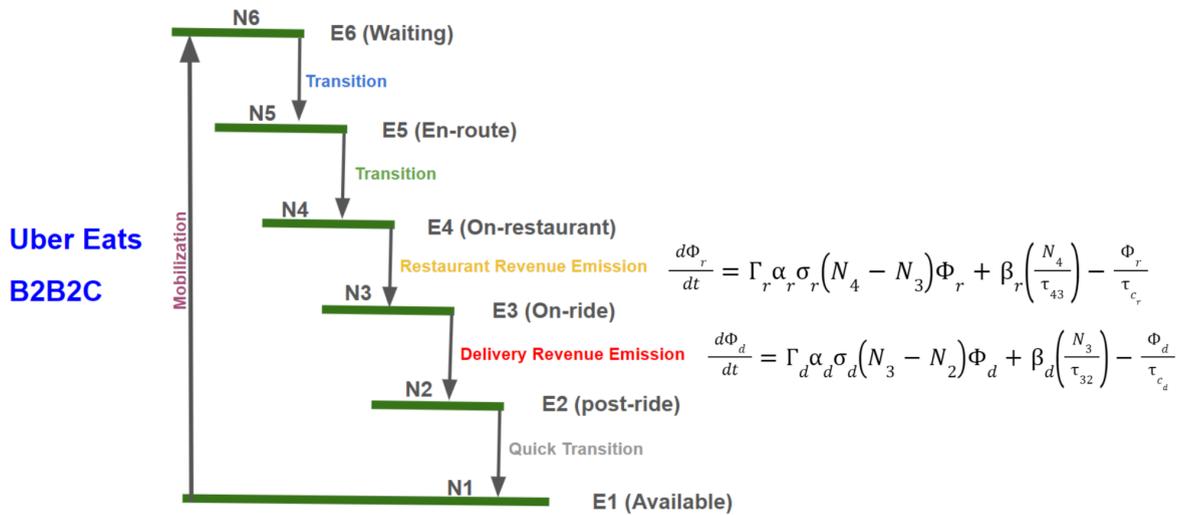

Figure 100. Illustration of the functionality of the Uber Eats PASER

### 5.8.2 Airbnb as a PASER With Longer Timescales

The PASER framework can also adapt to lodging platforms such as Airbnb, where time horizons differ dramatically from minute-level ride-hailing transactions. Airbnb may operate on a scale of hours or days. The booking process in Airbnb is akin to the matching process in Uber. The listings with their owners compose the revenue emitting medium in place of the vehicles and drivers in Uber. A listing can be online or offline and may transition in a closed loop through states like available, awaiting customer, check-in, on-stay, check-out, cleaning, available.

The revenue equation remains conceptually similar, with a "gain term" linked to the difference in population inversion indicating how the platform amplifies bookings once a critical mass is reached. The $\alpha$ and $\Gamma$, $\sigma$ parameters represent trust (reviews, verifications) and technical efficiency (booking interface quality, search functionality, algorithmic quality). Despite extended timescales, the fundamental logic of network effects amplification, population inversion, stimulated vs. spontaneous revenue, and revenue decay due to stays that do not amplify the mean of a gaussian fit to the stays distribution.

### 5.8.3 Goods Delivery Platforms

Beyond ride-hailing and food delivery, any goods-delivery cycle can similarly map onto the PASER framework. Consider a parcel delivery application connecting couriers (supply) with senders or recipients (demand). Typical states might be: awaiting-pickup, in-transit, delivered, return/logistics (analogous to post-ride in Uber). The essential physics-based principles and equations remain intact.

### 5.8.4 Autonomous taxi hailing as a single mode PASER

An exclusively AV-based ride-hailing system could be simulated by PASER with minimal adjustments of Uber's PASER. Setting $\alpha=1.0$ assumes maximum trust and immediate vehicle responsiveness, removing driver constraints. A simulation of an exclusively AV vehicle based Uber (September 2024 data driven) indicates a 28% revenue increase in a four-hour morning shift (figure 101). Figure 102 indicates a 81% revenue increase via a what-if scenario where the exclusively AVs base would also push the technical efficiency close to 100% due to the maximized trust and technical advancements.

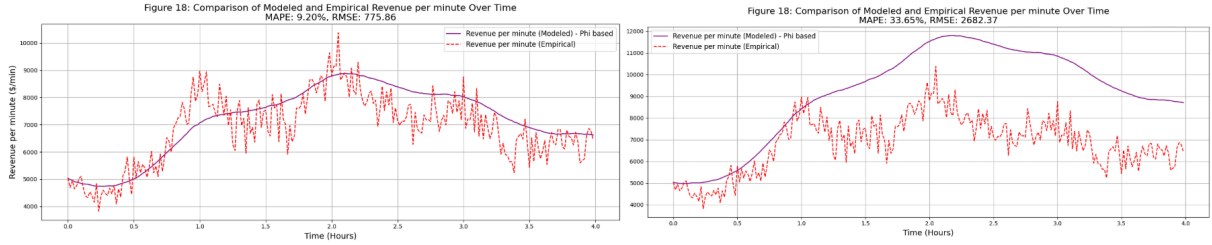

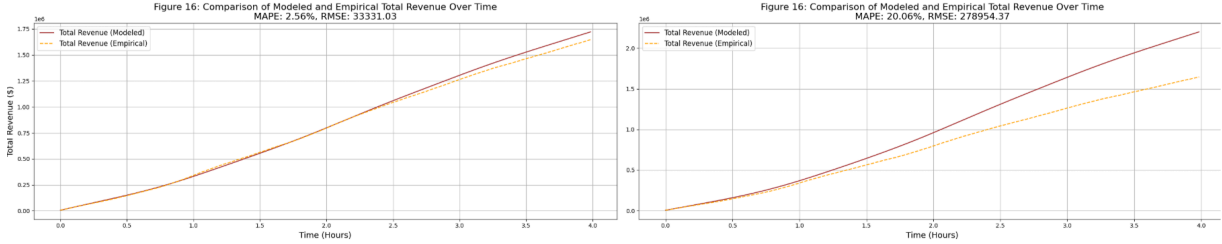

(Sep 4, 2024, 07:00-11:00) alpha=0.6 & gamma = 0.7→ Total revenue generated over simulation: $1721143.67

(Sep 4, 2024, 07:00-11:00) alpha=1.0 & gamma = 0.7 → Total revenue generated over simulation: $2200691.53

Figure 101. Uber's total revenue from 7 am to 11 am could increase by 28% if trust is maximized.

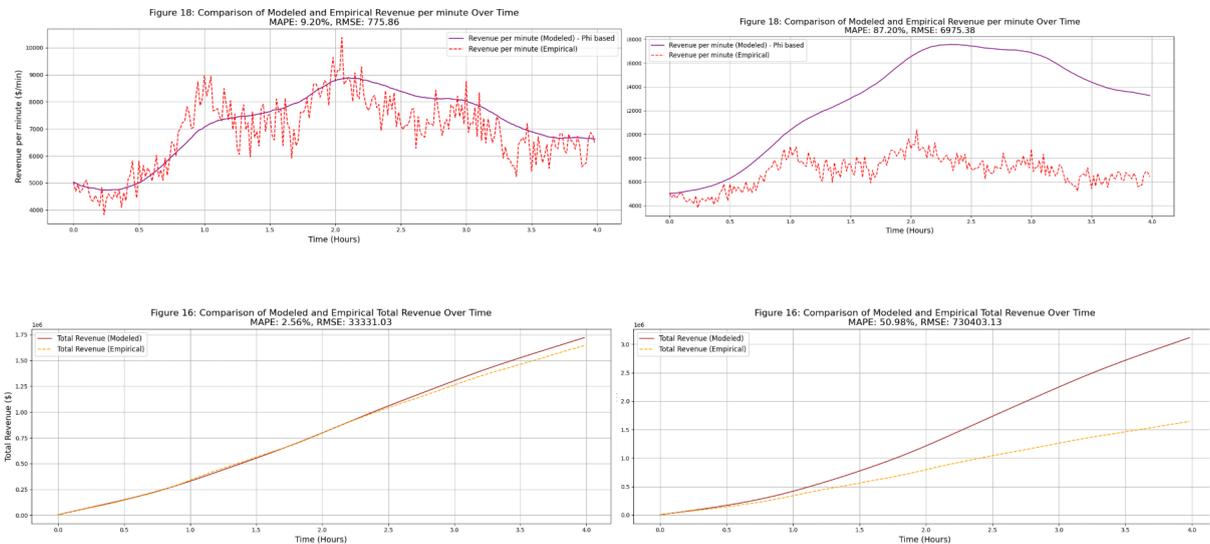

(Sep 4, 2024, 07:00-11:00) alpha=0.6 & gamma = 0.7→ Total revenue generated over simulation: $1721143.67

(Sep 4, 2024, 07:00-11:00) alpha=1.0 & gamma = 1.0 → Total revenue generated over simulation: $3117813.55

Figure 102. Uber's total revenue from 7 am to 11 am could increase by 81% if trust and efficiency are maximized.

### 5.8. Futuristic Domestic Robot Hailing

A PASER scenario for a futuristic on-demand platform could involve domestic robots hailing. AVs could be delivering the domestic robots to homes for housekeeping tasks. This could be running as a dual-revenue mode B2B2C platform where AV owners and robot owners are collaborating (a three-sided platform). The essential structure of stimulated network effects, mobilization, population inversion, and stimulated vs. spontaneous emission in each revenue mode remains identical.

In short, from Uber to Uber Eats, from Airbnb to goods delivery and beyond, the LABEL framework is flexible enough to integrate multiple states, multiple revenue streams, and widely differing time horizons. The underlying mathematics remain consistent offering generalizability.

### 5.9 SECTION SUMMARY

Section 5 confirms that a physics-driven model can unify the complexities revealed by the empirical data. It can reliably predict platform behaviors and guide interventions like surge pricing. These findings point toward a new paradigm for platform design, They provide a bridge between intangible trust mechanisms and purely algorithmic solutions. The final section next integrates these insights. It also addresses the broader contributions, limitations, and future directions that emerge from the paper.

## 6.     CONCLUSION

This final section draws together the research aims, findings, and theoretical advancements presented throughout this paper. It reflects on how each research question was addressed and what conclusions can be claimed. It also discusses the limitations that emerged during the investigation and points to areas for further research.

## 6.1 ADDRESSING THE RESEARCH QUESTIONS AND OBJECTIVES

**Central Aim and Research Questions.**

From the outset, this work sought to demystify the inner workings of digital, on-demand B2C and B2B2C platforms by developing a physics-inspired, data-driven framework capable of quantifying their core dynamics. This overarching vision can be distilled into the question: How do such platforms monetize the intricate interactions of supply, demand, trust, and positive network effects in real time? Section 4 offered emergent insights from raw data and section 5 provided the data induced PASER theory and its thorough validation. Showing that the PASER model indeed offers a powerful lens for explaining platform growth, real-time revenue generation, and stability thresholds across both normal and shock conditions.

**O1: Review the Literature.**

Section 2 synthesized a broad body of scholarship surrounding digital platforms. It highlighted the need for a unifying macro-level perspective of dynamic revenue generation that integrates intangible factors like trust with the micro-level dynamics of network effects, and resource allocation. The literature review confirmed a gap that the PASER framework uniquely addresses: while queueing models, agent-based simulations, and dynamic programming techniques each tackle valuable sub-problems, they often lack a unifying principle akin to conservation laws and threshold effects found in physics.

**O2: Quantify Key Catalysts.**

Operational and socio-technical variables were identified and measured. Such as enroute, on-scene, on-ride durations, ride requests, state transitions, cross platform network effects, and transient surges. Measured dynamic network effects depicted that they act as catalysts for platform growth. Statistical distributions of ride durations, revenue patterns, and supply mobilization (Sections 4.1–4.2) confirmed the platform's tendency toward "supplier population inversion," a physics-like threshold phenomenon that triggers rapid profitability once a critical mass is reached.

**O3: Construct and Validate the Physics-Inspired Theory.**

An analog to laser physics complete with a system of coupled core ODEs emerged inductively from the data in section 5. The two sided Uber platform was quantified by seven ODEs while the three sided Uber Eats was seamlessly quantified by eight coupled ODEs. Extensive data-driven validations  were conducted based on five years of high-volume trip records carefully retrieved with near 100% certainty. The model was rigorously validated through longitudinal testing, sensitivity testing, and stress testing. This proved that the principles of population inversion, amplified revenue emission via stimulated vs. spontaneous network effects, and a virtual "cavity" (formed by the apps and platform's infrastructure) have strong explanatory and predictive power.

**O4: Predictive and Policy-Relevant Modeling.**

It was shown that PASER delivers near-real-time predictive capabilities and serves as a substrate for scenario testing (Sections 5.6–5.8). Surge pricing emerges naturally from the equations, and "shock tests" highlight how critical mass can collapse or rebuild following disruptions. This level of detail addresses immediate industry concerns (such as how to calibrate trust-building mechanisms or driver incentives) and also guides policymakers on issues like labor transitions, especially in future scenarios where autonomous vehicles or domestic robots may shift labor patterns in the on-demand economy.

**O5: Enabling Generalization.**

Finally, Section 5.8 extended the framework beyond Uber ride-hailing to illustrate its applicability across multiple on-demand services, such as Airbnb, goods delivery, and futuristic robotics. The same essential mathematics (dynamic network effects, conservation laws, threshold-based gain, and revenue lifetime) appear to hold across platforms. Adapting time constants, parametric efficiencies, and the structure of transitions enables PASER to simulate a variety of B2C and B2B2C platforms.

## 6.2 KEY CONTRIBUTIONS AND CONCLUSIONS

A principal contribution of this work is the demonstration that a physics-inspired, minimalistic model can reliably capture the multi-faceted reality of on-demand digital platforms. The model offers a synergy with ML models by informing them with synthetic data while bounding them within physical realities. It is also compatible with being data driven by them. By linking micro-level states (driver transitions, trust signals) with macro-level profitability and platform stability, PASER consolidates disparate operational and strategic dimensions into a coherent analytical whole.

### 6.2.1 Contribution to Theory.

From the outset, this work sought to demystify the inner workings of digital, on-demand B2C and B2B2C platforms by developing a physics-inspired, data-driven framework capable of quantifying their core dynamics.

- Physics-based Unification: Borrowing concepts like population inversion and stimulated emission provides a unifying lens that quantifies and clarifies when and how network effects become self-reinforcing, under what conditions they collapse, and how intangible constructs like trust factor (α) integrate into the model.

- Scalable and Adaptable: PASER's ODE system remains computationally tractable at per-second or per-minute granularity. It also scales down to model geographically-small areas so as to model a big city via a swarm of PASERs based on CFD or swarm intelligence technologies while is also able to macroscopically model a mega city with a single PASER. This scalability is essential for bridging micro-level (agent states) with macro-level (overall platform revenue) phenomena.

- Translatability to Multiple Markets: By requiring minimal structural adjustments (just changing parameters or adding states) PASER readily adapts to new contexts such as Airbnb's lodging cycles or restaurant-courier multi-sided dynamics.

### 6.2.2 Contribution to Practice

- Real-Time Predictive Power: The model's capacity to forecast revenue via what-if scenarios, or revenue surges in real time can inform advanced dispatch, pricing, and driver mobilization strategies, potentially improving service quality and profitability.

- Transparency and Interpretability: The physics analogy lends interpretive clarity for managers, policymakers, and even drivers or couriers, going beyond the "black box" constraints of purely AI-driven systems. Moreover, it synergistically complements them by offering physics grounding, reliable synthetic data, and hallucination avoidance.

- What-if Scenario Analysis: PASER, by virtue of its small set of physically grounded parameters, offers a straightforward tool for hypotheticals: e.g., "What if trust α suddenly plummets by 20%?" or "What if half the fleet is replaced with AVs?". As demonstrated in Sections 5.6–5.8, these forecasts illuminate the system's sensitivity and potential pivot points.

From these results, it is concluded that digital platforms can often operate in a "laser-like" manner, harnessing positive feedback loops to amplify demand and supply into profitable revenue emission. This perspective can help managers demarcate stable operational regimes from precarious ones, spotlighting the role of trust, technical efficiency, and platform design quality.

## 6.3 LIMITATIONS

Despite encouraging validation outcomes, several limitations have been identified:

- Spatial Granularity: While PASER can embed spatial variables indirectly (e.g., by adjusting β to reflect spatially incoherent rides and $\tau_c$ for the outlier trips), the current implementation treats the city or region as a single "cavity." It is evident though that a single PASER can geographically scale down via forming a swarm of smaller area PASERs collectively modeling the whole city.

- Homogenization of Agents: PASER conceives of drivers, vehicles, or listings as homogeneous populations. In reality, user preferences and driver behavior can be more diverse (e.g., part-time vs. full-time drivers). A more advanced PASER could integrate segmented or "layered" populations with distinct behaviors.

- Parametric Tuning: Calibrating parameters like γ, α, and σ requires historical data of adequate size and certainty.

- Trust Modeling Simplifications: Although α acts as a social efficiency factor, real-world trust is multifaceted. A more granular trust sub-model could enhance predictive power..

- High-Frequency Shocks: The ODE framework sometimes "smooths out" abrupt transitions, like a curfew step-change. While it was shown that finer time steps (per-second) can reduce smoothing, filtering out the short lived noisy fluctuations may be seen as noise reduction.

These constraints are a reminder that PASER, while robust and generalizable, is an approximation whose parameter selection and state definitions must be adapted carefully to each new platform or scenario. They also highlight open questions about how best to integrate advanced features (such as spatio-temporal segmentation or deeper trust mechanisms) into future models.

## 6.4 DIRECTIONS FOR FUTURE RESEARCH

Several future directions suggest themselves:

- Multi-Area PASER Simulation: Dividing a metropolis into districts (each with a local sub-cavity) could illuminate how large platforms balance localized surpluses and deficits. Wrapping a swarm of PASER's in a CFD implementation for example could be simulating local spatial influx and outflux of drivers in their 'available'.

- Augmented Trust Modules: Building on α, one might incorporate dynamic user reviews, interpersonal trust cues, or pricing transparency to refine how trust shapes demand.

- Multi-Service Interaction: Section 5.8 hinted at concurrent modes of revenue in Uber Eats; an advanced extension might consider interactions among multiple concurrency modes (e.g., rides, food deliveries, and goods transport all within the same platform).

- Decentralized or Cooperative Models: With the rise of autonomous fleets owned by independent drivers, one might explore how partial decentralization impacts the key parameters (gamma, alpha) or modifies the "pumping rate" Wp (supplier mobilization).

- Integration with Machine Learning: While PASER is deliberately not "black box," it could supply synthetic training data for ML algorithms or serve as a transparent layer guiding deep reinforcement learning in driver repositioning tasks.

In each case, the fundamental architecture of PASER (dynamic network effects, conservation laws, threshold-based gain, conceptually minimal ODEs) would remain an asset, encouraging a modular approach where new states or submodels can be seamlessly added without sacrificing interpretability.

## 6.4 CONCLUDING REMARKS

This paper set out to quantify network effects and rigorously unify the operational logic of B2C and B2B2C digital platforms under a physics-inspired lens that synergises with isolated, algorithmic decision-making. By analogizing rides to photons and supplier mobilization to pumping electrons into excited states, the PASER framework offers a fresh viewpoint that has proven itself empirically, conceptually, and pragmatically. It not only clarifies how short-term shocks propagate through supply and demand but also explains why robust positive feedback loops can result in amplified, sustained, and coherent profit flows.

Looking forward, the rapid evolution of on-demand platforms (whether in transport, food delivery, lodging, goods distribution, or even domestic robotics) reinforces the importance of theoretical and practical tools that can adapt to new modes and new timeframes. PASER provides a structured approach: refine the states, calibrate the parameters, and capture how stimulated network effects amplify revenue generation dynamically and naturally. In doing so, it underscores a guiding principle of universal relevance: once a critical mass is reached, and once supplier population inversion (or profitable supplier positioning) is sustained, a platform can "lase" with minimal friction, scaling up profit generation dramatically.

Thus, while many challenges remain (from the intricacies of multi-area spatio-temporal modeling to the nuances of trust and policy) the PASER framework presented here can serve as a reliable foundation for both researchers and practitioners aiming to understand, design, and optimize the next generation of digital, on-demand platforms.

## ACKNOWLEDGEMENTS

I would like to thank Dr. Uma Urs (Oxford Brookes Business School) for her continuous support and invaluable feedback on my work. I also extend my gratitude to Dr. Birgit den Outer (Oxford Brookes Business School) for her encouragement and optimism throughout my research.

## APPENDIX A
## TABLES

**Table 2: Comparative Analysis of Light Emission and Revenue Generation**

| Aspect | Traditional Lamps (Businesses) | LASERs (Platforms like Uber) |
|---|---|---|
| **Energy Conversion** | ~10% to light, ~90% to losses | ~70% to light, ~30% to losses |
| **Emission Type** | **Diffuse**: Scattered in all directions | **Coherent**: Focused and directional |
| **Light Spectrum** | **Broad Spectrum**: Multiple wavelengths | **Monochromatic**: Single wavelength |
| **Resource Utilization** | **Inefficient**: High energy loss as heat | **Efficient**: Minimal energy loss |
| **Output Consistency** | **Unpredictable**: Inconsistent brightness | **Predictable**: Stable and consistent output |
| **Revenue Generation** | **Diffuse**: Scattered and unfocused efforts | **Coherent**: Focused and optimized efforts |
| **Profit Streams** | **Inconsistent**: Unpredictable and varied | **Scalable**: Consistent and amplified |

**Table 3: Cavity analogies between a LASER and a PASER (Uber, Lyft, Uber Eats, ...)**

| LASER Component | Uber Component |
|---|---|
| Laser Cavity | Platform Infrastructure |
| Gain Medium | Drivers and Vehicles in Uber |
| Excitation Input Energy | Ride Requests in Uber |
| Two Mirrors | The Two Uber Apps (Customer App and Driver App) |
| Energy Source | Ride Requests in Uber |
| Electrons | Drivers in Uber |
| Atoms | Vehicles in Uber |
| Electrons Start Cycling | Drivers Start Engaging |
| Photons | Rides that Generate Revenue |
| Gain in Photon Energy Increases | Intra-Cavity Revenue Increases |
| Gain Surpasses Losses in the Cavity | Critical Masses or Tipping Point |
| Partially Reflective Mirror | Uber Profits by Decoupling a Portion of Intra-Cavity Revenue |

**Table 4. Summary of analogies between LASER Components and Uber Platform Elements**

| LASER Component | Uber Platform Counterpart |
|---|---|
| Photons (period, wavelength, energy) | Rides (duration, distance, cost/revenue) |
| Gain Medium | Supply Agents (Drivers) |
| Laser Beam Coherence | Platform Profit Coherence |
| Lasing Threshold | Critical Mass |

| Quenching | Driver Oversupply |
|---|---|
| Stimulated Emission (Photons) | Stimulated Emission of Revenue (Rides) |
| Electron Population Inversion | Driver Population Inversion |
| Electron Energy Levels | Driver Excitation States |
| Pumping of Electrons | Driver Mobilization |
| Threshold Pump Power | Threshold Ride-Requests |
| Pumping Efficiency | Driver Mobilization Efficiency |
| Homogeneous Broadening | Homogeneous Driver Transitions |
| Linewidth (Spectral Purity) | Ride Mode Purity |
| Mirror Quality | App Quality |
| Modal Confinement | Main Ride Mode Confinement |
| Cavity Length & Mode Structure | App Tuning for Multiple Service Modes |
| Mode Competition | Competition Between Service Modes |
| Loss Mechanisms (Diffraction, Reabsorption) | Operational Inefficiencies (Slow Responses, Low Trust) |
| Thermal Management | Friction Management |
| Quality Factor | Platform Quality Factor |
| Intracavity Photon Density | Intra-Platform Revenue Density |
| Output Coupling | Revenue Share/Commission |
| Sustainability | Stable Platform Operation |
| Sustained & Amplified Lasing | Exponential Platform Growth |
| Laser Cavity | Platform Infrastructure |
| External Factors (Environment) | External Shocks (e.g., Regulation, Pandemic) |
| Alignment of Mirrors & Stability of Cavity | App Alignment & Algorithmic Stability |
| LASER Systemic Behavior | PASER Systemic Behavior |